\documentclass{jfm}
\usepackage[]{natbib}
\usepackage{graphicx}
\usepackage{epstopdf}
\usepackage{float}
\usepackage{wrapfig}
\restylefloat{figure} 
\usepackage{subcaption}
\usepackage[utf8]{inputenc}

\usepackage{float}
\usepackage{xcolor}
\usepackage{amsmath,rotating}
\usepackage{bm}
\usepackage{siunitx}
\usepackage{url}
\usepackage{hyperref}
\hypersetup{
    colorlinks=true,       
    linkcolor=blue,          
    citecolor=blue,        
    filecolor=blue,      
    urlcolor=black          
}

\usepackage{amssymb}

\newtheorem{lemma}{Lemma}

\newtheorem{definition}{Definition}

\usepackage{footmisc}
\usepackage{comment}
\usepackage{eurosym}
\usepackage{cleveref}
\usepackage{tikz}
\usepackage[percent]{overpic}
\usepackage[normalem]{ulem}
\usepackage{flafter}
\usepackage{xfrac}
\usepackage{textcomp}

\definecolor{rottengreen}{rgb}{0, 0.49, 0.0}

\newcommand{\fh}[1]{{\color{black} #1}} 

\newcommand{\Jav}{\langle\mathcal{J}\rangle}
\newcommand{\J}{\mathcal{J}}

\usetikzlibrary{backgrounds}
\usetikzlibrary{calc}
\DeclareCaptionFormat{suboverlay}{\gdef\subcapoverlay{(\thesubfigure) #3\par}}
\DeclareCaptionStyle{suboverlay}{format=suboverlay}
\newcommand{\subcaptionOverlay}[1]{%
  \subcaption{}%
  \begin{tikzpicture}
    \node [inner sep=0] (image) {#1};
    \draw node [black] at (current bounding box.north west) {\subcapoverlay};
  \end{tikzpicture}%
}
\DeclareRobustCommand*{\tikzbar}[1]{%
   \begin{tikzpicture}
     \useasboundingbox (-.15em,-.25em) rectangle (.15em,\ht0);
     \filldraw[color=#1,fill=#1] (-0.15em,-0.25em) rectangle (0.15em, 0.25em);
   \end{tikzpicture}%
}

\DeclareRobustCommand*{\tikzline}[1]{%
   \begin{tikzpicture}
     \useasboundingbox (-0.8em,0) rectangle (0.8em,\ht0);
     \filldraw[color=#1,fill=#1] (-0.8em,0.22em) rectangle (0.8em, 0.28em);
   \end{tikzpicture}%
}

\definecolor{blue1}{HTML}{4A98C9}
\definecolor{blue3}{HTML}{08306B}
\definecolor{c1}{HTML}{377EB8}
\definecolor{c2}{HTML}{E41A1C}
\definecolor{c3}{HTML}{4DAF4A}
\definecolor{c4}{HTML}{984EA3}

\shorttitle{Stability, sensitivity and optimisation of chaotic acoustic oscillations}
\shortauthor{F. Huhn and L. Magri}

\title{Stability, sensitivity and optimisation of chaotic acoustic oscillations}
\author{Francisco Huhn\aff{1}
 \and Luca Magri\aff{1}\corresp{\email{lm547@cam.ac.uk}}}

\affiliation{\aff{1}Cambridge University Engineering Department, Trumpington Street, Cambridge CB2 1PZ, UK}

\begin{document}

\maketitle

%
\begin{abstract}
In an acoustic cavity with a heat source, such as a flame in a gas turbine, the thermal energy of the heat source can be converted into acoustic energy, which may generate a loud oscillation. 
If uncontrolled, these nonlinear acoustic oscillations, also known as thermoacoustic instabilities, can cause large vibrations up to structural failure.
Numerical and experimental studies showed that thermoacoustic oscillations can be chaotic. It is not yet known, however, how to minimise such chaotic oscillations.   
We propose a strategy to analyse and minimise chaotic acoustic oscillations, for which traditional stability and sensitivity methods break down. We investigate the acoustics of a nonlinear heat source in an acoustic resonator. 
First, we propose covariant Lyapunov analysis as a tool to calculate the stability of chaotic acoustics making connections with eigenvalue and Floquet analyses. We show that covariant Lyapunov analysis is the most general flow stability tool.
Second, covariant Lyapunov vector analysis is applied to a chaotic system. The time-averaged acoustic energy is investigated by varying the heat-source parameters. Thermoacoustic systems can display both hyperbolic and non-hyperbolic chaos, as well as discontinuities in the time-averaged acoustic energy.
Third, we embed sensitivities of the time-averaged acoustic energy in an optimisation routine. This procedure achieves a significant reduction in acoustic energy and identifies the bifurcations to chaos. 

The analysis and methods proposed enable the reduction of chaotic oscillations in thermoacoustic systems by optimal passive control. The techniques presented can be used in other unsteady fluid dynamics problems with virtually no modification.
\end{abstract}
\section{Introduction}
\label{sec:intro}
%

Gas-turbine and rocket-motor manufacturers strive to design engines that do not experience thermoacoustic instabilities  \citep{Lieuwen2005,Culick2006,Dowling2015,Poinsot2017,juniper_sujith_2018}. Thermoacoustic instabilities occur when the heat released by the flame is sufficiently in phase with the acoustic pressure~\citep{rayleigh_1878} such that the thermal energy of the flame that is converted into acoustic energy exceeds dissipation mechanisms. 
The first objective of manufacturers is to design a thermoacoustic system in which small acoustic perturbations decay after some time, i.e. all the eigenvalues are stable. Eigenvalue analysis is routinely used in industrial preliminary design and parametric studies because it can be run quickly~\citep[e.g.][]{Lieuwen2005,Magri2019_amr}. 
%
However, when nonlinearities become active, thermoacoustic systems exhibit rich behaviours both via supercritical bifurcations, i.e. when an eigenvalue becomes unstable, and subcritical bifurcations, i.e., when eigenvalues are stable but the nonlinearity is triggered by finite-amplitude perturbations~\citep{Subramanian2011}. When the bifurcation parameter is varied, thermoacoustic systems may display periodic, quasi-periodic and   chaotic oscillations~\citep{Kabiraj2011,Gotoda2011,Gotoda2012,Kabiraj2012,Kashinath2013,Waugh2014,Nair2014,Nair2015,Orchini2015a}. Whereas methods to investigate the stability and sensitivity of fixed points and periodic solutions are well established~\citep[e.g.][]{juniper_sujith_2018,Magri2019_amr}, a stability and sensitivity framework to tackle chaotic acoustic oscillations is not available yet. This paper proposes a framework for stability and sensitivity analysis of chaotic acoustic oscillations.  

%
In thermoacoustics, chaotic acoustic oscillations originate from two main physical nonlinearities, which are deterministic. 
First, the heat released by the flame is a nonlinear function of the acoustic perturbations at the flame's base, i.e. the flame saturates nonlinearly~\citep{Dowling1997,Dowling1999}.  
Both experimental investigations \citep{Gotoda2011,Kabiraj2011, Gotoda2012, Kabiraj2012} and numerical studies \citep{Waugh2014,Kashinath2013,Orchini2015a} showed that the nonlinear flame saturation may cause a periodic acoustic oscillation to become chaotic, by either period doubling, or Ruelle--Takens--Newhouse, or intermittency scenarios~\citep{Nair2014,Nair2015}, which are common to  other fluid dynamics systems~\citep{Eckmann1981,Miles1984,Eckmann1985}. The numerical studies of~\citet{Waugh2014,Kashinath2013,Orchini2015a} showed that the nonlinear flame saturation may generate chaotic acoustic oscillations even in laminar flame models, where the turbulent hydrodynamics is not modelled. 
Second, the geometry of the combustor promotes hydrodynamic instabilities, such as vortex shedding and shear-layer instabilities~\citep{Lieuwen2012}, which result in energetic coherent structures. In turbulent combustors, turbulence unpredictably modulates the dynamics of coherent structures, which, in turn, unpredictably modulate the flame dynamics, thereby changing the heat release that feeds into the acoustics. This paper investigates the chaotic acoustics generated by the nonlinear response of the heat source. 
Although oscillations in thermoacoustic systems may be nonlinear and chaotic, industrial preliminary design is based on linear analysis~\citep{Lieuwen2005,juniper_sujith_2018}: the first objective is to design eigenvalue-stable thermoacoustic systems.  
%
Sensitivity methods have recently been developed to calculate the effect that a small change to the system has on the eigenvalue, as reviewed by~\citet{Magri2019_amr}.
Sensitivity analysis quantitatively informs the practitioner, among others, on  
(i) how to optimally change design parameters, such as geometric quantities~\citep{Magri2013c}; 
(ii) which passive device is most stabilising~\citep{magri_juniper_2013}; 
and (iii) how large the uncertainty of the stability calculations is~\citep{Magri2016c,Silva2016,Mensah2018}. 
When the gradient provided by sensitivity analysis is embedded in an optimisation routine, it is possible to calculate the optimal arrangement of acoustic dampers~\citep{Mensah2017a} and a stable set of geometric parameters~\citep{Aguilar2018_asme}.  
%
%
%
However, 
eigenvalue analysis is necessary but not sufficient to prevent large acoustic oscillations. This is the case of subcritical bifurcations, where the system can self-sustain finite-amplitude oscillations in the bistable region, where all eigenvalues are stable. 
%
%
In this paper, we provide a method to calculate the sensitivity of chaotic acoustic oscillations, which is the most general nonlinear scenario, to minimise their energy. 
First, we need to define a quantity of interest of which we want to calculate the sensitivity in a chaotic oscillation. In aperiodic flows, a quantity of interest is the time average of a cost functional, $\J$, 
\begin{align}
    \label{eq:Jav}
    \langle \J(\bm{s}) \rangle \triangleq \lim_{T \rightarrow \infty} \frac{1}{T} \int_0^T \J(\bm{s}; \bm{q}(t)) dt,
\end{align}
where $\bm q$ is the state vector, $t$ is the time, $\langle\cdot\rangle$ represents the time average operation, which is equal to the expected value in ergodic systems (Birkhoff ergodic theorem \citep{Birkhoff1931}), and $\bm s$ is the parameters' vector. 
Physically, $\J$ may be an acoustic energy, which we want to minimise to make the combustor operate in stable conditions.
Therefore, the objective is to calculate the sensitivity of the time-averaged cost functional given a perturbation to the parameters' vector, i.e. $\nabla_{\bm{s}}\langle \J \rangle$. 
Whereas the sensitivity analysis of eigenvalues is robust, traditional sensitivity methods fail in chaotic systems because of the butterfly effect~\citep{lea_2000}, see \S\ref{sec:shadowing_lemma}.
%
%
%
Shadowing methods have recently been proposed~\citep{wang_2013,wang_2014,ni_2017} to carry out sensitivity analysis of chaotic systems as a more efficient alternative to ensemble methods~\citep{Eyink2004}. 
By noting that changing a parameter of a chaotic system has a similar effect to changing the initial condition, shadowing methods find a perturbed (shadow) trajectory that does not diverge from the unperturbed trajectory. 
Such a trajectory is guaranteed to exist by the shadowing lemma \citep[e.g.][]{katok_hasselblat,Holmes1996,pilyugin2006shadowing} and the sensitivity calculation is enabled because the expectation~\eqref{eq:Jav} is a smooth function of the parameters in hyperbolic dynamical systems, as explained in Ruelle's linear theory~\citep{ruelle_2009}.  
A hyperbolic strange attractor is an invariant set whose tangent space can be decomposed into stable, unstable and neutrally stable subspaces at almost every point.  One basis for this decomposition consists of the covariant Lyapunov vectors \citep{Ginelli2007,ginelli_2013}. Hyperbolic attractors are also ergodic and, importantly, they have differentiable expectations~\citep{ruelle_2009}, $\Jav$, whereas non-hyperbolic systems may not. Thus, the sensitivities of time-averaged cost functionals are well defined in hyperbolic systems, but may be ill defined in non-hyperbolic systems. For chaotic sensitivity methods to work in thermoacoustics, it is crucial that the hyperbolicity assumption is verified. In this paper, first, we introduce covariant Lyapunov vector analysis as a generalisation of traditional flow stability analysis. 
It is mathematically and numerically shown that covariant Lyapunov vector  analysis becomes eigenvalue analysis when the attractor is a fixed point, and becomes Floquet analysis when the attractor is a periodic orbit. 
Third, we show that the system admits both hyperbolic and non-hyperbolic chaotic solutions.
Fourth, by combining covariant Lyapunov vector analysis and the non-intrusive least squares shadowing method \citep{ni_2017}, we embed the sensitivities of the time-averaged acoustic energy to the heat-source parameters in a gradient-based optimisation algorithm to minimise the energy of the oscillation.
Fifth, we suggest how the methods we propose can be used for the suppression of acoustic oscillations in high-fidelity design.

The paper is structured in two parts. 
The first part is theoretical and is kept as general as possible. 
\S\ref{sec:theory} recalls the concept of Lyapunov exponents, covariant Lyapunov vectors and the numerical algorithm for computing them. 
In \S\ref{sec:eigen} and \S\ref{sec:floquet}, we show analytically that fixed-point and Floquet analyses are subsets of covariant Lyapunov vector  analysis, which are general results.
The second part applies the theory to a chaotic acoustic system with a heat source (\S\ref{sec:systems}).
The covariant Lyapunov vector analysis of the thermoacoustic model is presented in \S\ref{sec:results:rijke}. 
Finally, a gradient-based optimisation is performed in \S\ref{sec:optim} to minimise a time-averaged cost functional. 
The paper ends with suggestions for future work and a summary of the main results in~\S\ref{sec:conclusion}. 
%
%
\section{Covariant Lyapunov vector analysis}
\label{sec:theory}
This section introduces the key concepts to perform stability and sensitivity analysis of chaotic thermoacoustic systems. %
In particular, we present the key results of Oseledets' theorem~\citep{oseledets_1968} to lay out the fundamentals of covariant Lyapunov vector analysis~\citep{ginelli_2013}, which has recently seen interest from the fluid dynamics community~\citep{Inubushi2015,Schubert2015,Xu2016}.
(Note that covariant Lyapunov vector analysis has nothing to do with Lyapunov stability analysis based on Lyapunov functions, which is used, for example, in control theory.)

\subsection{Lyapunov exponents}
\label{sec:theory:le}
The thermoacoustic problem is governed by partial differential equations, i.e. the compressible Navier-Stokes equation with equations for the chemistry,  and mass and energy conservation. After spatial discretisation, the thermoacoustic problem is formally an autonomous dynamical system 
\begin{equation}
    \label{eq:system}
	\begin{cases}
		\dot{\bm{q}}(t) = \bm{F}(\bm{q}(t)) \\
        \bm{q}(0) = \bm{q}_0
	\end{cases}
\end{equation}
where the overdot $\dot{(\,)}$ is Newton's notation for time differentiation; $\bm{q}\in\mathbb{R}^{N}$ is the state vector (e.g. pressure and velocity at each discrete location), where the integer $N$ denotes the discrete degrees of freedom; the subscript $0$ denotes the initial condition; and $\bm F: \mathbb{R}^N \rightarrow \mathbb{R}^N$ is a 
nonlinear smooth function, which encapsulates the discretised boundary conditions. 
We are interested in the evolution of small perturbations, therefore  we split the solution as 
\begin{align}\label{eq:split_sol}
\bm{q}(t) = \bar{\bm{q}}(t) + \bm{q}'(t), 
\end{align}
where $\bar{\bm{q}}(t)$ is the unperturbed solution of~\eqref{eq:system} such that $\lvert\lvert\bar{\bm{q}}(t)\rvert\rvert\sim O(1)$, and $\bm{q}'(t)$ is the small perturbation such that $\lvert\lvert\bar{\bm{q}}'(t)\rvert\rvert\sim O(\epsilon)$, where $\epsilon\rightarrow0$. The perturbation is governed by the tangent equation 
\begin{align}\label{eq:tangent_eq}
\begin{cases}
\dot{\bm{q}}' =  \bm{J}(t) \bm{q}', \\
\bm{q}'(0) = \bm{q}'_0, 
\end{cases}
\end{align}
where $\bm{J}(t)\equiv\frac{d\bm{F}}{d\bm{q}}\big|_{\bar{\bm{q}}(t)}$ is the Jacobian. 
To define the Lyapunov exponents, it is convenient to introduce the tangent propagator, which maps the perturbation, $\bm{q}'$, from time $t$ to time $\tilde{t}$, as
\begin{align}
    \bm{q}'(t+\tilde{t}) = \bm{M}(t,\tilde{t}) \bm{q}'(t).  
\end{align}
The tangent propagator is governed by the matrix equation 
\begin{equation}
\begin{cases}
	\frac{d\bm{M}}{d\tilde{t}} = \bm{J}(\tilde{t}) \bm{M}, \\
	\bm{M}(t,0) = \bm{I}, 
	\end{cases}
    \label{eq:tangent_system}
\end{equation}
where $\bm{I}$ is the identity matrix. 
%
%
Setting $t=0$ without loss of generality, the norm of an infinitesimal perturbation, $\bm{q}_0'$, to the initial condition, $\bar{\bm{q}}_0$, asymptotically grows (or decays) as  
\begin{equation}
	||\bm{q}'(\tilde{t})|| \cong ||\bm q_0'|| e^{\lambda(\bm{q}'_0 , \bar{\bm q}_0) \tilde{t}},
\end{equation}
where $\cong$ means ``asymptotically equal to", and
\begin{equation}
	\label{eq:lyapunov_exponent}
	\lambda(\bm{q}'_0 , \bar{\bm{q}}) = \lim_{\tilde{t} \rightarrow \infty} \frac{1}{\tilde{t}} \log \frac{||\bm{M}(0, \tilde{t}) \bm{q}'_0 ||}{||\bm{q}'_0 ||} 
\end{equation}
is the characteristic Lyapunov exponent. 
%
Oseledets' theorem~\citep{oseledets_1968} shows that there exist $m \leq N$ distinct Lyapunov exponents $ \lambda_1(\bar{\bm{q}}) > \lambda_2(\bar{\bm{q}}) > \cdots > \lambda_m(\bar{\bm{q}})$, which provide a filtration of the tangent space $\mathcal{T}_{\bar{\bm{q}}}$, into subspaces $S_i$, i.e. $\mathcal{T}_{\bar{\bm q}} \equiv S_1 \supset S_2 \supset \cdots \supset S_m$, such that $\bm{q}'_0 \in S_j \backslash S_{j+1} \Leftrightarrow \lambda(\bm{q}'_0 , \bar{\bm{q}}) = \lambda_j(\bar{\bm{q}})$. 
Furthermore, Oseledets' theorem shows that $\lambda_j(\bar{\bm q})$ are constants of the attractor $\bar{\bm{q}}$, and, in ergodic systems, $\lambda_j$ do not depend on the initial condition, $\bar{\bm{q}}_0$. 
Physically, the Lyapunov exponents are the average exponential contraction/expansion rates of an infinitesimal volume of the phase space moving along the attractor.  
For example, as shown in \S\ref{sec:eigen}, if the attractor is a fixed point, the Lyapunov exponents are equal to the real part of the eigenvalues of the Jacobian at the fixed point. Similarly, as shown in \S\ref{sec:floquet}, if the attractor is a limit cycle, the Lyapunov exponents are equal to the real part of the Floquet exponents. 
%
%
\subsection{Oseledets splitting and covariant Lyapunov vectors}
\label{sec:theory:clv}
\begin{figure}[tb]
	\centering
	\includegraphics[width=0.7\textwidth]{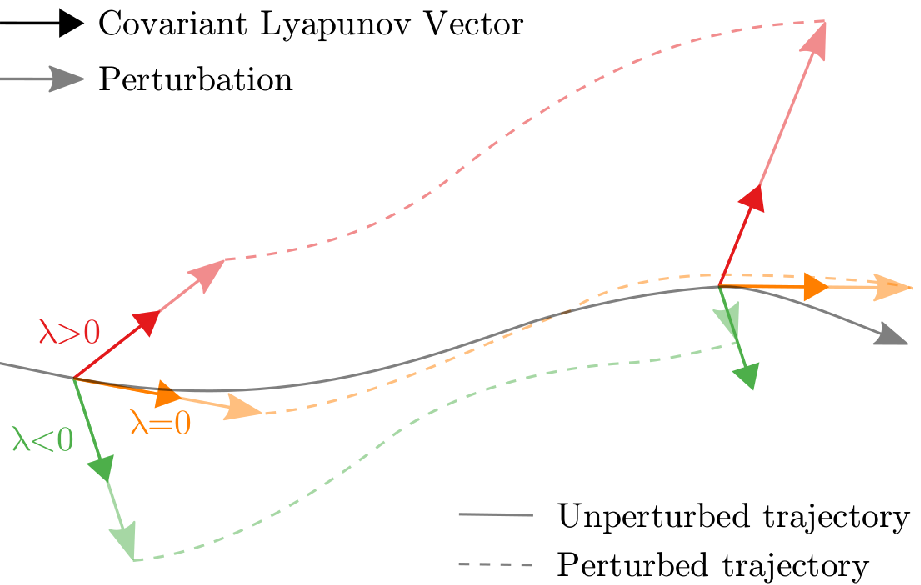}
    \caption{Schematic diagram of covariant Lyapunov vectors and perturbations on an unperturbed trajectory (solid grey line). Three covariant Lyapunov vectors are shown at two different instants, each associated with a different Lyapunov exponent, which can be positive, zero or negative. The decay/growth of three perturbations, along the stable (green), neutral (orange), unstable (red) covariant Lyapunov vector, respectively. The resulting perturbed trajectories (dashed lines), converge, remain at a constant distance, or diverge, respectively, to/from the unperturbed trajectory, depending on the sign of the Lyapunov exponent. This explains why trajectories emanating from two very close initial conditions will almost surely diverge in chaotic systems -- it is highly unlikely for the vector connecting the two initial conditions not to have a component in the direction of the unstable covariant Lyapunov vector.}
    \label{fig:pert_growth_sketch}
\end{figure}

The Lyapunov exponents are invariant measures of the attractor, however, they do not inform on the directions along which the infinitesimal volume of the phase space contracts/expands. 
%
Such directions are provided by the Oseledets splitting, which is composed by the Lyapunov subspaces, which are, in turn, spanned by the covariant Lyapunov vectors. First, the Oseledets matrix~\citep{oseledets_1968} is defined as
\begin{equation}
	\label{eq:oseledets_operator}
    	\bm{\Xi^{\pm}}(t) = \lim_{\tilde{t} \rightarrow \pm \infty} \frac{1}{2\tilde{t}} \log \left[ \bm{M}(t,\tilde{t})^T  \bm{M}(t,\tilde{t}) \right]. 
\end{equation}
This matrix is called ``forward'' if $\tilde{t} \rightarrow +\infty$ or ``backward'' if $\tilde{t} \rightarrow -\infty$. The spectrum of this matrix contains $m \leq N$ distinct eigenvalues, which are the Lyapunov exponents of the system $\lambda_1 > \lambda_2 > \cdots > \lambda_m$. However, the eigenvectors of the forward and backward matrices differ from each other and are not invariant under time reversal. 
To gain more insight in the Oseledets matrix,  
consider the singular value decomposition $\bm M(t, \tilde t) = \bm U \bm \Sigma \bm V^T$, where $\bm U$ and $\bm V$ are orthogonal matrices and $\bm \Sigma$ is a diagonal matrix with non-negative real entries (the singular values). We can obtain an eigenvalue decomposition of the argument of the logarithm of \eqref{eq:oseledets_operator}, $\bm M^T \bm M = \bm V (\bm \Sigma^T \bm \Sigma) \bm V^T = \bm V \bm \Sigma^2 \bm V^T$, which, after applying the logarithm, becomes $\bm V \log(\bm \Sigma^2) \bm V^T = 2 \bm V \log(\bm \Sigma) \bm V^T$. Thus, equation~\eqref{eq:oseledets_operator} can be rewritten as 2$\bm \Xi^\pm(t) = \lim_{\tilde{t} \rightarrow \pm \infty} \bm V (\log(\bm \Sigma(t,\tilde{t}))/\tilde{t}) \bm V^T$, which shows that the eigenvalues of $\bm \Xi^\pm$ are the Lyapunov exponents, which are equal to the exponential average of the singular values of $\bm M(t, \tilde{t})$.
%
Let $V_j^\pm(t)$ be the $j$-th eigenspace of the forward (backward) Oseledets matrix, then the Oseledets subspaces are defined as
\begin{align}
	\Gamma_j^+(t) &= V_j^+(t) \oplus \cdots \oplus V_m^+(t), \\
    \Gamma_j^-(t) &= V_1^-(t) \oplus \cdots \oplus V_j^-(t),
\end{align}
where $\oplus$ is the direct sum. The Oseledets subspaces have the property
\begin{equation}
	\lim_{\tilde{t} \rightarrow \infty} \frac{1}{\tilde{t}} \log \frac{||\bm{M}(t, \tilde{t}) \bm{q}'(t) ||}{||\bm{q}'(t) ||} = \lambda_j, \quad \text{for } \bm{q}'(t) \in \Gamma_j^\pm(t) \backslash \Gamma_{j+1}^\pm(t)
\end{equation}
and a nested structure $\mathbb{R}^N = \Gamma_1^+(t) \supset \Gamma_2^+(t) \supset \cdots \supset \Gamma_m^+(t) \supset \Gamma_{m+1}^+(t) \equiv \emptyset$ and $\mathbb{R}^N = \Gamma_m^-(t) \supset \Gamma_{m-1}^-(t) \supset \cdots \supset \Gamma_1^-(t) \supset \Gamma_0^-(t) \equiv \emptyset$.
By intersecting the Oseledets subspaces, we obtain the Lyapunov subspaces
\begin{equation}
	\Omega_j(t) = \Gamma_j^+(t) \cap \Gamma_j^-(t), \quad j \in \{1, \ldots, m\},
    \label{eq:oseledets_splitting}
\end{equation}
which compose the Oseledets splitting. The Lyapunov subspaces are (i) generally non-orthogonal to each other, (ii) covariant with the dynamics, i.e. $\bm{M}(t, \tilde{t}) \Omega_j(t) = \Omega_j(t+\tilde{t})$, and (iii) invariant under time reversal.
Each vector $\bm \phi_j(t)$ of a set that spans one of the Lyapunov subspaces is a covariant Lyapunov vector associated with the Lyapunov exponent $\lambda_j$. If a trajectory is infinitesimally perturbed at some time $t_1$ along a covariant Lyapunov vector, the perturbation will grow at an exponential rate dictated by the associated Lyapunov exponent and will stay aligned with that same covariant Lyapunov vector  (figure~\ref{fig:pert_growth_sketch}).
%
%
%

We derive the equation that governs the covariant Lyapunov vectors. First, because the Lyapunov subspaces are covariant with the dynamics, the following definition holds
\begin{align}
    \label{eq:covariant_eq} \bm M(t, \tilde{t}) \bm \phi(t) &= \eta(t, \tilde{t}) \bm \phi(t + \tilde{t}), \\
   \textrm{where}\;\;\;\;\;\;\;\label{eq:propagator} \bm M(t, \tilde{t})& = e^{\int_t^{t+\tilde{t}} \bm J(\chi) d\chi},
\end{align}
where $\eta(t, \tilde{t})$ is a scalar that measures the asymptotic growth of the norm and allows $\bm \phi(t+\tilde{t})$ to have any desired bounded norm, and the exponential is to be interpreted as the path-ordered exponential -- notice however that the following derivation does not require an expression for $\bm M(t, \tilde t)$, but rather its derivative, $d \bm M(t, \tilde{t})/d \tilde{t}=\bm J(\tilde t) \bm M(t, \tilde t)$. Substituting \eqref{eq:propagator} in \eqref{eq:covariant_eq} and differentiating with respect to $\tilde{t}$ results in
\begin{equation}
    \bm J(t + \tilde{t}) e^{\int_t^{t+\tilde{t}} \bm J(\chi) d\chi} \bm \phi(t) = \frac{d\eta(t, \tilde{t})}{d\tilde{t}} \bm \phi(t + \tilde{t}) + \eta(t, \tilde{t}) \frac{d\bm \phi(t + \tilde{t})}{d\tilde{t}}.
\end{equation}
By setting $t=0$ and omitting the explicit dependence on $t=0$, we obtain
\begin{equation}
    \label{eq:i_clv_eq}
    \frac{d \bm \phi}{d \tilde{t}} = \bm J(\tilde{t}) \bm \phi(\tilde{t}) - \frac{1}{\eta(\tilde{t})}\frac{d\eta(\tilde{t})}{d\tilde{t}} \bm \phi(\tilde{t}), 
\end{equation}
for any $\eta(\tilde{t})\not=0$. 
Moreover, we know from Oseledets' theorem that
\begin{equation}
    \left \lvert \left \lvert \bm M(\tilde{t}) \bm \phi(0) \right \rvert \right \rvert \cong e^{\lambda \tilde{t}} ||\bm{\phi}(0)||,
\end{equation}
which shows that $\eta(\tilde{t}) \cong e^{\lambda \tilde{t}}$. 
If we choose to have a bounded non-zero covariant Lyapunov vector, i.e. $0<||\bm \phi||<\infty$, \eqref{eq:i_clv_eq} becomes
\begin{equation}
    \label{eq:clv_eq}
    \frac{d \bm \phi}{d \tilde{t}} = \bm J \bm \phi - \lambda \bm \phi.
\end{equation}
It is easier to mathematically manipulate and numerically solve \eqref{eq:clv_eq} than \eqref{eq:oseledets_splitting}. Moreover, equation~\eqref{eq:clv_eq} provides a clear picture of the evolution of a covariant Lyapunov vector: the vector is evolved by the tangent dynamics $\bm J \bm \phi$, while the extra term $-\lambda \bm \phi$ guarantees that its norm is bounded. It can be shown that if the attractor is periodic or chaotic, there is a neutral mode ($\lambda=0$), where $\bm \phi$ is collinear to $\dot{\bar{\bm q}}$~\citep{katok_hasselblat}.

In the remainder of this paper, $t=0$ without loss of generality and the tilde, $\tilde{()}$, is dropped for brevity.

\subsection{Numerical computation of Lyapunov exponents and covariant Lyapunov vectors}
\label{sec:theory:qr}
%
We use a robust algorithm~\citep{Ginelli2007,ginelli_2013}, called the QR algorithm for brevity, to calculate the Lyapunov spectrum and covariant vectors. The algorithm evolves a set of $m$ column vectors $\bm{g}_j,\, j=\{1,\ldots,m\}$ of an $N \times m$ matrix $\bm M$, via the tangent equation \eqref{eq:tangent_system}. 
Because the $\bm{g}_j$ will likely have a component in the most unstable (least stable) space $\Omega_1$, their norm will exponentially grow (decay) at rate $\lambda_1$, which is bound to numerically overflow (underflow). To overcome this numerical instability, the QR algorithm executes periodic orthonormalisations of $\bm M$. 
By denoting the time step with a superscript, the calculation of the Lyapunov exponents and covariant Lyapunov vectors is enabled by the following algorithm.

\vspace{10pt}

\begin{enumerate}
	\item Set the initial condition $\bar{\bm{q}}^0$ and initialise $\bm{M}^0$ to a random orthonormal set of vectors $[\bm{g}_1 \ldots \bm{g}_m]$.
    \item Evolve $\bar{\bm{q}}^0, \bm{M}^0$ using \eqref{eq:system}, \eqref{eq:tangent_system} for $n_{su}$ iterations, where $n_{su}$ is called the spinup time, which must be sufficiently large such that $\bar{\bm{q}}^{n_{su}}$ is in the attractor (to some numerical tolerance).
    \item Evolve $\bar{\bm{q}}^j, \bm{M}^j$ for $n_{QR}$ iterations. \label{item:evolve}
    \item Perform QR decomposition on $\bm{M}^j$, obtaining $\bm{Q}^j, \bm{R}^j$. Store $\bm{Q}^j, \bm{R}^j$, and set $\bm{M}^j := \bm{Q}^j$. If $j < n_T$, where $n_T$ is the total number of iterations corresponding to the total simulation time, go back to Item~(\ref{item:evolve}). \label{item:qr}
    \item Randomly initialise an upper triangular matrix $\bm{C}^{n_T}$ of the same dimension as all $\bm{R}^j$.
    \item Evolve $\bm{C}^j$ backward by solving $\bm{R}^j \bm{C}^j = \bm{C}^{j+1}$ for $\bm{C}^j$ and subsequently normalise its columns, i.e.\ ensuring $\sum_k (\bm{C}^j_{lk})^2 = 1$.
    \item Compute Lyapunov exponents: $[\lambda_1 \ldots \lambda_m] = ((n_T-n_{su}) \Delta t)^{-1} \sum_{j=n_{su}}^{n_T} \log(|\text{diag}(\bm{R}^j)|)$, where $\Delta t$ is the time step.
    \item Compute covariant Lyapunov vectors: $[\bm \phi_1 | \ldots | \bm \phi_m]^j = \bm{Q}^j \bm{C}^j$, valid only for $j \in [n_{su}, n_T-n_{sd}]$, where $n_{sd}$ is the spindown time, which must be sufficiently large for $\bm{C}^j$ to converge to the covariant Lyapunov vector expansion coefficients.
\end{enumerate}

\subsection{Hyperbolicity}
\label{sec:hyperbolicity}

A strange attractor is hyperbolic if there is a splitting of the tangent space into stable, neutral and unstable subspaces at every point of the trajectory, $\bar{\bm{q}}(t)$. Formally, $\mathcal{T}_{\bar{\bm q}} = E_{\bar{\bm q}}^s \oplus E_{\bar{\bm q}}^n \oplus E_{\bar{\bm q}}^u$, where $E_{\bar{\bm q}}^s$ and $E_{\bar{\bm q}}^u$ are the stable and unstable subspaces of dimension $N^s$ and $N^u$, defined by the directions along which the derivative contracts and expands, respectively, and $E_{\bar{\bm q}}^n$ is the  one-dimensional neutral subspace.
\fh{(Consequently, a quasi-periodic solution is not hyperbolic because it has at least two zero Lyapunov exponents, i.e., $E_{\bar{\bm q}}^n$ is at least two-dimensional.)}
Hyperbolicity has profound implications on the behaviour of a dynamical system. The existence of unstable subspaces gives rise to exponentially diverging trajectories, which in turn gives rise to unpredictable dynamics in the long term. 
%
Furthermore, hyperbolicity often implies structural stability of the attractor, i.e. the qualitative behaviour of the attractor does not change if the system is slightly perturbed.  
In the problem of computing sensitivities, hyperbolicity is crucial because it determines whether the time-averaged cost functional responds smoothly to perturbations to the parameters~\citep{Ruelle1980}. Indeed, the most robust sensitivity algorithms~\citep[e.g.][]{wang_2013,Blonigan2014a,wang_2014,Blonigan2017,ni_2017} rely on the shadowing lemma~\citep{Bowen1975,pilyugin2006shadowing}, which is valid only in hyperbolic systems. 
Importantly, it has been hypothesised by~\citet{Gallavotti1995,Gallavotti2006} that most physical dynamical systems develop asymptotically on an attracting set, the dynamics of which can be regarded as hyperbolic. This is called the chaotic hypothesis, which stems from the measure theory of turbulence of~\citet{Ruelle1980}. 
%
%
In order to verify hyperbolicity in a numerical simulation, the method described in \citet{takeuchi_2011} is used here. The angles between the three pairs of subspaces, $\theta_{u,n} = \angle(E_{\bar{\bm q}}^u, E_{\bar{\bm q}}^n)$, $\theta_{u,s} = \angle(E_{\bar{\bm q}}^u, E_{\bar{\bm q}}^s)$, $\theta_{n,s} = \angle(E_{\bar{\bm q}}^n, E_{\bar{\bm q}}^s)$, are computed. These angles are computed by using the principal angles, $\cos \left( \theta_{A,B} \right) = \bar{\sigma}(\bm Q_A \bm Q_B)$, where matrices $\bm Q_A$ and $\bm Q_B$ define the orthonormal bases of any subspaces $A$ and $B$ (not only $E_{\bar{\bm q}}^u$, $E_{\bar{\bm q}}^n$, $E_{\bar{\bm q}}^s$), respectively, and $\bar{\sigma}$ is the largest singular value. Then, a probability density function of each of these angles is extracted via a histogram of the time series (e.g. figure~\ref{fig:rijke:hyp:pdf}). The system behaves hyperbolically if there are no tangencies between the subspaces, \. i.e. the value of the probability density functions at $\theta = 0$ is $0$.

\subsection{Shadowing lemma}
\label{sec:shadowing_lemma}
Shadowing-based sensitivity methods are centred around the shadowing lemma. 
The shadowing lemma exists both for discrete or continuous dynamical systems, but we will only present its discrete version because most engineering problems are numerically discretised to be solved. 
\begin{definition}[$\epsilon$-pseudo-orbit]
	An $\epsilon$-pseudo-orbit for the map $\bm f$ is a sequence of points $\{\bm y_n\}$, such that
	\begin{equation*}
		\left| \left| \bm y_{n+1} - \bm f(\bm y_n) \right| \right| < \epsilon .
	\end{equation*}
\end{definition}
An $\epsilon$-pseudo-orbit is thus a series $\{ \bm y_n \}$ where each point $\bm y_{n+1}$ is at most $\epsilon$ away from the true iterate $\bm f(\bm y_n)$ of the previous point $\bm y_n$.
\begin{definition}[$\delta$-shadow-orbit]
	\label{def:delta_shadow}
	An actual orbit $\{\bm x_n\}$, where $\bm x_n = \bm f^n(\bm x_0)$, is said to be a $\delta$-shadow-orbit of the $\epsilon$-pseudo-orbit $\{\bm y_n\}_{a < n < b}$ if
	\begin{equation*}
		\left| \left| \bm x_n - \bm y_n \right| \right| < \delta. 
	\end{equation*}
\end{definition}
In other words, an $\epsilon$-pseudo-orbit is $\delta$-shadowed by a true orbit if the true orbit is closer than $\delta$ from it at each point.  
\begin{lemma}[Shadowing Lemma]
	\label{lemma:shadowing}
	Let $\Lambda$ be a hyperbolic attractor for $\bm f$. Then, for every $\delta > 0$, there is an $\epsilon > 0$ such that every $\epsilon$-pseudo-orbit in $\Lambda$ is $\delta$-shadowed by the actual orbit of some point $\bm q \in \Lambda$~\citep[e.g.][]{Bowen1975}.
\end{lemma}
Intuitively, the shadowing lemma guarantees that, even in chaotic systems where infinitely close trajectories diverge, there is a trajectory of a slightly perturbed system that does not diverge from the unperturbed system. 
%

\subsection{Shadowing methods for sensitivity}
\label{sec:shadowing_methods}
%
%
The gradient of the time-averaged cost functional \eqref{eq:Jav} explicitly reads 
\begin{align}
\nabla_{\bm s}\langle\mathcal{J}\rangle \triangleq \frac{d}{d\bm s}\left(\lim_{T \rightarrow \infty} \frac{1}{T} \int_0^T \J(\bm{s}; \bm{q}(t)) dt\right). 
\end{align}
%
In a chaotic attractor, the operations of differentiation and time average do not commute, i.e. $\nabla_{\bm s}\langle\mathcal{J}\rangle\not=\langle\nabla_{\bm s}\mathcal{J}\rangle$, where 
\begin{align}\label{eq:shasha}
\langle\nabla_{\bm s}\mathcal{J}\rangle = \lim_{T \rightarrow \infty} \frac{1}{T} \int_0^T \left(\frac{\partial \mathcal{J}}{\partial \bm s} + \frac{\partial \mathcal{J}}{\partial \bm q}\frac{\partial \bm q}{\partial \bm s}\right)dt. 
\end{align}
Equation~\eqref{eq:shasha} is an unbounded quantity because $\partial\bm q/\partial\bm s$ lives in the tangent space, a subspace that is exponentially unstable in chaotic attractors. 
Shadowing-based sensitivity methods integrate the sensitivity of the cost functional along the shadow trajectory, which does not diverge from the attractor. This way, the quantity \eqref{eq:shasha} is bounded and equal to $\nabla_{\bm s}\langle\mathcal{J}\rangle$. 
The original shadowing method~\citep{wang_2013} achieves
 this by calculating the perturbation to $\bm F$ in \eqref{eq:system} due to a perturbation in a parameter $\bm s \rightarrow \bm s + \delta \bm s$, that is, $\partial \bm{F}/\partial \bm s \cdot \delta \bm s$. The perturbation is decomposed in the covariant Lyapunov vector  basis of the baseline trajectory to obtain a set of independent ordinary differential equations, one for each mode. The solutions of these equations are the components of the shadow trajectory in the covariant Lyapunov vector  basis. After obtaining the perturbed trajectory in the phase space, the sensitivities can be readily computed.
A major drawback of the original shadowing method is the need to compute all the covariant Lyapunov vectors for all time steps, which is computationally expensive. The least-squares shadowing  method~\citep{wang_2014} overcomes this by finding a trajectory of the system at parameter value $\bm s + \delta \bm s$ that is close to a trajectory of the system at parameter $\bm s$ via solving a least squares minimisation problem, which minimises the distance between the two trajectories at regular checkpoints. While the least-squares shadowing method is faster than the original shadowing method, it still carries high computational cost, as it requires solving a linear system of dimension equal to the dimension of the phase space times the number of checkpoints. \citet{ni_2017} developed the non-intrusive least-squares shadowing method, the computational cost of which scales only with the number of unstable covariant Lyapunov vectors. 
In this paper, we will apply the non-intrusive least-squares shadowing method to a chaotic thermoacoustic system.

\section{Eigenvalue and Floquet analyses as subsets of covariant Lyapunov analysis}
\label{sec:equiv}
Covariant Lyapunov vector analysis is the most general linear stability tool because it can be applied to aperiodic solutions (\S\ref{sec:theory}).
On the one hand, when covariant Lyapunov vector  analysis is applied to a fixed point, we recover eigenvalue analysis.
On the other hand, when covariant Lyapunov vector analysis is applied to a periodic solution, we recover Floquet analysis~\citep{Trevisan1998}.
We analytically show the limits of eigenvalue and Floquet analyses in \S\ref{sec:eigen} and \S\ref{sec:floquet}, respectively. These results are general -- they do not depend on the autonomous nonlinear system under investigation -- and can be applied to other problems in flow stability.
\subsection{Eigenvalue analysis of fixed points: connection with covariant Lyapunov vectors}
\label{sec:eigen}
Eigenvalue analysis determines the linear stability of a fixed point of $\bm{F}$. 
Mathematically, in decomposition~\eqref{eq:split_sol}, $\bm{\bar q}$ does not depend on time. The linearised dynamics around the fixed point $\bm{\bar q}$ is governed by \eqref{eq:tangent_eq} where the Jacobian $\bm{J}=\left. d \bm F/d \bm q \right|_{\bm{\bar q}}$ is constant. The formal solution for an initial condition reads 
\begin{equation}
	\label{eq:linearized_solution_fp}
	\bm{q}'(t) = e^{\bm J t} \bm{q}_0' .
\end{equation}
By assuming that $\bm J$ has a complete eigenbasis, i.e. it is not defective, $\bm{q}_0'$ can be decomposed in the eigenbasis $\{\hat{\bm q}_1, \dots, \hat{\bm q}_N\}$, where $\hat{()}_j$ is an eigenvector of $\bm J$, as 

\begin{equation}
	\label{eq:eigendecomp1}
	\bm{q}_0' = \sum_{j=1}^N \underbrace{(\bm{q}_0' \cdot \hat{\bm q_j})}_{\triangleq \alpha_j} \hat{\bm q}_j, 
\end{equation}
where, to keep a similar notation to covariant Lyapunov vector  analysis, the eigenpairs are sorted in descending order according to the real part of the corresponding eigenvalue $\sigma_j$, i.e. $j=1$ denotes the eigenpair with largest growth rate. 
Substituting~\eqref{eq:eigendecomp1} in~\eqref{eq:linearized_solution_fp} yields
\begin{align}
	\label{eq:eigendecomp2}
	\bm{q'}(t) &= e^{\bm J t} \sum_{j=1}^N \alpha_j \hat{\bm q}_j \nonumber \\
    &= \sum_{j=1}^N \alpha_j e^{\sigma_j t} \hat{\bm q}_j. 
\end{align}
%
Substituting the perturbation \eqref{eq:eigendecomp2} into the definition of Lyapunov exponent, \eqref{eq:lyapunov_exponent}, yields 
\begin{align}
	\label{eq:le_computation_fp}
    \lambda 
    &= \lim_{t \rightarrow \infty} \frac{1}{t} \log \frac{\left| \left| \sum_{j=1}^N \alpha_j e^{\sigma_j t} \hat{\bm q}_j \right| \right|}{||\bm{q}'_0||} \nonumber \\
    &=\lim_{t \rightarrow \infty} \frac{1}{t} \log \left( \frac{|e^{\sigma_k t}| \, ||\alpha_k \hat{\bm q}_k||}{||\bm{q}_0'||} \right) \nonumber \\
    &= \lim_{t \rightarrow \infty} \frac{1}{t} \left[\log \left( e^{\mathcal{R}(\sigma_k) t} \right) + \log \left( \frac{||\alpha_k \hat{\bm q}_k||}{||\bm{q}_0'||} \right) \right] \nonumber \\
    &= \mathcal{R}(\sigma_k),
\end{align}
where $k$ is the first index such that $\alpha_k \neq 0$, and $\mathcal{R}$ denotes the real part. Equation~\eqref{eq:le_computation_fp} shows that the $k$-th Lyapunov exponent, $\lambda_k$, of a linear system on a fixed point is the real part of the eigenvalue of the Jacobian, $\mathcal{R}(\sigma_k)$. Physically, the Lyapunov exponent is the growth (or decay) rate of small perturbations on top of the steady solution.  
This means that the Lyapunov exponent associated with the perturbation $\bm q_0'$ is $\mathcal{R}(\sigma_k)$ if $\bm q_0'$ does not belong to $\text{Span}(\hat{\bm q}_1, \dots, \hat{\bm q}_{k-1})$, which is the subspace spanned by the first $k-1$ eigenvectors. 
%
Because the covariant Lyapunov vector is an unsteady vector, we have to decompose it in a time-varying basis of the propagator $e^{\bm J t}$, which consists of the time-varying vectors $\hat{\bm{q}}_j e^{\mathrm{i} \omega_j t}$~\citep{Curtain1995}, where $\mathrm{i}$ is the imaginary unit. The angular frequency of the linear oscillation is denoted $\omega_j = \mathcal{I}(\sigma_j)$, where $\mathcal{I}$ is the imaginary part.  Abusing notation by re-using $\alpha_j$ as the coordinates in the new basis, we can decompose $\bm \phi$ as 
\begin{equation}
	\label{eq:eigendecomp3}
	\bm \phi(t) = \sum_{j=1}^N \underbrace{(\bm{\phi}_0 \cdot \hat{\bm q}_j)}_{\triangleq \alpha_j} \hat{\bm q}_j e^{\mathrm{i} \omega_j t}, 
\end{equation} 
%
%
such that its time derivative reads
\begin{align}
\dot{\bm \phi} &= \sum_{j=1}^N \alpha_j\left( \sigma_j - \mathcal{R}(\sigma_j) \right) \hat{\bm q}_j e^{\mathrm{i} \omega_j t}\nonumber \\
    &= \sum_{j=1}^N \alpha_j \bm J \hat{\bm q}_j e^{\mathrm{i} \omega_j t} - \sum_{j=1}^N \lambda_j \alpha_j \hat{\bm q}_j e^{\mathrm{i} \omega_j t} \nonumber \\
    &= \bm{J\phi} - \sum_{j=1}^N \lambda_j \alpha_j \hat{\bm q}_j e^{\mathrm{i} \omega_j t} . 
\end{align}
In order to factor out $\lambda_j$, we consider the set of eigenvectors that share the same growth rate, $\mathcal{R}(\sigma_j)$, although they may have different angular frequencies, $\omega_j$. Mathematically,
by considering $\alpha_k = 0: \, k \notin \{j: \, \lambda_j = \lambda\}$, the covariant Lyapunov vector  equation~\eqref{eq:clv_eq} is recovered
\begin{align}
	\dot{\bm \phi} &= \bm J \bm \phi - \sum_{j=1}^N \lambda_j \alpha_j \hat{\bm q_j} e^{\mathrm{i} \omega_j t} \nonumber \\
    &= \bm{J\phi} - \lambda \sum_{j, \alpha_j \neq 0} \alpha_j \hat{\bm q}_j e^{\mathrm{i} \omega_j t} \nonumber \\
    &= \bm{J\phi} - \lambda \bm \phi, \label{eq:clv_sum}
\end{align}
which shows that covariant Lyapunov vectors and eigenvectors, grouped by growth rates, span the same subspaces. Furthermore, any real linear combination of $\hat{\bm q}_j e^{\mathrm{i} \omega_j t}$ is a covariant Lyapunov vector with the different $\hat{\bm q}_j$ corresponding to eigenvalues that have the same growth rate (i.e., same Lyapunov exponents). 
On the one hand, if $\sigma_j\in\mathbb{R}$, there is only one such $j$ and thus $\bm \phi = \hat{\bm q_j}$ is a covariant Lyapunov vector. (In principle, there may be cases where the spectrum contains one real eigenvalue and two complex conjugates with the same real part as the real eigenvalue. Although we do not consider this special case, the conclusions we draw still hold.)  
On the other hand, if $\sigma_j\in\mathbb{C}$, there is a pair $\{\alpha_j \hat{\bm q_j} e^{\mathrm{i} \omega_j t}, \alpha^*_j \hat{\bm q}^*_j e^{\mathrm{i} \omega_j t}\}$ that defines a two-dimensional subspace in which all vectors are covariant Lyapunov vectors satisfying~\eqref{eq:clv_eq}, where $^*$ denotes the complex conjugate. 
Any two non-collinear vectors, e.g. $\alpha_j = \alpha^*_j = 1/2$ or $\alpha_j = -\alpha^*_j = -i/2$,  can be taken to define the Lyapunov subspace
%
%
\begin{equation}
    \label{eq:theoretical_clv}
\begin{bmatrix}
	\cos (\omega_j t) & -\sin (\omega_j t) \\ 
	\sin (\omega_j t) &  \cos (\omega_j t)
\end{bmatrix}
\begin{bmatrix}
	\mathcal{R}(\hat{\bm q_j}) \\ 
	\mathcal{I}(\hat{\bm q_j})
\end{bmatrix} .
\end{equation}
On a fixed point, we showed that the plane spanned by the covariant Lyapunov vectors does not change in time because the plane spanned by $\mathcal{R}(\hat{\bm q_j})$ and $\mathcal{I}(\hat{\bm q_j})$ is constant. In other words, while the angles between covariant Lyapunov vectors in the different Lyapunov subspaces vary in time, the angles between different Lyapunov subspaces are constant.  This is in contrast to the chaotic case, where the angles between Lyapunov subspaces vary in time. Using the subspaces instead of the covariant Lyapunov vectors is crucial for the analysis of chaotic thermoacoustic systems, as shown in~\S\ref{sec:results:rijke:ch}.   

\subsection{Floquet analysis of limit cycles: connection with covariant Lyapunov vectors}
\label{sec:floquet}

Similarly to \S\ref{sec:eigen}, in this section we show that if the attractor is a limit cycle, the Lyapunov exponents correspond to the real part of the Floquet exponents and that the covariant Lyapunov vectors correspond to the eigenvectors of the monodromy matrix.
Consider the tangent problem \eqref{eq:tangent_eq}. We assume that the solution is a limit cycle, i.e. the solution is $T$-periodic, i.e. $\bar{\bm q}(t+T) = \bar{\bm q}(t)$, hence, the Jacobian is $T$-periodic, i.e. $\bm{J} \equiv \left. d\bm F/d\bm q \right|_{\bar{\bm q}(t)}$.
Let $\bm Q(t) = \left[ \bm Q_1 | \dots | \bm Q_N \right]$ be the fundamental matrix and $\bm B$ the monodromy matrix~\citep{Guck1983}, i.e.
\begin{align}
	& \bm{q}'(t) = \bm Q(t) \bm c, \\ 
    & \bm Q(t+T) = \bm Q(t) \bm B,
\end{align}
where $\bm c$ is the initial condition, $\bm q'(0)$, in the basis $\{\bm{Q}_1(0), \dots, \bm{Q}_N(0)\}$.
Let $\bm{b}_j$ be the eigenvector of $\bm B$ corresponding to the Floquet multiplier $\rho_j = e^{\nu_j T}$, where $\nu_j$ is the $j$-th Floquet exponent, sorted in descending order according to its real part, i.e. $j=1$ denotes the Floquet exponent with largest growth rate. Although the Floquet multipliers, which are the eigenvalues of the linearised Poincar\'e map, $\bm{Q}(t)$, are not unique because of the periodicity of the complex exponential, the Floquet exponents are unique. Noting that $\bm{q}'(t) = \bm{q}'(t^+ + mT) = \bm Q(t^+) \bm B^m \bm c$, with $0 \leq t^+ < T$ and $m = {0,1,2,\dots}$, and decomposing $\bm c$ in the eigenbasis $\{\bm{b}_1, \ldots, \bm{b}_N\}$, where we abuse notation and re-use the symbol $\alpha_j$ to represent the coordinates in the local basis, yields
\begin{align}
	\bm{q}'(t) &= \bm{q}'(t^+ + mT) \nonumber \\
    &= \bm Q(t^+) \bm B^m \bm c \nonumber \\
    &= \bm Q(t^+) \bm B^m \sum_{j=1}^N (\underbrace{\bm c \cdot \bm{b}_j}_{\triangleq \alpha_j}) \bm{b}_j \nonumber \\
    &= \bm Q(t^+) \sum_{j=1}^N \alpha_j \rho_j^m \bm{b}_j \nonumber \\
    &= \bm Q(t^+) \sum_{j=1}^N \alpha_j e^{\nu_j m T} \bm{b}_j. 
\end{align}
%
Using \eqref{eq:lyapunov_exponent}, we can calculate the Lyapunov exponent restricted to times $t=t^+ + mT$
\begin{align}
	\label{eq:le_computation_lc}
	\lambda(t^+) 
	&= \lim_{m \rightarrow \infty} \frac{1}{t^+ + mT} \log \left( \frac{||\bm{q}'(t^+ + mT)||}{||\bm Q(0) \bm c||} \right) \nonumber \\
    &= \lim_{m \rightarrow \infty} \frac{1}{t^+ + mT} \log \left( \frac{\left| \left| \bm Q(t^+) \sum_{j=1}^N \alpha_j e^{\nu_j m T} \bm{b}_j \right| \right|}{\left|\left| \bm Q(0) \bm c \right|\right|} \right) \nonumber \\
    &= \lim_{m \rightarrow \infty} \frac{1}{t^+ + mT} \log \left( \frac{|e^{\nu_k mT}| \, ||\alpha_k \bm Q(t^+) \bm{b}_k ||}{||\bm Q(0) \bm c||} \right)\nonumber \\
    &= \lim_{m \rightarrow \infty} \frac{1}{t^+ + mT} \left[\log \left( e^{\mathcal{R}(\nu_k) mT} \right) + \log \left( \frac{|| \alpha_k \bm Q(t^+) \bm{b}_k ||}{||\bm Q(0) \bm c||} \right) \right] \nonumber \\
    &= \lim_{m \rightarrow \infty} \frac{m T}{t^+ + mT} \mathcal{R}(\nu_k) \nonumber \\
    &= \mathcal{R}(\nu_k),
\end{align}
which shows that the $k$-th Lyapunov exponent, $\lambda_k$, is equal to the real part of the $k$-th Floquet exponent, $\mathcal{R}(\nu_k)$. The result is independent of $t^+$ and valid for any $t^+ \in [0,T)$.
%

Consider a covariant Lyapunov vector  $\bm \phi$ associated with the Lyapunov exponent $\lambda = \mathcal{R}(\nu)$. \textit{A priori}, we do not know the shape of $\bm \phi$. Notwithstanding, we can express it in terms of a slightly modified eigenbasis of $\bm B$ 
\begin{equation}
    \label{eq:floquet_ansatz}
    \bm \phi(t) = \bm Q(t) \sum_{j=1}^N \alpha_j \bm b_j e^{-\mathcal{R}(\nu_j) t}.
\end{equation}
%
By differentiating \eqref{eq:floquet_ansatz} in time, we obtain
\begin{align}
	\dot{\bm \phi} &= \dot{\bm Q} \sum_{j=1}^N \alpha_j  \bm b_j e^{-\mathcal{R}(\nu_j) t} - \bm Q \sum_{j=1}^N \mathcal{R}(\nu_j) \alpha_j \bm b_j e^{-\mathcal{R}(\nu_j) t} \nonumber \\
	&= \bm J \bm \phi - \bm Q \sum_{j=1}^N \mathcal{R}(\nu_j) \alpha_j \bm b_j e^{-\mathcal{R}(\nu_j) t}.
\end{align}
Similarly to \S\ref{sec:eigen}, we consider linear combinations of modes that have the same value of Lyapunov exponent, i.e. $\alpha_k = 0 : \lambda_k \neq \lambda$. Thus
\begin{align}
	\dot{\bm \phi} &= \bm J \bm \phi - \bm Q \sum_{j=1}^N \mathcal{R}(\nu_j) \alpha_j \bm b_j e^{-\mathcal{R}(\nu_j) t} \nonumber \\
	&= \bm J \bm \phi - \mathcal{R}(\nu) \bm Q \sum_{j=1}^N \alpha_j \bm b_j e^{-\mathcal{R}(\nu_j) t}\nonumber \\
	&= \bm J \bm \phi - \lambda \bm \phi,
\end{align}
which recovers the covariant Lyapunov vector  equation \eqref{eq:clv_eq}. In the same vein as in the fixed-point case (\S\ref{sec:eigen}), any real linear combination of Floquet modes that have the same growth rate is a covariant Lyapunov vector. Finally, notice that $\bm \phi(t)$ need not be periodic, except if it spans a one-dimensional Lyapunov subspace. In fact, if it is a linear combination of two complex conjugate Floquet modes, we have
\begin{align}
	\bm \phi(t+T) &= e^{-\mathcal{R}(\nu_j) (t+T)} \bm Q(t+T) \left( \alpha_j \bm b_j + \alpha_j^* \bm b_j^* \right) \nonumber \\
	&= e^{-\mathcal{R}(\nu_j) (t+T)} \bm Q(t) \bm B \left( \alpha_j \bm b_j + \alpha_j^* \bm b_j^* \right) \nonumber \\
	&= e^{-\mathcal{R}(\nu_j) (t+T)} \bm Q(t) \left( \alpha_j \bm b_j e^{\nu_j T} + \alpha_j^* \bm b_j^* e^{\nu_j^* T} \right) \nonumber \\
	&= e^{-\mathcal{R}(\nu_j) t} \bm Q(t) \left( \alpha_j \bm b_j e^{\mathcal{I}(\nu_j) T} + \alpha_j^* \bm b_j^* e^{\mathcal{I}(\nu_j^*) T} \right) \nonumber \\
	&\neq \bm \phi(t), 
\end{align}
which shows that $\bm \phi$ is not $T$-periodic. Although this result might seem odd at first, $\bm \phi$ behaves similarly to the covariant Lyapunov vectors in \eqref{eq:theoretical_clv} because, in both cases, the imaginary part dictates the rate at which they rotate in the plane spanned by the corresponding mode.
Although the mathematics is more involved, the connection between Floquet analysis and covariant Lyapunov vector analysis naturally follows the connection with eigenvalue analysis of fixed points (\S\ref{sec:eigen}):  a limit cycle can be viewed as a fixed point of a Poincar\'e map.
In summary, on the one hand, covariant Lyapunov vector analysis provides the same  linear dynamics as eigenvalue (Floquet) analysis when $\bm{\bar{q}}$ is a fixed-point (periodic solution).  On the other hand, covariant Lyapunov vector analysis provides the  linear dynamics when $\bm{\bar{q}}$ is a chaotic attractor, where eigenvalue and Floquet analyses cannot be applied.

The general theoretical analysis we have presented, which can be applied to other problems in flow stability, concludes the first part of this paper. From now on, we focus on a chaotic thermoacoustic system, which is a multi-physical problem in thermo-fluid dynamics that is relevant to aeronautical propulsion and clean power generation. 
\section{Thermoacoustic model}
\label{sec:systems}
We present a model of a thermoacoustic system that can exhibit rich dynamics, such as fixed points, limit cycles, quasi-periodic solutions and chaotic attractors. The acoustics are longitudinal and governed by the linearised Euler equations. 

The stability, sensitivity and optimisation framework presented in this paper can be used in more realistic models, for example by also solving for the flame, with virtually no modification. 
\subsection{Acoustics and heat source}
\label{sec:systems:rijke}
\begin{figure}
    \centering
    \begin{subfigure}[b]{0.49\textwidth}
        \subcaptionOverlay{\includegraphics[width=\textwidth]{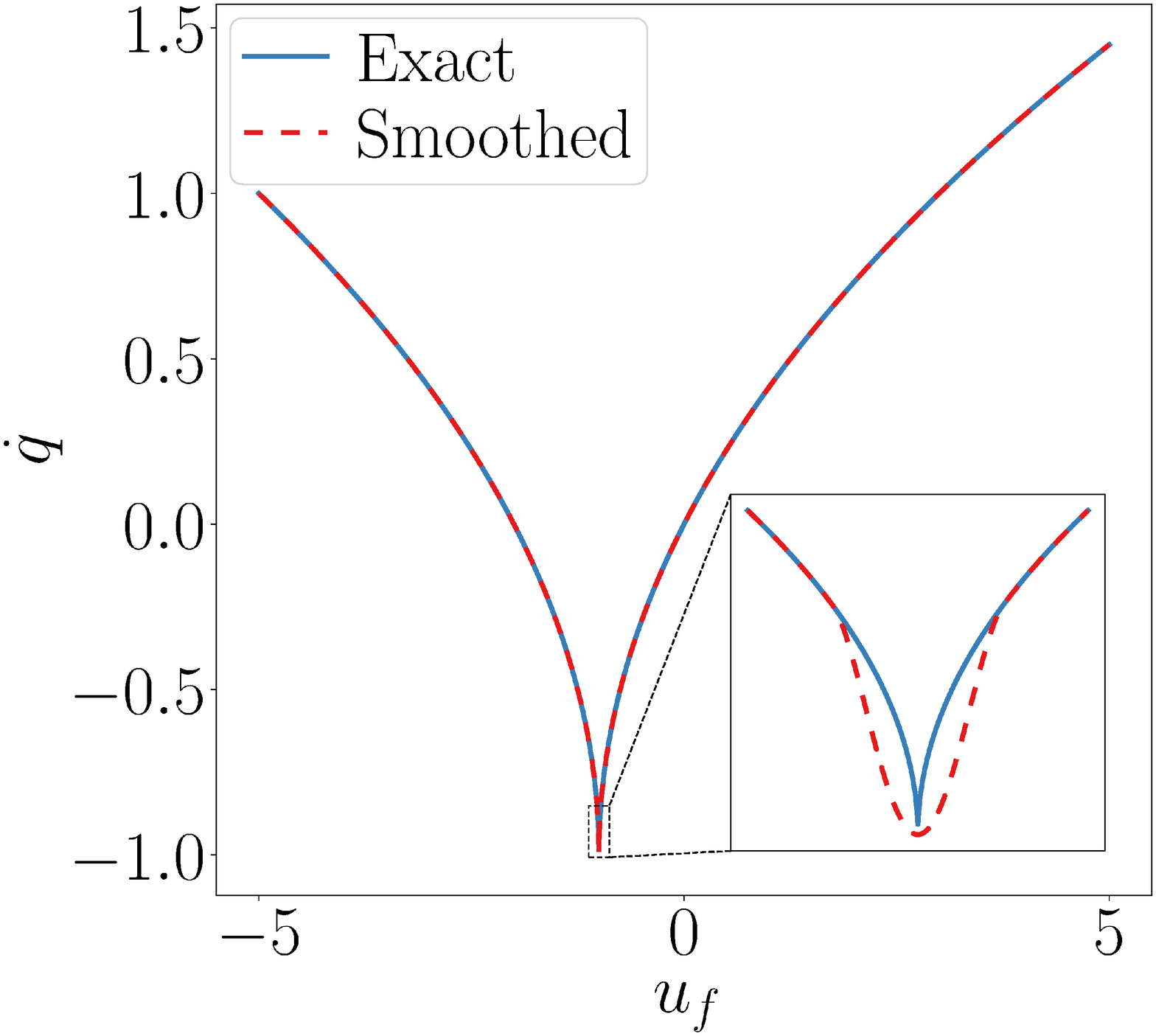}}
        \label{fig:kings:kings}
    \end{subfigure}
    \begin{subfigure}[b]{0.49\textwidth}
        \subcaptionOverlay{\includegraphics[width=\textwidth]{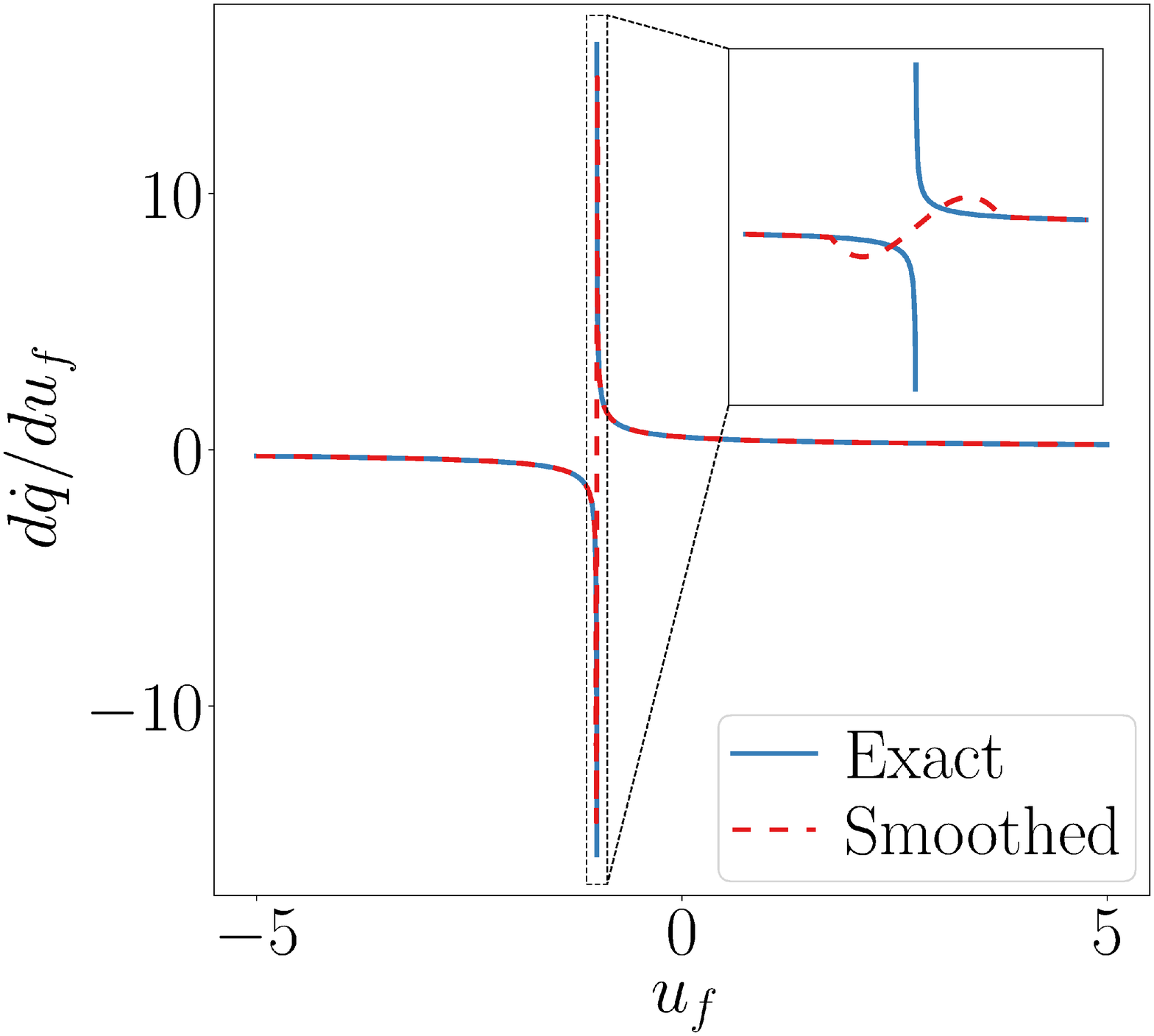}}
        \label{fig:kings:kings_tan}
    \end{subfigure}
    \caption{Comparison between King's law and our smoothed version. The smoothed version is exactly equal to King's law outside the range $|1+u_f| < \epsilon$, inside of which it is approximated by a fourth-degree polynomial, which enables continuity of both the function and its derivative.}
    \label{fig:kings}
\end{figure}

%
%
%
%
%
A thermoacoustic system consists of three subsystems that interact with each other: the acoustics, flame and hydrodynamics~\citep{Lieuwen2012,Magri2019_amr}. 
The acoustics strongly depend on the geometry of the configuration and the boundary conditions.
The flame is governed by chemistry mechanisms and their interaction with the turbulent environment.
The hydrodynamics is governed by the geometry of the inlets and flame holders, which generate large coherent structures due to flow instabilities (vortex shedding, shear layer instabilities, etc.), which, in turn, are modulated by turbulence. 
To accurately model thermoacoustic instabilities,  high-fidelity simulations can be employed~\citep[e.g.][]{Poinsot2017}. 
However, in this fundamental paper, we aim at capturing the essential physical mechanisms of chaotic thermoacoustic instabilities. Therefore, we choose a prototypical time-delayed thermoacoustic system with a longitudinal acoustic cavity and a heat source modelled with a nonlinear time-delayed model~\citep{Subramanian2010a}. %
The main assumptions are:  
(i) the acoustics are small perturbations onto a mean flow at rest with uniform density; 
(ii) viscosity and diffusivity are negligible; and 
(iii) the acoustics are one-dimensional, i.e. the cut-on frequency of the duct is much higher than the frequency of the instability. Under these assumptions, the linearisation of the inviscid momentum and energy equations yields, respectively~\citep{Balasubramanian2008a,juniper_2011,magri_juniper_2013}  
\begin{align}
	& \frac{\partial u}{\partial t} + \frac{\partial p}{\partial x} = 0 \label{eq:rijke:mom} \\
	& \frac{\partial p}{\partial t} + \frac{\partial u}{\partial x} + \zeta p - \dot{q}\delta(x - x_f)= 0 \label{eq:rijke:ene},
\end{align}
where $u$, $p$, $\dot{q}$, $x$ and $t$ are the non-dimensional velocity, pressure, heat-release rate, axial coordinate and time, respectively. The reference scales for speed, pressure, length and time are the mean-flow convection velocity, the mean-flow Mach number multiplied by the heat capacity factor, the length of the tube and the length of the tube divided by the mean-flow speed of sound, respectively. 
%
The duct has open ends, which means that the acoustic pressure is zero at the boundaries. The damping coefficient, $\zeta$, takes into account all the acoustic dissipation (\S\ref{sec:numericaldiscretization}). 
%
%
The spatial extent of the heat source is assumed to negligible as compared to the acoustic wavelength~\citep{Dowling1997}, thus, it is modelled as a compact source of acoustic energy through a Dirac delta (generalised) function, $\mathit{\delta(x-x_f)}$ localised at $x_f=0.2$.
The heat-release rate is provided by a modified King's law~\citep{King1914,Heckl1988, Heckl1990, Polifke2001, orchini_2016}
\begin{equation}
	\label{eq:modified_kings}
	\dot{q}(t) = \beta \left[ \left(1+u_f(t-\tau)\right)^{\frac{1}{2}} - 1 \right], 
\end{equation}
which is a nonlinear time-delayed model for an electrically heated mesh of wires. 
This model has similar features to flame models, such as the $n$-$\tau$ model~\citep[e.g.][]{juniper_sujith_2018}. 
%
%
%
\fh{In future work, the heat-release rate, $\dot{q}(t)$, can be obtained, for example, from the dynamics of premixed flames~\citep{Kashinath2013c, Kashinath2013b,Waugh2014,Orchini2015a,Yu2019_jcp} or diffusion flames ~\citep{Tyagi2007b,Magri2013c}. Solving for the flame dynamics adds many numerical degrees of freedom to the state vector, which makes the calculations computationally more expensive, but it does not change the framework we propose.}
Because Lyapunov analysis is valid only for smooth dynamical systems, we approximate the heat-release law~\eqref{eq:modified_kings} by a fourth-degree polynomial in a small neighbourhood of $u_f(t-\tau)=-1$ to make the derivative smooth (figure~\ref{fig:kings}). 

The heat parameter, $\beta$, and time delay, $\tau$, encapsulate all information about the heat source, base velocity and ambient conditions.
\fh{We transform the time-delayed problem into an initial value problem. (This operation is not mandatory, however, it enables us to use the adaptive initial value problem time integrator \texttt{scipy.integrate.odeint} with no modification.) Thus, we model the advection of a dummy variable $v$ with velocity $\tau^{-1}$ as  
\begin{align}
    \label{eq:advection}
    \frac{\partial v}{\partial t} + \frac{1}{\tau}\frac{\partial v}{\partial X} &= 0,\quad 0 \le X\le 1, \\
    \label{eq:advection_bc}
    v(X=0,t) &= u_f(t).  
\end{align}
\fh{The dummy variable $v$ takes time $\tau$ to travel from the left to the right boundary. 
Therefore, the time-delayed acoustic velocity is provided by the value of $v$ at the right boundary, i.e. $u_f(t-\tau) = v(X=1,t)$. The calculation of the time-delayed acoustic velocity via~\eqref{eq:advection} adds only a few numerical degrees of freedom (\S\ref{sec:numericaldiscretization}). } 
The time-delayed problem~\eqref{eq:rijke:mom}-\eqref{eq:modified_kings} is mathematically equivalent to the initial value problem~\eqref{eq:rijke:mom}-\eqref{eq:rijke:ene} and~\eqref{eq:advection}-\eqref{eq:advection_bc}~\citep{Jarlebring2008}}.
%
%
%
%
%
\subsubsection{Numerical discretisation}\label{sec:numericaldiscretization}
Equations~\eqref{eq:rijke:mom},~\eqref{eq:rijke:ene} are discretised by a Galerkin method~\citep{Zinn1971}. 
First, the acoustic variables are separated in time and space as 
\begin{align}
    & u(x,t) = \sum_{j=1}^{N_g} \eta_j(t)\cos(j \pi x), \label{eq:galerkin_decomp_u}\\
    & p(x,t) =  -\sum_{j=1}^{N_g} \mu_j(t) \sin(j \pi x), \label{eq:galerkin_decomp_p}
\end{align}
where each spatial function is a natural acoustic mode of the open-ended duct.
The partial differential equations~\eqref{eq:rijke:mom},~\eqref{eq:rijke:ene} are projected onto the Galerkin spatial basis $\{ \cos(\pi x), \cos(2 \pi x),$ $\dots, \cos(N_g \pi x)\}$ to yield 
\begin{align}
	& \dot{\eta}_j - j \pi \mu_j = 0 \label{eq:rijke:gal_mom} \\
	& \dot{\mu}_j + j \pi \eta_j + \zeta_j \mu_j + 2 \dot{q} \sin(j \pi x_f) = 0 \label{eq:rijke:gal_ene}.
\end{align}
The system has $2N_g$ degrees of freedom. 
The time-delayed velocity becomes 
\begin{equation}
    \label{eq:delayed_uf}
    u_f(t-\tau) = \sum_{k=1}^{N_g} \eta_k(t-\tau) \cos(k \pi x_f),
\end{equation}
and the damping, $\zeta_j$, is modelled by a modal expression that damps out higher-frequency oscillations, $\zeta_j = c_1 j^2 + c_2 j^{1/2}$,
where $c_1=0.1$ and $c_2=0.06$~\citep{Subramanian2010a}. This damping model originates from physical principles, as explained in~\citet{Landau1987}.
We have assumed that the mean flow is sufficiently slow such that it can be neglected. Adding a mean flow may quantitatively change the phases between acoustic waves~\citep{Dowling2005}, but the  conclusions of this paper are qualitatively unaffected. With a mean flow, a wave approach can be used instead~\citep{Dowling2005}. 

The linear advection equation~\eqref{eq:advection} is discretised using $N_c+1$ points with a Chebyshev spectral method~\citep{trefethen2000spectral}. This discretisation adds $N_c$ degrees of freedom. The resulting discretised system is time integrated using \texttt{scipy.integrate.odeint}, which in turn calls \texttt{lsoda} from the \texttt{odepack} library. This method detects numerical stiffness and switches automatically between the Adams method in the non-stiff case and backward differentiation formula in the stiff case~\citep{petzold_1983}.
On numerical discretisation, the thermoacoustic state vector is $\bar{\bm q}$, with $\bar{\bm q}^T = \left( \eta_1, \dots, \eta_{N_g}, \; \mu_1, \dots, \mu_{N_g}, \; v_1, \dots, v_{N_c}\right)$. 
\subsubsection{Effect of numerical discretisation}
\begin{figure}[tb]
    \centering
    \begin{subfigure}[b]{\textwidth}
        \subcaptionOverlay{\includegraphics[width=\textwidth]{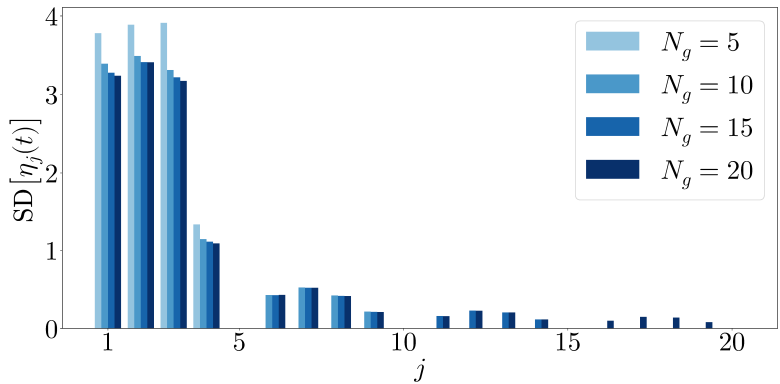}}
        \label{fig:systems:ng_conv_q}
    \end{subfigure}

    \begin{subfigure}[b]{\textwidth}
        \subcaptionOverlay{\includegraphics[width=\textwidth]{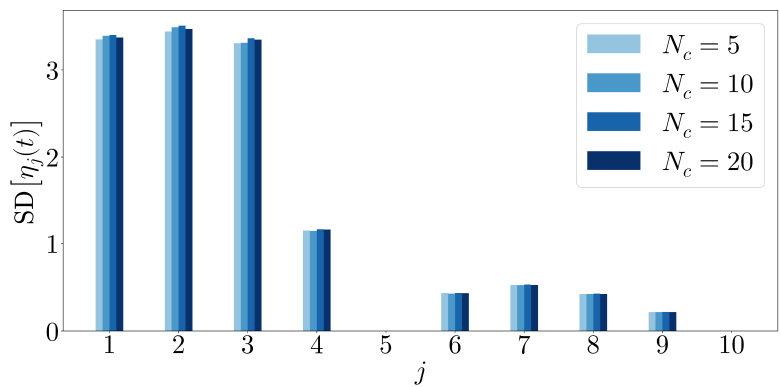}}
        \label{fig:systems:nc_conv_q}
    \end{subfigure}
    \caption{Convergence study on a chaotic attractor ($\beta=7.0$, $\tau=0.2$). \subref{fig:systems:ng_conv_q} Standard deviation of the Galerkin modes $\eta_j(t)$ on a chaotic attractor, varying the number of Galerkin modes, $N_g$, with the number of Chebyshev points fixed to $N_c+1=11$. The fifth and tenth modes are of the order of machine precision. \subref{fig:systems:nc_conv_q} Same as panel~\subref{fig:systems:ng_conv_q}, but with the number of Galerkin modes fixed to $N_g=10$ and number of Chebyshev points, $N_c+1$, varying.}
    \label{fig:systems:conv_ch}
\end{figure}

To investigate the effect of the numerical discretisation, we perform two types of tests. Figure~\ref{fig:systems:conv_ch} depicts the first type of test, which is done on a chaotic attractor. Figure~\ref{fig:systems:ng_conv_q} shows the standard deviation of $\eta_j(t), \, j = 1, \cdots, N_g$ for different values of $N_g$ with fixed $N_c=10$.  The dominant unstable mode is the first mode because the heat source is located at $x_f=0.2$, where most of the energy excites the first mode. Apart from $N_g=5$, where a large difference is observed, the calculations with small $N_g$ correctly capture the energy associated with each of the Galerkin modes that they compute. When increasing the number of Galerkin modes, the accuracy on the modes that were previously included does not improve. The benefit of increasing $N_g$ is to increase the spatial resolution by including higher wavenumbers. The magnitude of the standard deviation decays sharply at the beginning up to $j=10$, followed by a slower decay. Therefore, capturing modes of lower intensity, e.g. $O\left( {\text{SD}}\left[\eta_j\right] \right) \sim 10^{-3}$, requires a large increase in the number of Galerkin modes. Thus, a good compromise between accuracy and computational cost is obtained with $N_g = 10$\fh{, which is the number of Galerkin modes used throughout the rest of this paper}.

\begin{figure}[tb]
    \centering
    \begin{subfigure}[b]{0.49\textwidth}
        \subcaptionOverlay{\includegraphics[width=\textwidth]{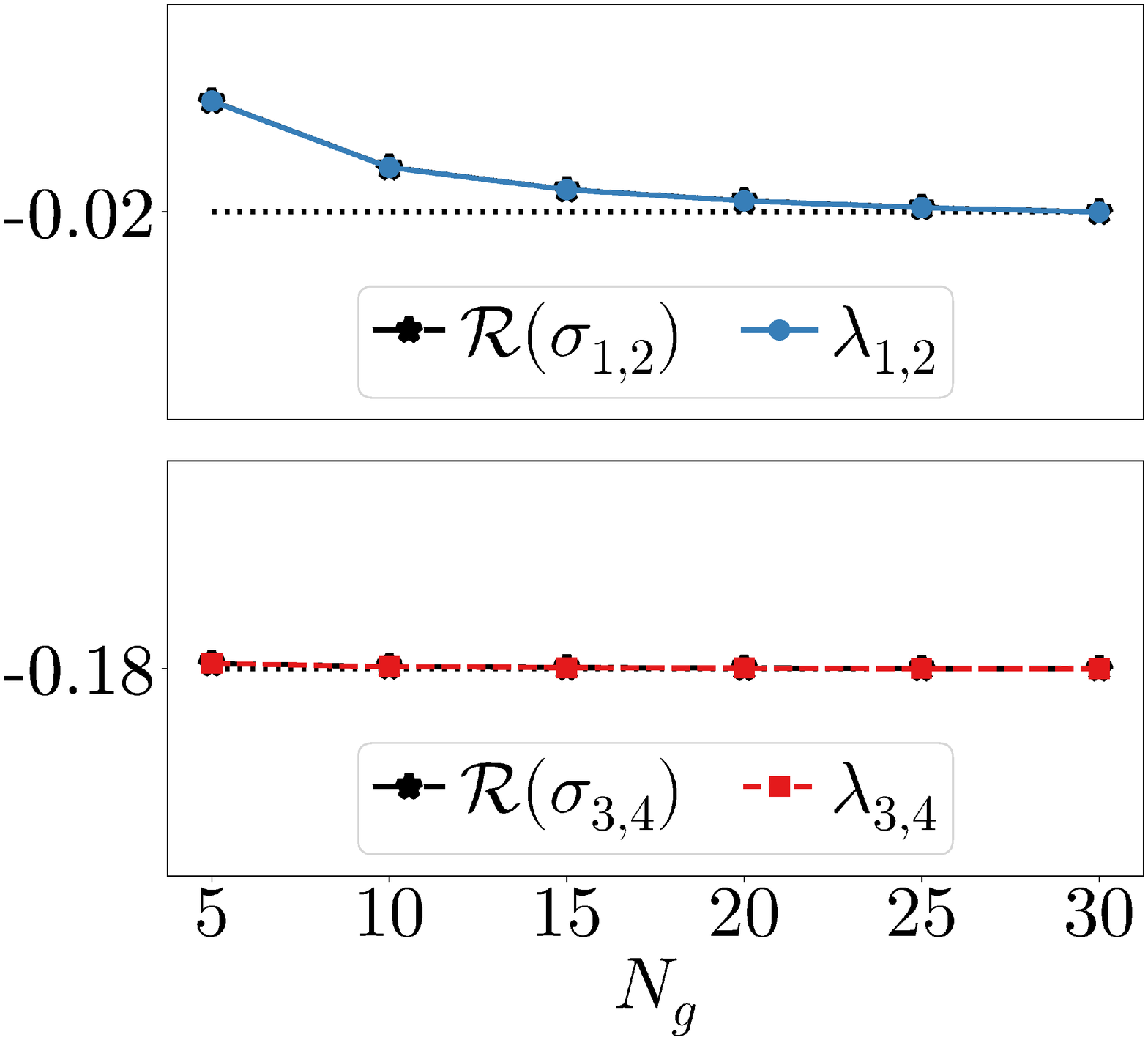}}
        \label{fig:systems:ng_conv_le}
    \end{subfigure}
    \begin{subfigure}[b]{0.49\textwidth}
        \subcaptionOverlay{\includegraphics[width=\textwidth]{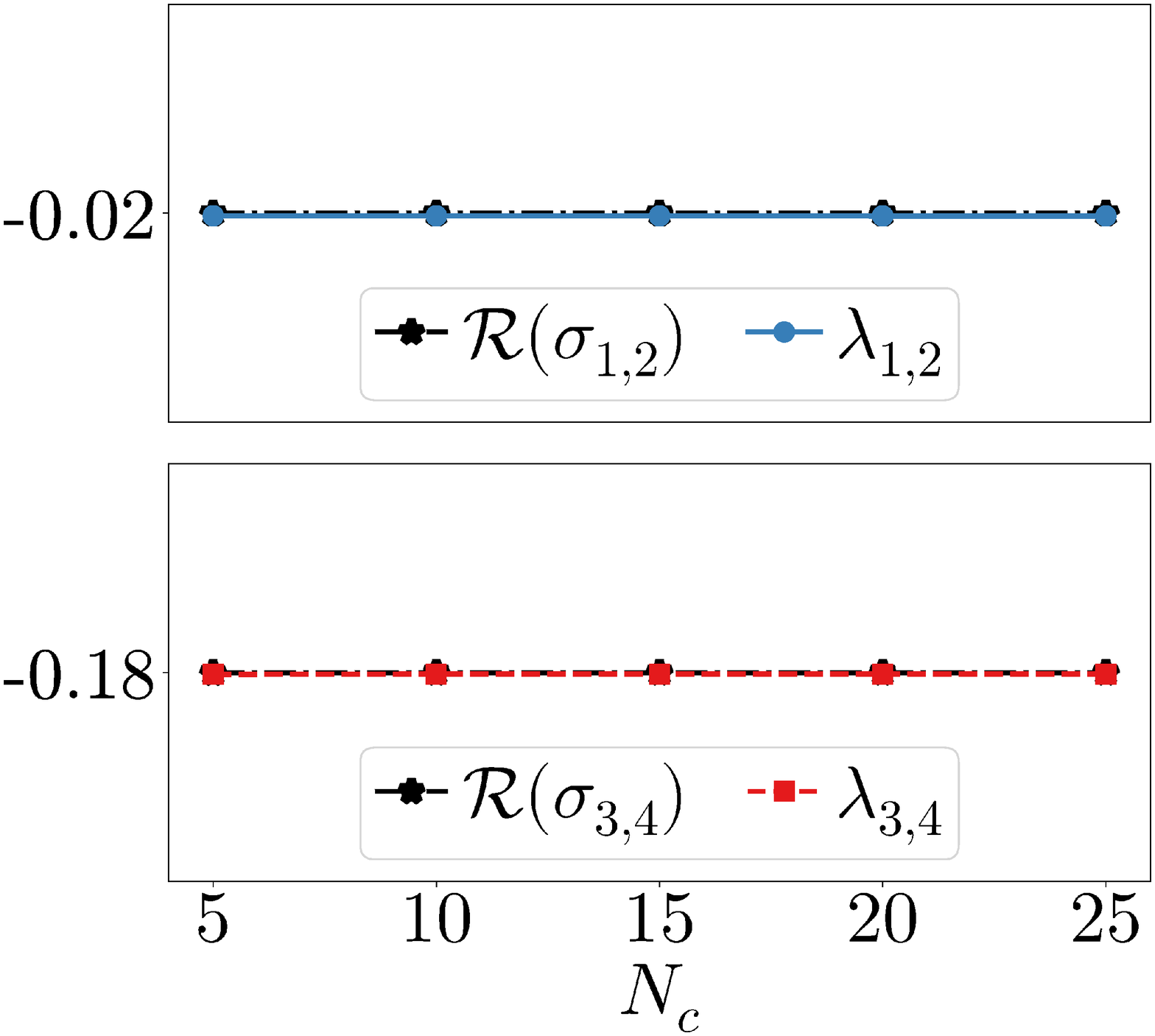}}
        \label{fig:systems:nc_conv_le}
    \end{subfigure}
    \caption{Convergence study on a fixed point ($\beta=0.4$, $\tau=0.2$). \subref{fig:systems:ng_conv_le} First four Lyapunov exponents (since they come in pairs, their mean is plotted) and real part of the corresponding eigenvalues, varying the number of Galerkin modes, $N_g$, with the number of Chebyshev points fixed to $N_c+1=11$. The vertical axes' ranges correspond to $\pm 0.5\%$ of the converged value (dotted line). \subref{fig:systems:nc_conv_le} Same as panel~\subref{fig:systems:ng_conv_le}, but with the number of Galerkin modes fixed to $N_g=10$ and number of Chebyshev points, $N_c+1$, varying. The vertical axes' ranges correspond to $\pm 5\%$ of the converged value (dotted line).}
    \label{fig:systems:conv_fp}
\end{figure}

In \S\ref{sec:eigen}, we showed analytically that, if the attractor is a fixed point, the Lyapunov exponents are equal to the real part of the eigenvalues of the Jacobian at the fixed point. Therefore, the difference between the two is a metric that can be used to assess the quality of the numerical solution, which is the second test. Figure~\ref{fig:systems:ng_conv_le} shows the evolution of the first four Lyapunov exponents, the real part of the corresponding eigenvalues and their converged values as the number of Galerkin modes is increased to \num{30}. As explained in \S\ref{sec:results:rijke}, the Lyapunov spectrum of this system is composed of Lyapunov exponents with double multiplicity. The residuals of these Lyapunov exponents are of the order of $10^{-4}$ at the end of the simulations. Once again, a good compromise between the accuracy of the first four Lyapunov exponents and computational cost is obtained with $N_g=10$. A similar analysis is run by fixing the Galerkin modes to $N_g=10$ and varying the Chebyshev points, $N_c+1$, (figures~\ref{fig:systems:nc_conv_q}, \ref{fig:systems:nc_conv_le}), which shows that the influence of the number of Chebyshev points is not significant. In this paper, we use $N_c+1=11$.

\section{Covariant Lyapunov vector analysis of nonlinear thermoacoustics}
\label{sec:results:rijke} 
\begin{figure}[tb]
	\centering
	\includegraphics[width=\textwidth]{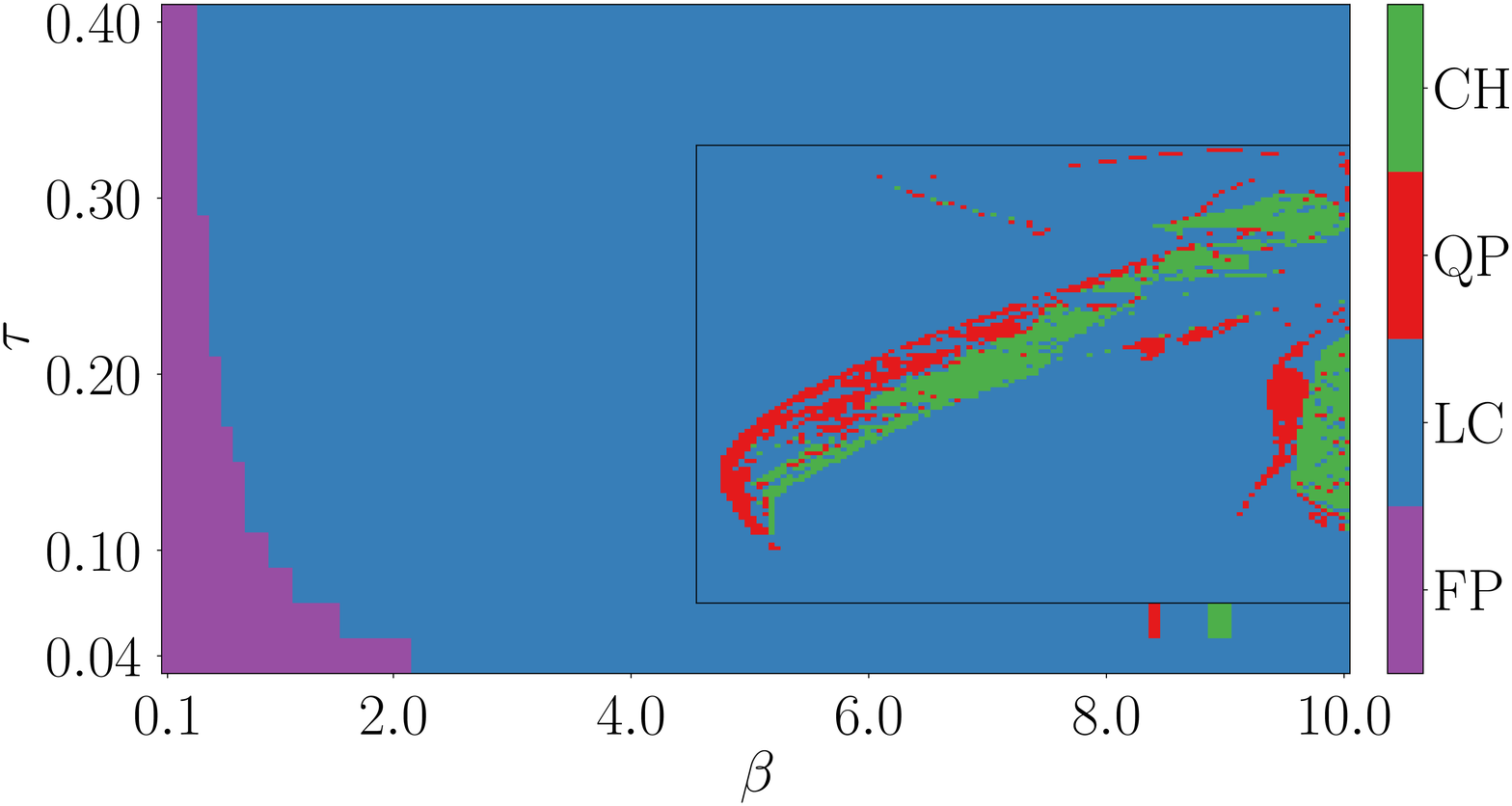}
    \caption{Bifurcation diagram of the thermoacoustic system with respect to the  parameters $\beta$ and $\tau$. The attractor classification is obtained by using the Lyapunov exponents of each solution \eqref{eq:classification}. The area marked by the black rectangle corresponds to a refined sweep. The coarse sweep is done with $\Delta \beta = 0.1$ and $\Delta \tau = 0.02$, while the fine sweep is done with $\Delta \beta = 0.05$ and $\Delta \tau = 0.002$.}
    \label{fig:rijke:bifurcation}
\end{figure}

\begin{figure}[tb]
    \centering
    \begin{subfigure}[b]{0.98\textwidth}
        \subcaptionOverlay{\includegraphics[width=\textwidth]{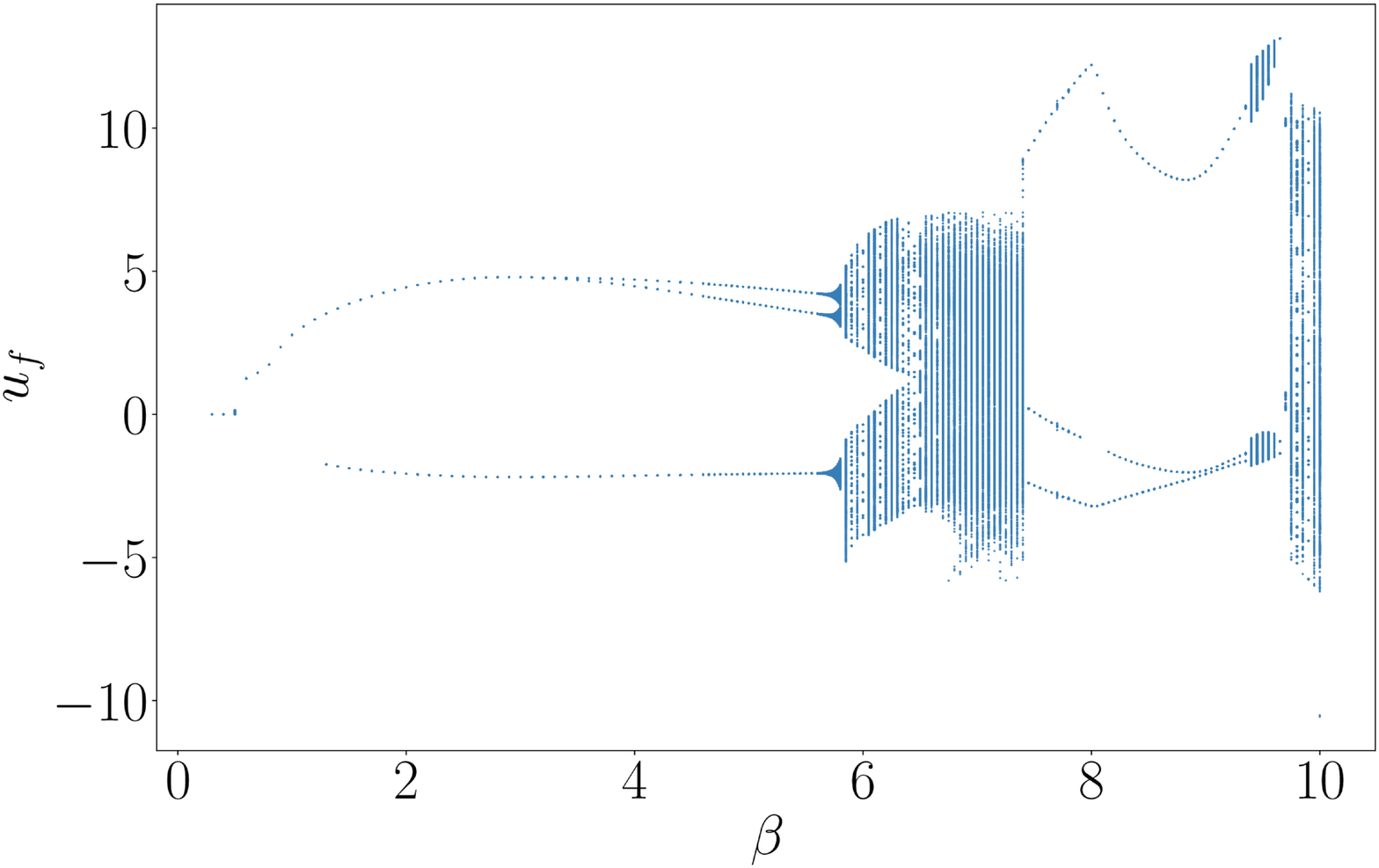}}
        \label{fig:rijke:bifurcation_uf}
    \end{subfigure}

    \begin{subfigure}[b]{0.98\textwidth}
        \subcaptionOverlay{\includegraphics[width=\textwidth]{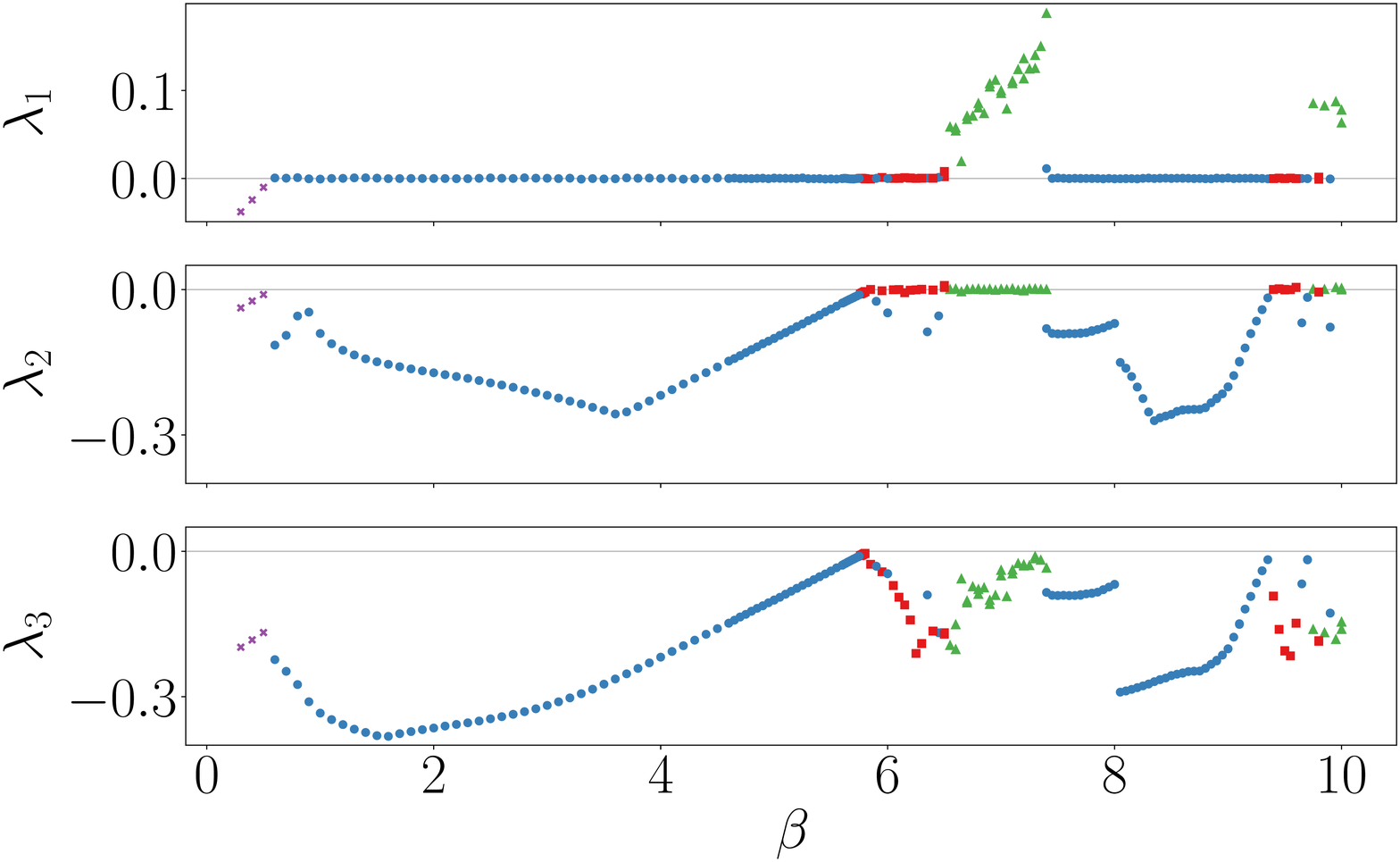}}
        \label{fig:rijke:bifurcation_le}
    \end{subfigure}
    \caption{Bifurcation diagrams of the thermoacoustic system versus $\beta$ ($\tau=0.2$). \subref{fig:rijke:bifurcation_uf} Local maxima of the time series of the acoustic velocity at the flame location, $u_f(t)$. \subref{fig:rijke:bifurcation_le} The first three Lyapunov exponents, $\lambda_1$, $\lambda_2$, $\lambda_3$, which determine the type of solution: fixed point (\textcolor{c4}{$\bm{\times}$}), limit cycle (\textcolor{c1}{\fullcirc}), quasi-periodic (\textcolor{c2}{\fullsquare}), chaotic (\textcolor{c3}{\fulltriangle}).}
    \label{fig:rijke:bifurcation1d}
\end{figure}

We study the time-delayed thermoacoustic system (\S\ref{sec:systems:rijke}) with $\tau = 0.2$ and $\bar{\bm q}_0 = [1 \, 0 \cdots 0]$, unless stated otherwise. The two-dimensional bifurcation diagram, which took seven days to be computed on a 16-core machine (Intel\textsuperscript{\textregistered} Xeon\textsuperscript{\textregistered} CPU E5-2620 v4 (2.10GHz)), is shown in figure~\ref{fig:rijke:bifurcation}. The solutions are classified according to their Lyapunov exponents (sorted in descending order)
\begin{equation}
    \label{eq:classification}
    \begin{cases}
        \text{Chaotic} & \text{if } \lambda_1 > 0 \\
        \text{Quasi-periodic} & \text{if } \lambda_1 = 0 \wedge \lambda_2 = 0 \\
        \text{Limit-cycle} & \text{if } \lambda_1 = 0 \wedge \lambda_2 < 0 \\
        \text{Fixed point} & \text{if } \lambda_1 < 0
    \end{cases}.
\end{equation}
With low $\beta$ and $\tau$, the system converges to a fixed point because  the energy input from the flame is not sufficient to overcome the damping for oscillations to persist. On the one hand, with constant low $\beta$, the solution bifurcates from fixed point to limit cycle as $\tau$ is increased. Physically, at this bifurcation point the Rayleigh's criterion (see \S\ref{sec:sens}) is fulfilled: the pressure and heat release are sufficiently in phase to balance the dissipation and give rise to a self-excited oscillation. On the other hand, when fixing $\tau$ between $0.08$ and $0.36$ and increasing $\beta$, different types of solution appear, such as quasi-periodic or chaotic attractors, especially at higher values of $\beta$, which suggests that a saturation of heat release is responsible for the emergence of these rich dynamics, as also observed in~\cite{Subramanian2010a}. This is further investigated with the one-dimensional bifurcation diagram (fixed $\tau=0.2$) of figure~\ref{fig:rijke:bifurcation1d}. Starting from $\beta=0.1$, the system bifurcates from fixed point to limit cycle. At first, the limit cycle becomes more stable as $\lambda_2$ decreases until $\beta \approx 3.6$. With further increasing $\beta$, the trend reverses and the limit cycle becomes less stable until $\beta \approx 5.8$, where a bifurcation from limit cycle to quasi-periodic attractor occurs. The zone $5.8 \lesssim \beta \lesssim 6.5$ corresponds to a quasi-periodic attractor, with the occasional limit cycle due to frequency locking. At the upper bound of this region, a new bifurcation occurs, passing from quasi-periodic to chaotic solutions. Thus, the system undergoes a Ruelle--Takens--Newhouse route to chaos in the case of increasing $\beta$ while $\tau = 0.2$. The chaotic attractor becomes ``more chaotic" in the sense that its leading Lyapunov exponent, $\lambda_1$, increases. This trend stops abruptly at $\beta \approx 7.4$, which coincides with $\lambda_3$ becoming approximately 0 and the system bifurcates from chaotic to periodic. This most likely represents a period-doubling route to chaos in reverse (from high to low $\beta$), which is not captured due to lack of resolution in $\beta$. Two more major bifurcations occur as $\beta$ is increased to 10: limit cycle to quasi-periodic and quasi-periodic to chaotic.

\subsection{Analysis on a fixed-point attractor}
\label{sec:results:rijke:fp}
\begin{figure}
	\centering
    \includegraphics[width=0.98\textwidth]{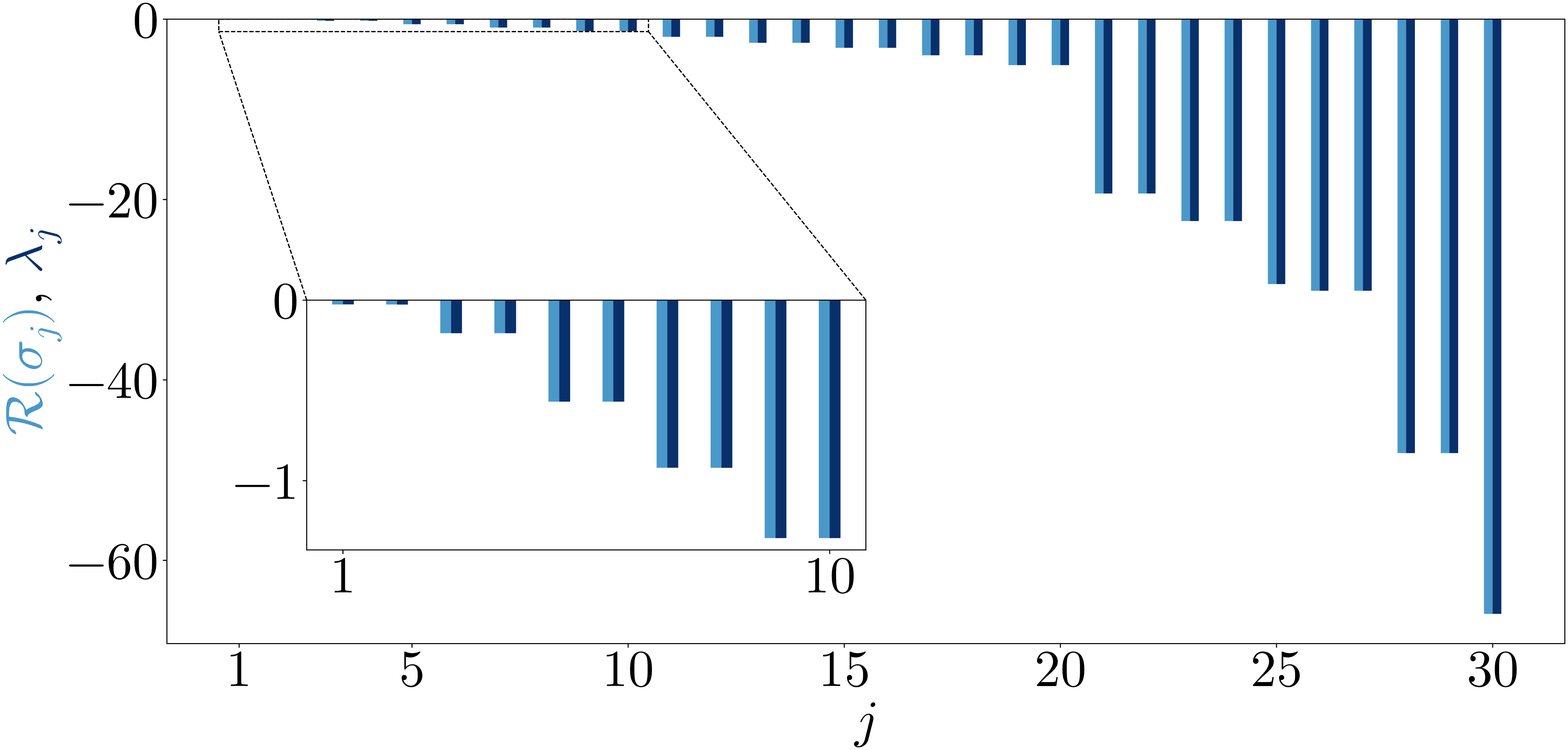}
    \caption{Real part of the eigenvalues (\tikzbar{blue1}) and Lyapunov spectrum (\tikzbar{blue3}) match on the fixed point solution $\bar{\bm q}=0$ ($\beta=0.4$).}
    \label{fig:rijke:fp:spectrum}
\end{figure}

\begin{figure}[!ht]
	\centering
    \begin{subfigure}[b]{0.49\textwidth}
        \subcaptionOverlay{\includegraphics[width=\textwidth]{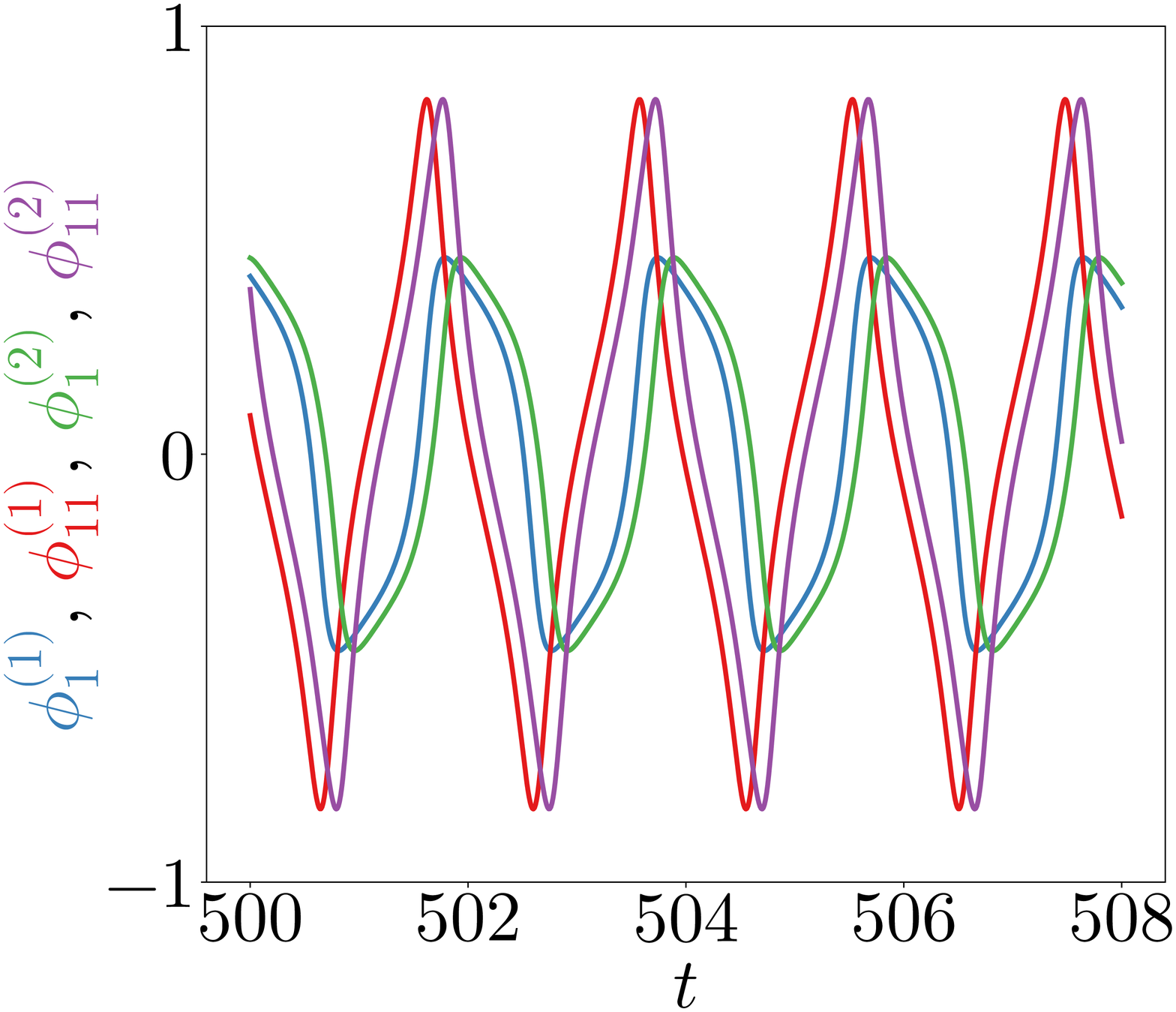}}
        \label{fig:rijke:fp:clv}
    \end{subfigure}
    \begin{subfigure}[b]{0.49\textwidth}
        \subcaptionOverlay{\includegraphics[width=\textwidth]{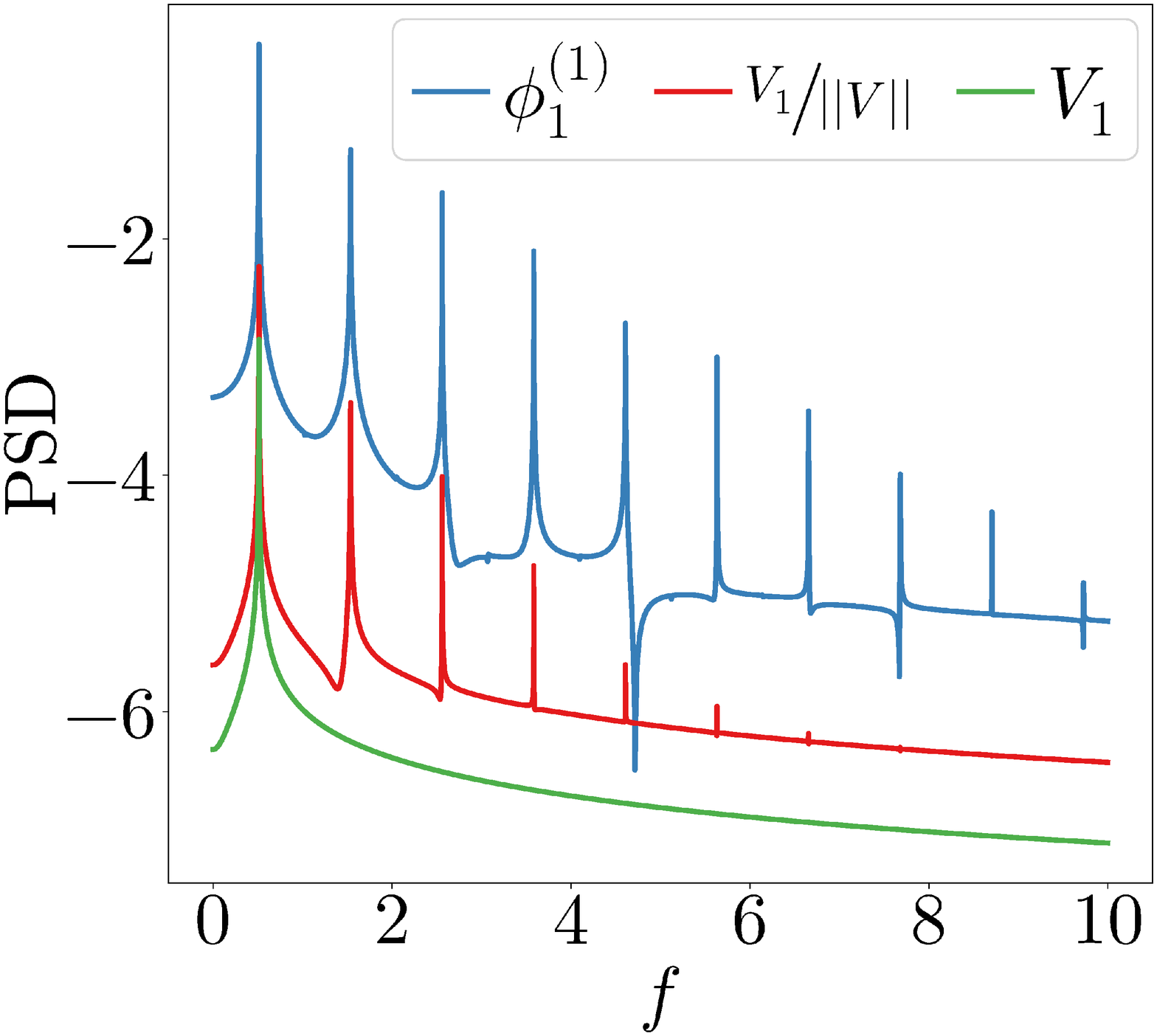}}
        \label{fig:rijke:fp:frequency}
    \end{subfigure}
    \caption{Fixed point solution $\bar{\bm q}=0$ ($\beta=0.4$).
    \protect\subref{fig:rijke:fp:clv} Velocity and pressure components of the first Galerkin mode, $\eta_1$ and $\mu_1$, of the covariants Lyapunov vectors $\bm \phi^{(1)}$ and $\bm \phi^{(2)}$: $\bm \phi_{1}^{(1)}$ (\tikzline{c1}), $\bm \phi_{11}^{(1)}$ (\tikzline{c2}), $\bm \phi_{1}^{(2)}$ (\tikzline{c3}) and $\bm \phi_{11}^{(2)}$ (\tikzline{c4}) versus time -- each component oscillates between the corresponding component of $\pm \sqrt{\mathcal{R}(\hat{\bm q})^2 + \mathcal{I}(\hat{\bm q})^2}$, where $\hat{\bm q}$ is the corresponding eigenvector;
    \protect\subref{fig:rijke:fp:frequency} Power spectral density of $\bm \phi_1^{(1)}$ (\tikzline{c1}), $\sfrac{\bm V_1}{||\bm V||}$ (\tikzline{c2}) and $\bm V_1$ (\tikzline{c3}), where $\bm V$ is the first vector of \eqref{eq:theoretical_clv}. The vertical axis is logarithmic of base 10.}
\end{figure}

\begin{figure}[tb]
	\centering
	\includegraphics[width=0.49\textwidth]{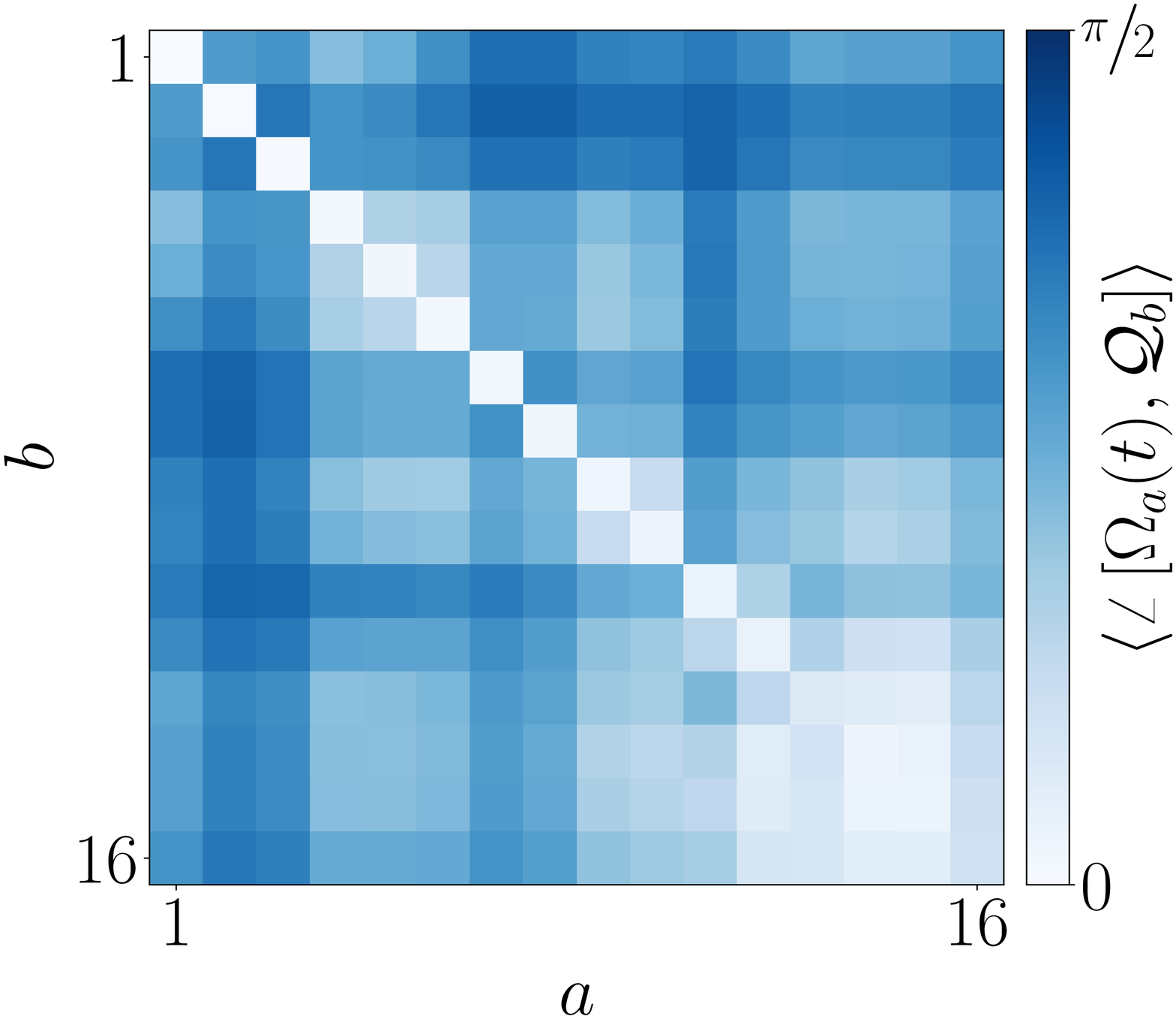}
	\caption{Mean angle between Lyapunov subspaces and eigensubspaces, $\angle(\Omega_a, \mathcal{Q}_b)$, on the fixed point solution $\bar{\bm q}=0$ ($\beta=0.4$). (Standard deviation not shown because it is \num{0} to precision.) The main diagonal, which corresponds to the angles between Lyapunov subspaces and eigenspaces of the same growth rate (same index), is \num{0}, showing that covariant Lyapunov vectors or eigenvectors of the same growth rate span the same subspaces.}
    \label{fig:rijke:fp:angles_clv_eigvecs}
\end{figure}

By setting $\beta=0.4$, the solution converges to the fixed point $\bar{\bm q} = 0$. The spinup and spindown times are the same and equal to $200$ time units, while the simulation time is $1000$ with a time segment of $0.01$, yielding $[200,800]$ as the interval of study. The spinup time is chosen \textit{a posteriori}, by choosing a time such that the system is past the transient regime.
%
Figure~\ref{fig:rijke:fp:spectrum} shows the Lyapunov exponents and the real part of the eigenvalues of the Jacobian of the system at $\bar{\bm q} = 0$, demonstrating that the Lyapunov spectrum matches the real part of the eigenvalues. There are \num{16} distinct values of Lyapunov exponents, \num{14} of which have multiplicity two, corresponding to Lyapunov subspaces of dimension \num{2}, while $\lambda_{25}$ and $\lambda_{30}$ correspond to one-dimensional Lyapunov subspaces, as described in \S\ref{sec:eigen}, for a total of \num{16} Lyapunov subspaces: $\Omega_1, \cdots, \Omega_{16}$
The velocity and pressure components of the first Galerkin mode, $\eta_1$ and $\mu_1$, of the covariant Lyapunov vectors $\bm \phi^{(1)}$ and $\bm \phi^{(2)}$ are denoted $\bm \phi_{1}^{(1)}$, $\bm \phi_{11}^{(1)}$, $\bm \phi_{1}^{(2)}$ and $\bm \phi_{11}^{(2)}$, respectively (note that the $11$-th component of the state vector corresponds to $\mu_1$ in the arrangement of \S\ref{sec:numericaldiscretization}). They are plotted in figure~\ref{fig:rijke:fp:clv}. The time series are not purely sinusoidal, as predicted by the analytical result \eqref{eq:theoretical_clv} of \S\ref{sec:eigen}, because the covariant Lyapunov vectors are defined up to a time-varying factor. This time-varying factor is the normalisation that is imposed in the Gram--Schmidt orthonormalization (i.e. the QR decomposition) in step \ref{item:qr} in~\S\ref{sec:theory:qr}. This normalisation varies in time because it is repeated at every time segment. Therefore, the time-varying normalisation generates higher harmonics in the power spectral density. This can be seen by comparing the power spectral densities of $\bm \phi_1^{(1)}$, $\bm V_1$ and $\bm V_1/||\bm V||$ (figure~\ref{fig:rijke:fp:frequency}), where
$\bm V$ is the first vector of~\eqref{eq:theoretical_clv}. While $\bm V_1$ expectedly presents one mode only at $f_1 = \omega_1/2\pi \approx 0.6$, $\sfrac{\bm{V_1}}{||\bm V||}$ has peaks at frequencies of the form $k f_1, \,  k \in \{1, 3, 5, \dots\}$, exactly like $\bm \phi_1^{(1)}$.
%
The mean of the angles between the Lyapunov subspaces and the eigensubspaces are shown in figure~\ref{fig:rijke:fp:angles_clv_eigvecs}. (Eigensubspaces are subspaces that are spanned by eigenvectors corresponding to eigenvalues with the same real part, e.g. pair of complex conjugates, as described in \S\ref{sec:eigen}.) The main diagonal, which compares Lyapunov subspaces to eigensubspaces of the same growth rate, is 0 (to precision), showing that the Lyapunov subspaces are indeed equal to the (constant) eigensubspaces.
The analysis of this section numerically shows the equivalence between eigenvectors and covariant Lyapunov vectors on stable fixed points, as analytically explained in~\S\ref{sec:eigen}.
\subsection{Analysis on a periodic attractor}
\label{sec:results:rijke:lc}
\begin{figure}[tb]
	\centering
    \begin{subfigure}[b]{0.49\textwidth}
        \subcaptionOverlay{\includegraphics[width=\textwidth]{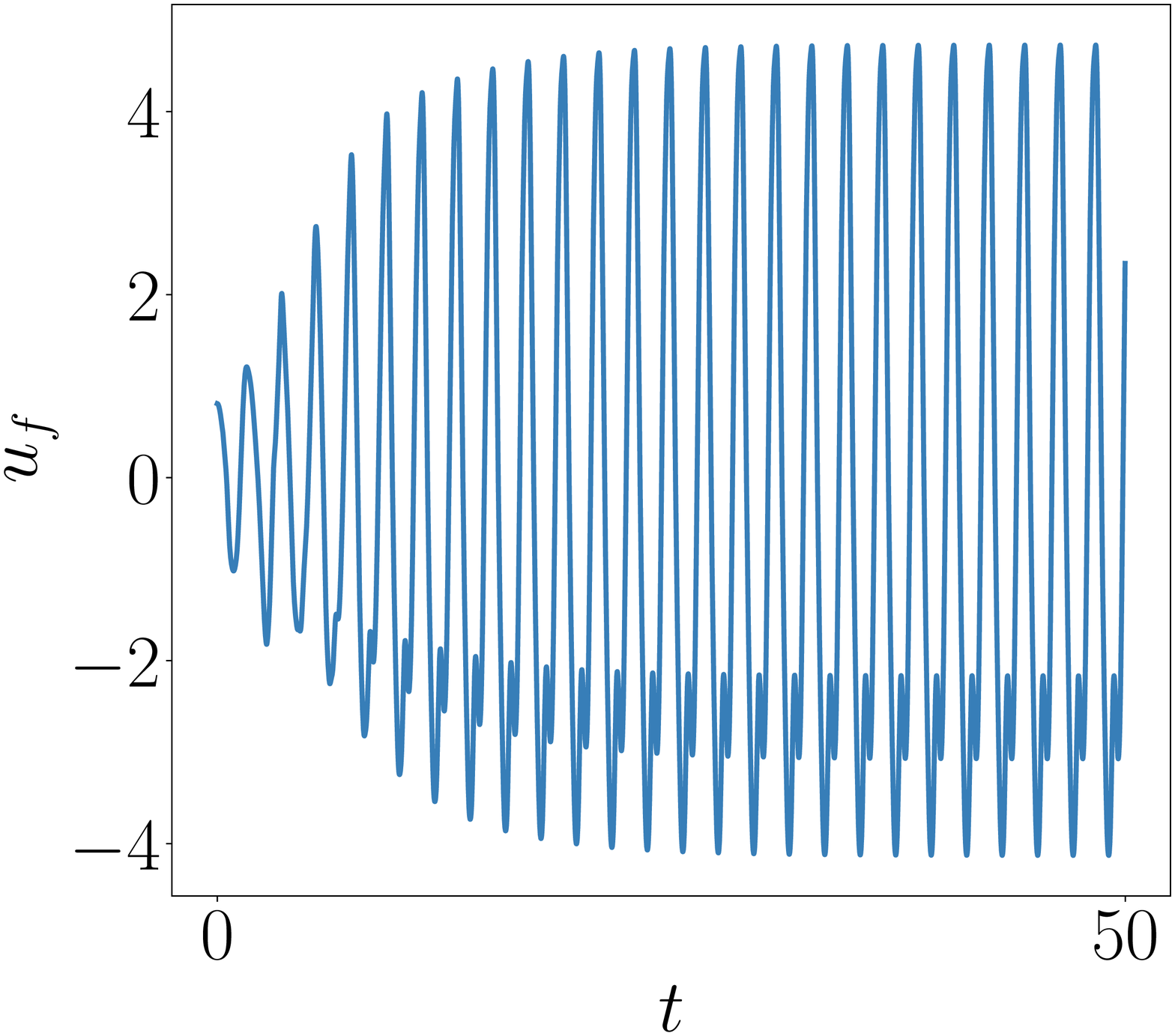}}
        \label{fig:rijke:lc:uf_time}
    \end{subfigure}
    \begin{subfigure}[b]{0.49\textwidth}
        \subcaptionOverlay{\includegraphics[width=\textwidth]{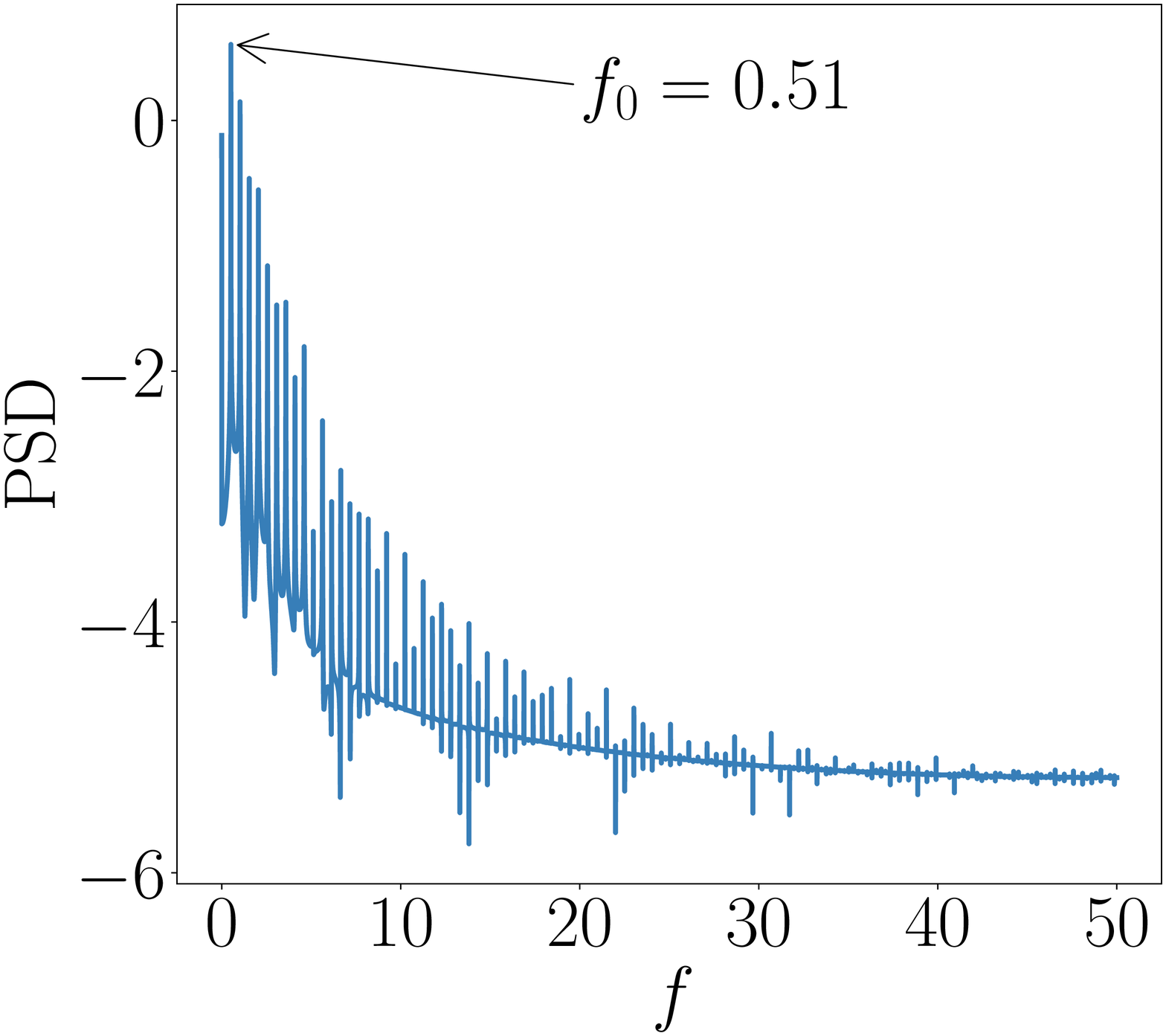}}
        \label{fig:rijke:lc:uf_freq}
    \end{subfigure}
    \caption{Acoustic velocity at the heat-source location, $u_f(t)$, of a limit cycle ($\beta=2.5$). \subref{fig:rijke:lc:uf_time} Time series with initial transient (the full simulation period (1000) is not depicted). $u_f(t)$ becomes periodic with period $1.95$ at $t \approx 30$. \subref{fig:rijke:lc:uf_freq} Power spectral density. The vertical axis is logarithmic of base 10. The global maximum is at $f_0=0.51$ and the other local maxima are its higher harmonics.}
\end{figure}

By increasing the heat-release intensity parameter to $\beta = 2.5$, the thermoacoustic system converges to a limit cycle. The spinup and spindown times are the same and equal to $200$ time units, while the simulation time is $1000$ with a time segment of $0.01$, yielding $[200,800]$ as the interval of study. The velocity at the heat source $u_f(t)$ (figure~\ref{fig:rijke:lc:uf_time}) oscillates within $[-4.13, 4.73]$. The fact that the minima and maxima of $u_f(t)$ are not equal in absolute value can be explained by the asymmetry of the heat-release law~\eqref{eq:modified_kings}. The period is $T_0=1.95$, corresponding to a frequency of $f_0=0.51$, which appears in the power spectral density of $u_f(t)$ (figure~\ref{fig:rijke:lc:uf_freq}) as a maximum peak. The subsequent peaks occur at $k f_0, \, k \in \{2, 3, \dots\}$. The frequency $f_0=0.51$ has a value that is close to the natural acoustic frequency of the first mode of the duct, which is $f=\pi/(2\pi)=0.5$ (\S\ref{sec:numericaldiscretization}). The frequency shift is physically due to the effect of the heat release and damping. 

\begin{figure}[tb]
	\centering
    \includegraphics[width=0.98\textwidth]{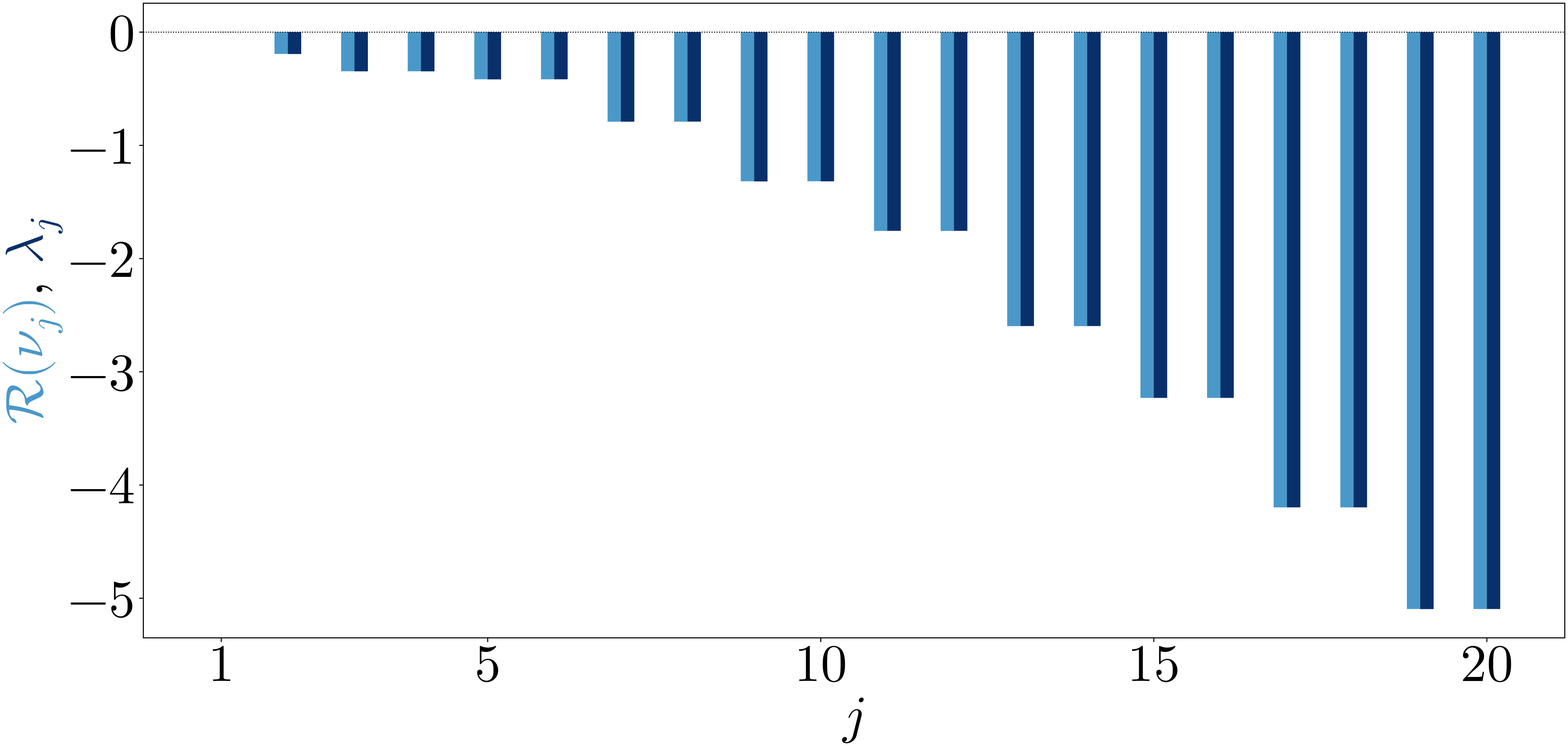}
    \caption{Real part of the Floquet exponents (\tikzbar{blue1}) and Lyapunov spectrum (\tikzbar{blue3}) (first 20) match on a limit cycle solution ($\beta=2.5$). The remaining 10 Floquet exponents are not shown because their values are large in absolute value and the accuracy of their calculation is limited by machine precision in the computation of the monodromy matrix.}
    \label{fig:rijke:lc:spectrum}
\end{figure}

Figure~\ref{fig:rijke:lc:spectrum} shows the real part of the first \num{20} Floquet exponent and the corresponding Lyapunov exponents. The remaining 10 Floquet exponents are not shown because their values are large in absolute value and the accuracy of their calculation is limited by machine precision in the computation of the monodromy matrix (e.g. $e^{\lambda_{21} T} \approx e^{-20 \times 2} \sim 10^{-18}$). These modes are non-physical and correspond to the Chebyshev discretisation of the advection equation, as described \S\ref{sec:numericaldiscretization}.
The first Lyapunov and Floquet exponents are zero (to a small numerical error), i.e. they are the neutral modes, and they correspond to vectors tangent to the limit cycle.
%
The second Lyapunov exponent is the least stable mode, which corresponds to a one-dimensional Lyapunov subspace. 
The remaining Lyapunov exponents have multiplicity two and match the real part of the Floquet exponents, the latter of which come as pairs of complex conjugates. 
%
%
%
As explained in \S\ref{sec:results:rijke:fp} and the present section, the fact the Lyapunov subspaces have double multiplicity has a physical interpretation. The thermoacoustic dynamics is driven by the nonlinear saturation of the thermoacoustic eigenfunctions, which come as complex conjugate pairs. As shown in \S\ref{sec:results:rijke:ch}, this physical mechanism is dominant even when the system is chaotic.

\begin{figure}[tb]
	\centering
    \begin{subfigure}[b]{0.49\textwidth}
        \subcaptionOverlay{\includegraphics[width=\textwidth]{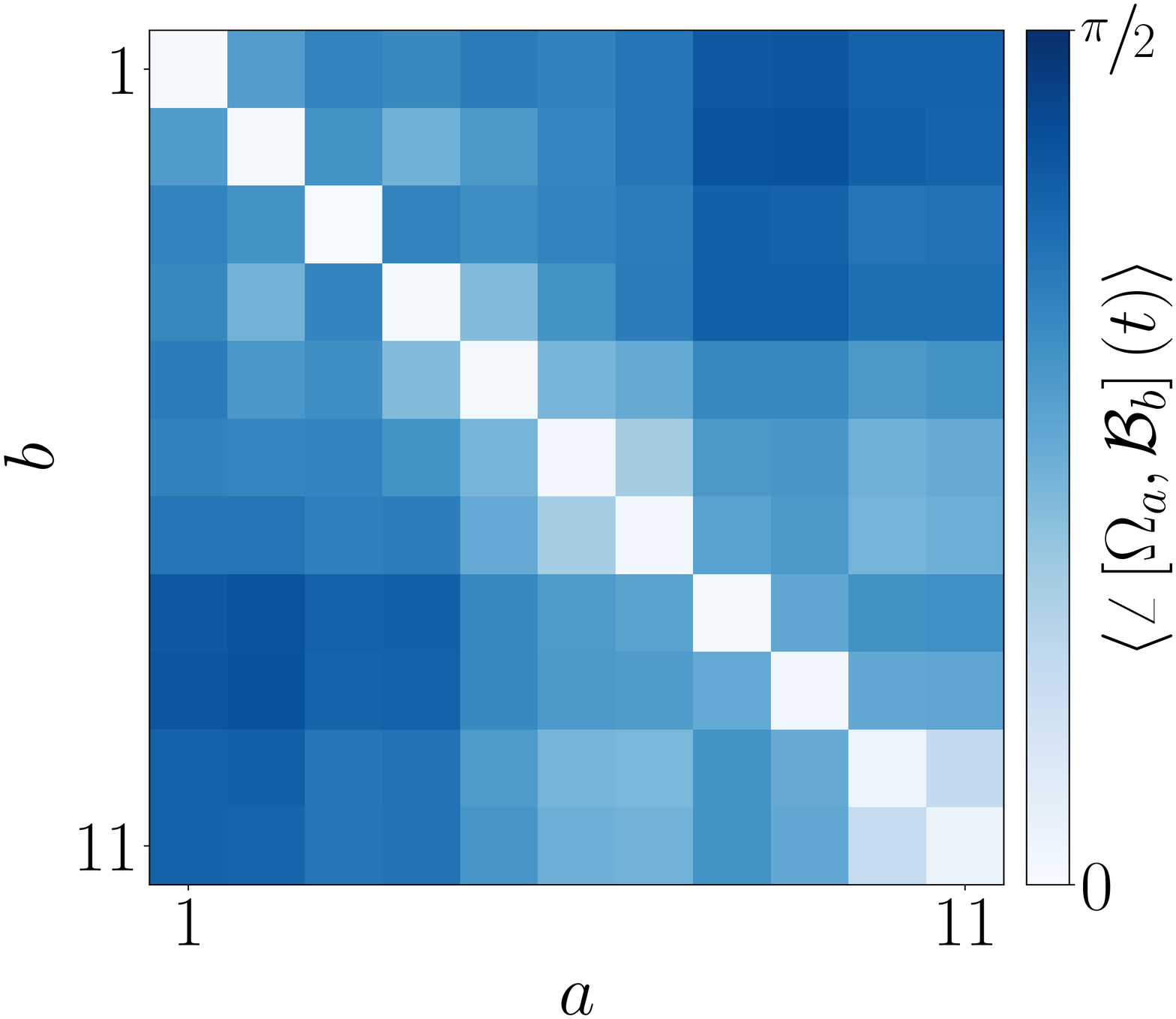}}
        \label{fig:rijke:lc:angles_mean}
    \end{subfigure}
    \begin{subfigure}[b]{0.49\textwidth}
        \subcaptionOverlay{\includegraphics[width=\textwidth]{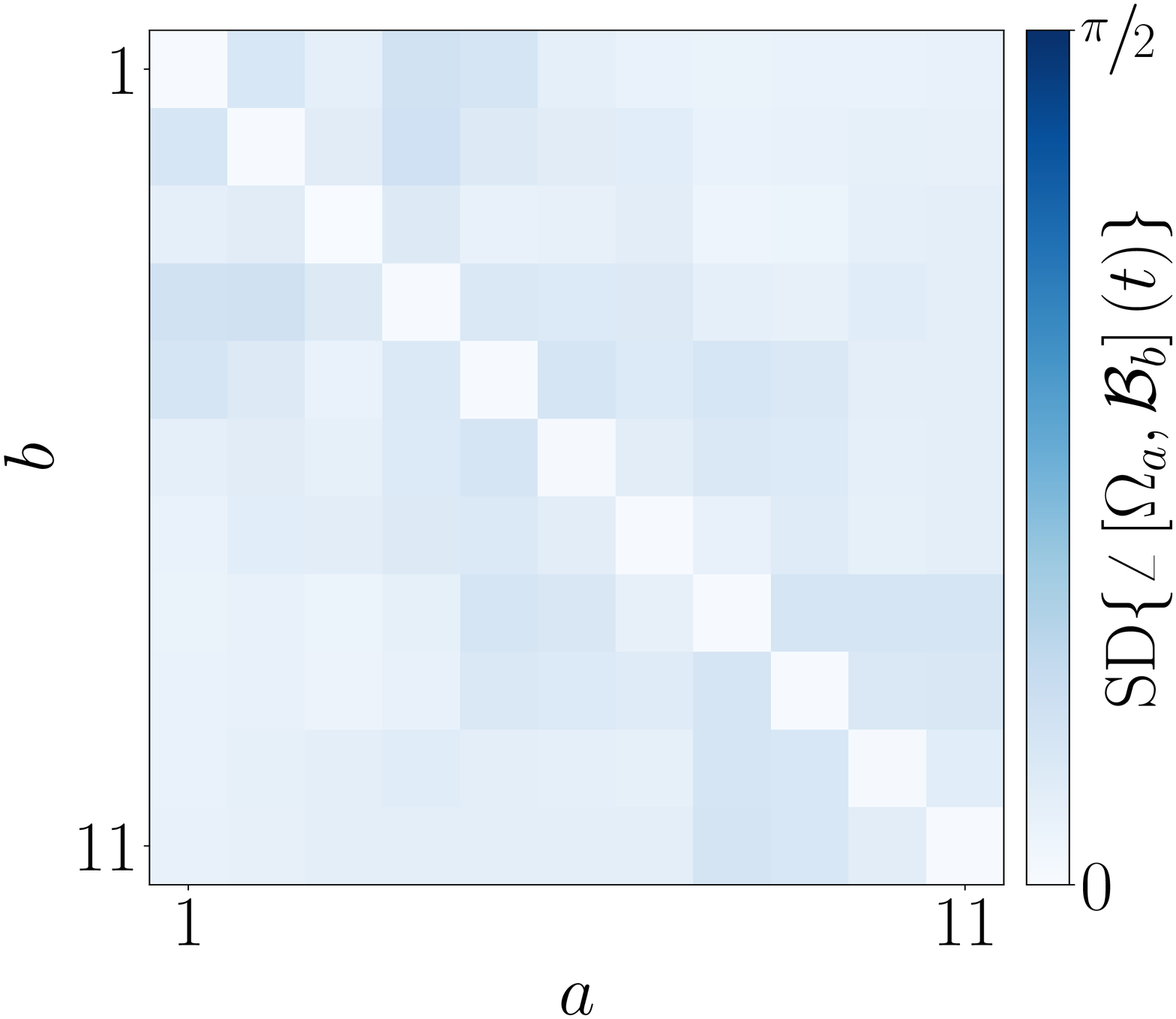}}
        \label{fig:rijke:lc:angles_std}
    \end{subfigure}
    \caption{Angles between Lyapunov subspaces and Floquet subspaces, $\angle(\Omega_a, \mathcal{B}_b)$, on a limit cycle solution ($\beta=2.5$). \protect\subref{fig:rijke:lc:angles_mean} Mean. \protect\subref{fig:rijke:lc:angles_std} Standard deviation. The main diagonals of both figures are \num{0}, showing that Floquet vectors (eigenvectors of the monodromy matrix) or covariant Lyapunov vectors of the same growth rate span the same subspaces.}
    \label{fig:rijke:lc:angles}
\end{figure}

Figures~\ref{fig:rijke:lc:angles_mean}, \ref{fig:rijke:lc:angles_std} show the mean and the standard deviation, respectively, of the angle between the Lyapunov subspaces and the Floquet subspaces (subspaces spanned by groups of eigenvectors of the monodromy matrix that have the same real part of the Floquet exponent). The fact that the main diagonal of figure~\ref{fig:rijke:lc:angles_std} is \num{0} demonstrates that these angles are constant, while the fact that the main diagonal of figure~\ref{fig:rijke:lc:angles_mean} is also \num{0} shows that the constant angles are \num{0}. Thus, the Lyapunov subspaces are equal to the Floquet subspaces.
Similarly to \S\ref{sec:results:rijke:fp}, the analysis of this section numerically shows the equivalence between Floquet vectors (eigenvectors of the monodromy matrix) and covariant Lyapunov vectors on stable limit cycles, as analytically explained in~\S\ref{sec:floquet}.
%
%
\subsection{Analysis on a chaotic attractor}
\label{sec:results:rijke:ch}
\begin{figure}[tb]
	\centering
    \begin{subfigure}[b]{0.49\textwidth}
        \subcaptionOverlay{\includegraphics[width=\textwidth]{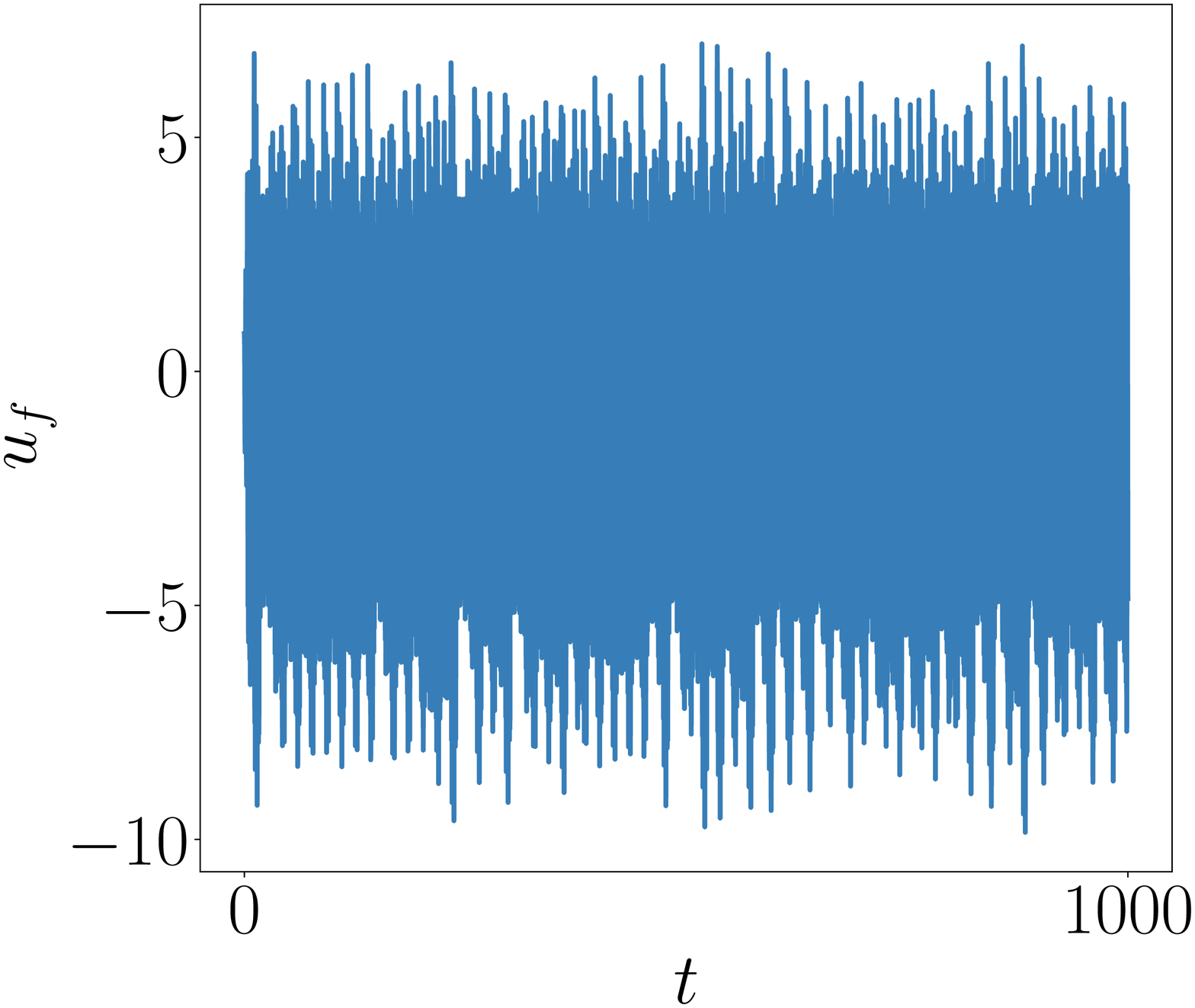}}
        \label{fig:rijke:ch:uf_time}
    \end{subfigure}
    \begin{subfigure}[b]{0.49\textwidth}
        \subcaptionOverlay{\includegraphics[width=\textwidth]{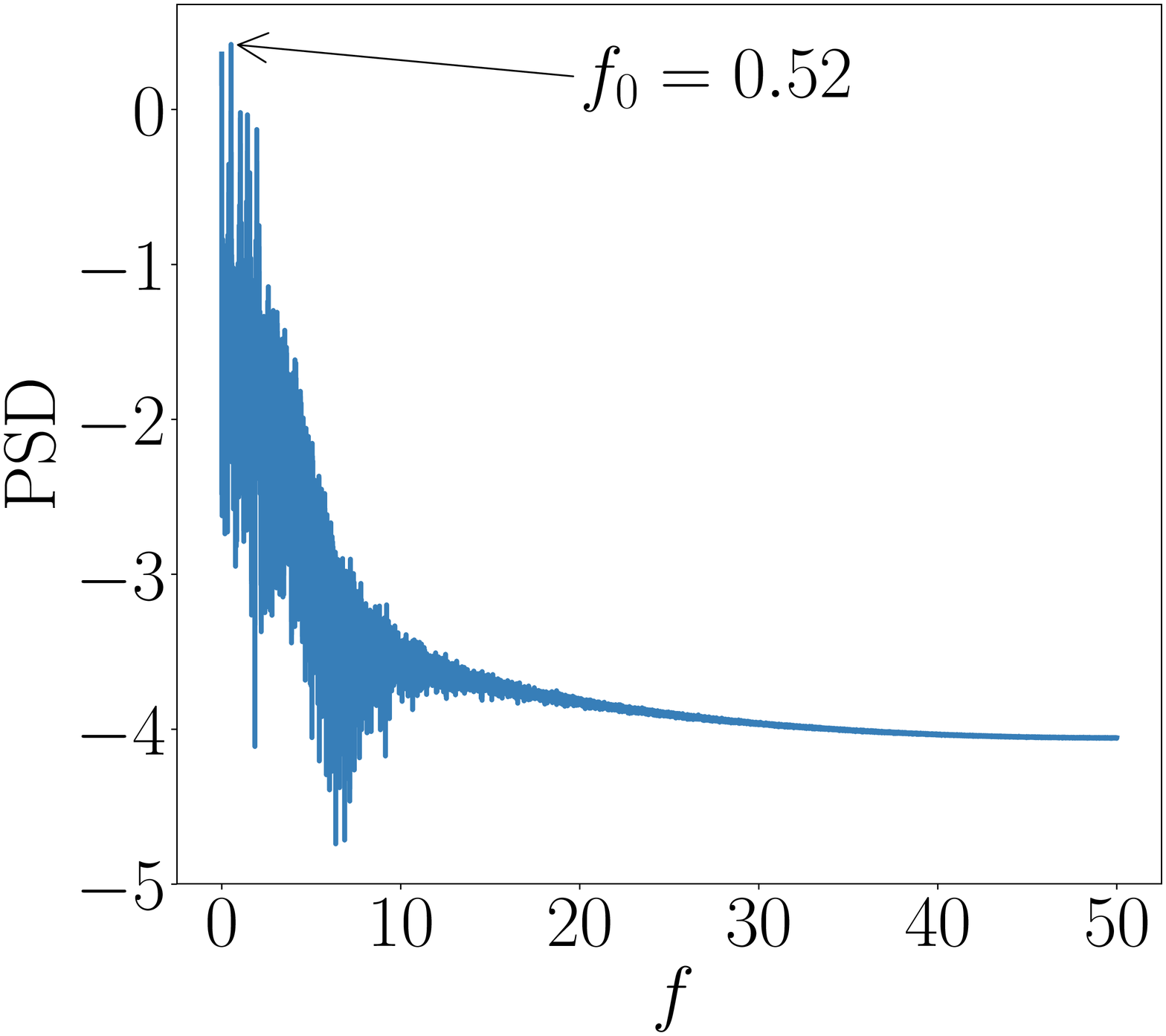}}
        \label{fig:rijke:ch:uf_freq}
    \end{subfigure}
    \caption{Acoustic velocity at the heat-source location, $u_f(t)$, on a chaotic solution ($\beta=7.0$). \subref{fig:rijke:ch:uf_time} Time series. $u_f(t)$ is oscillatory, but aperiodic. \subref{fig:rijke:ch:uf_freq} Power spectral density. The global maximum is at $f_0 = 0.52$, which is close to the global maximum found in the limit cycle case of \S\ref{sec:results:rijke:lc}.}
\end{figure}

\begin{figure}[tb]
	\centering
    \includegraphics[width=\textwidth]{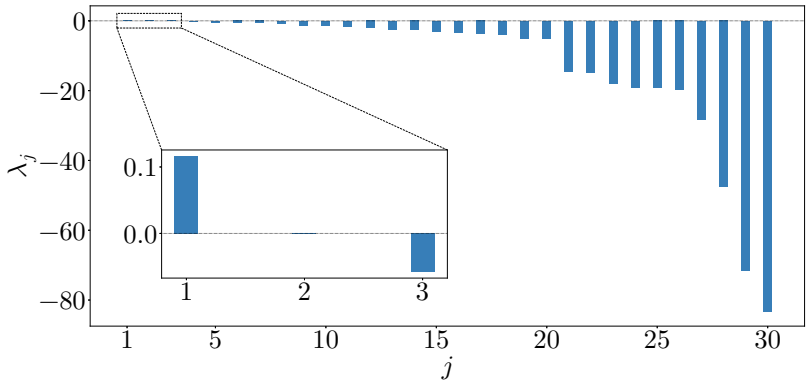}
	\caption{Complete Lyapunov spectrum, with a closeup of the first \num{3} Lyapunov exponents, on a chaotic attractor ($\beta=7.0$). $\lambda_1 \approx 0.12$ and $\lambda_2 = 0$ to numerical error (neutral mode). The remaining Lyapunov exponents are negative.}
	\label{fig:rijke:ch:spectrum}
\end{figure}

The heat-release intensity parameter is further increased to $\beta=7.0$, with the system converging to a chaotic attractor. In a chaotic solution, only covariant Lyapunov vector analysis can calculate the linear dynamics of the attractor. Eigenvalue and Floquet analyses are no longer valid.
The acoustic velocity at the base of the heat source, $u_f(t)$, is oscillatory (figure~\ref{fig:rijke:ch:uf_time}) but aperiodic. While the dominant peak of the power spectral density (figure~\ref{fig:rijke:ch:uf_freq}) is largely unaltered from \S\ref{sec:results:rijke:lc}, the frequency spectrum is now denser and shows multiple peaks at several frequencies, which indicates the presence of chaos. However, quasi-periodic solutions can exhibit the same behaviour and can be mistaken for chaotic. It is hard thus to classify the attractor from figures~\ref{fig:rijke:ch:uf_time},~\ref{fig:rijke:ch:uf_freq} alone.
Instead, we use the Lyapunov spectrum (figure~\ref{fig:rijke:ch:spectrum}). The first Lyapunov exponent, $\lambda_1 \approx 0.12$, is positive, thus confirming that the attractor is indeed chaotic: $\bm \phi_2(t)$ is the neutral covariant Lyapunov vector because $\lambda_2=0$ to numerical error. The remaining Lyapunov exponents correspond to one-dimensional modes, except for the pairs $(\lambda_9$, $\lambda_{10})$, $(\lambda_{19}$, $\lambda_{20})$, $(\lambda_{21}$, $\lambda_{22})$ and $(\lambda_{24}, \lambda_{25})$, each corresponding to two-dimensional Lyapunov subspaces. Thus, we conclude that
\begin{align}
    E_q^u &= \text{Span}(\bm \phi_1), \\
    E_q^n &= \text{Span}(\bm \phi_2), \\
    E_q^s &= \text{Span}(\bm \phi_3, \cdots, \bm \phi_{30}) .
\end{align}
For sensitivities to exist, the angles between these subspaces must be bounded away from \num{0} (), that is, the attractor must be hyperbolic. In the next section, this question is investigated on multiple design points where the system exhibits chaotic behaviour.

\subsection{Are chaotic acoustic attractors hyperbolic?}
\label{sec:results:rijke:hyp}
\begin{figure}[!ht]
	\centering
    \begin{subfigure}[b]{0.49\textwidth}
        \subcaptionOverlay{\includegraphics[width=\textwidth]{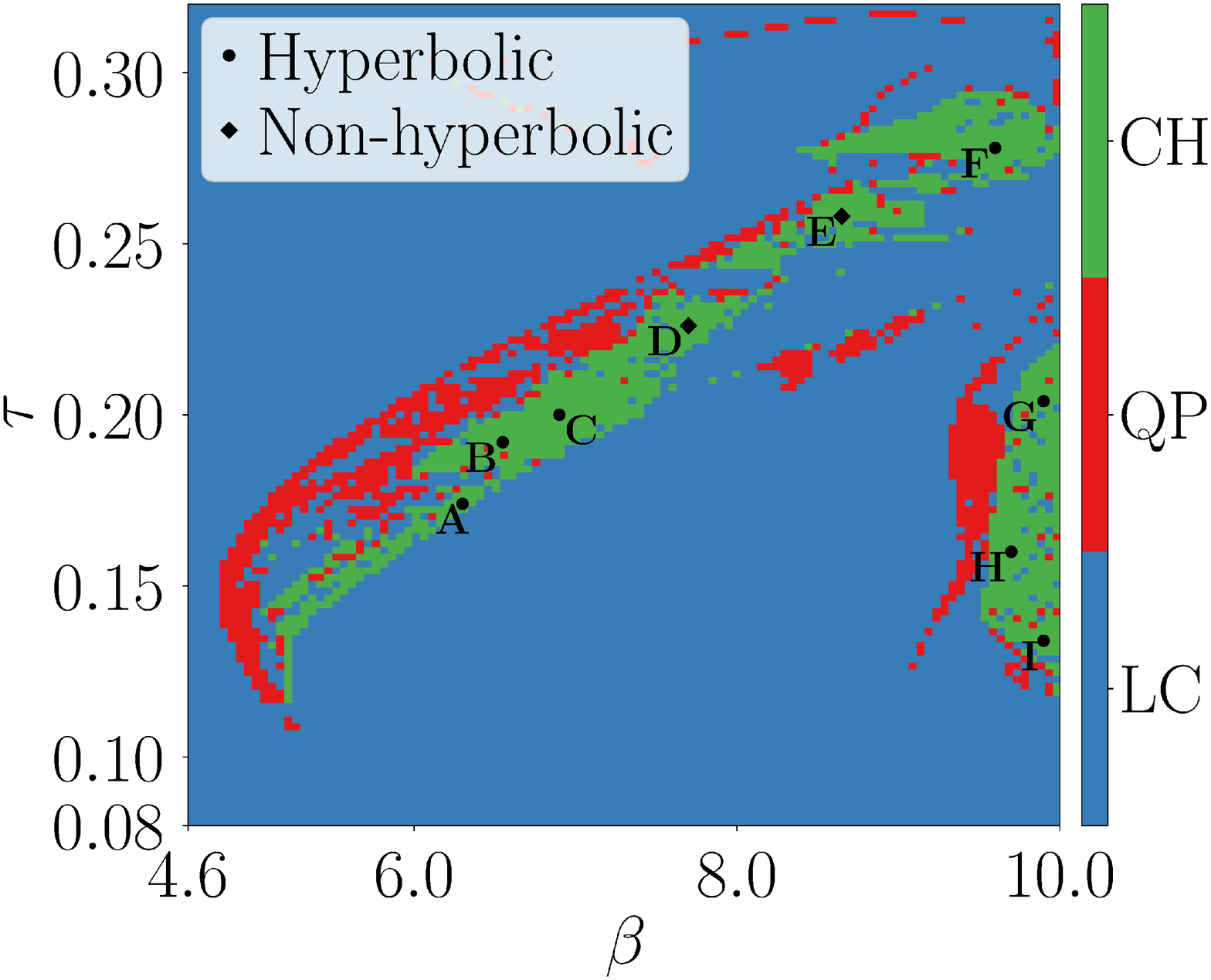}}
        \label{fig:rijke:hyp:probes}
    \end{subfigure}
    \begin{subfigure}[b]{0.49\textwidth}
        \subcaptionOverlay{\includegraphics[width=\textwidth]{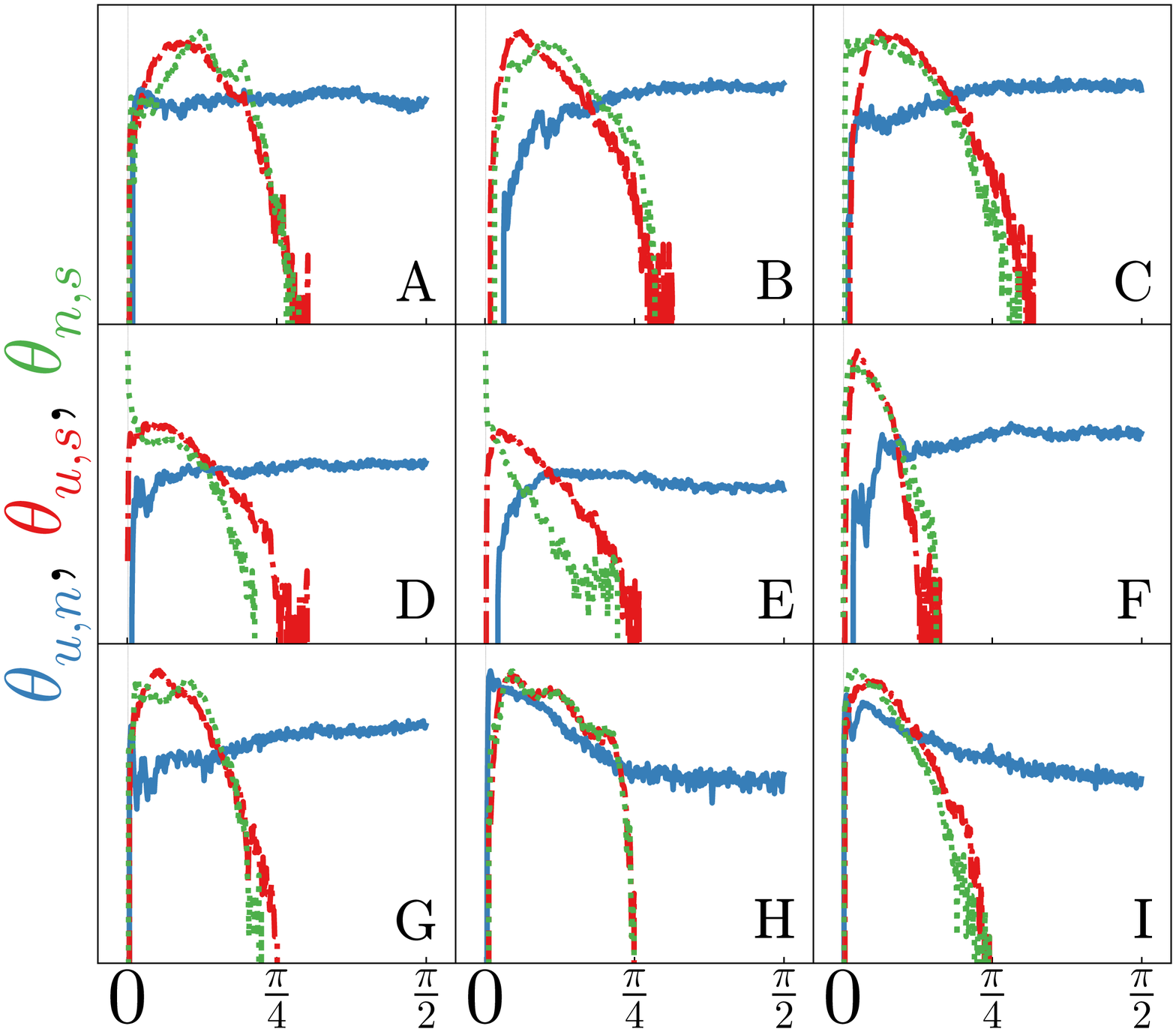}}
        \label{fig:rijke:hyp:pdf}
    \end{subfigure}
	\caption{\subref{fig:rijke:hyp:probes} Design points where the Probability Density Functions are calculated. \subref{fig:rijke:hyp:pdf} Probability density function of angles between $E^u$, $E^n$, $E^s$. Tangencies are observed for points \textbf{D} and \textbf{E}, demonstrating that the chaotic attractors in these positions are not hyperbolic.}
\end{figure}

In \S\ref{sec:optim}, the sensitivities of the time-averaged acoustic energy with respect to the heat-source parameters are embedded in an optimisation routine to minimise the size of acoustic oscillations. 
However, as discussed in \S\ref{sec:shadowing_methods}, for such sensitivities to exist, the thermoacoustic chaotic attractor must be hyperbolic (\S\ref{sec:hyperbolicity}), otherwise the sensitivities of \eqref{eq:Jav} might not exist. Here, we seek to determine the hyperbolicity of the system.
To determine whether a system is hyperbolic, the complete spectrum should be computed to construct the unstable, neutral and stable subspaces. This is possible with the reduced-order model of this paper, but it could be prohibitively expensive in high-dimensional systems. For the latter, only a portion of the Lyapunov spectrum and covariant vectors is typically calculated~\citep[e.g.][]{Blonigan2016ctr,Fernandez2017}. 
Because it is computationally expensive to determine hyperbolicity everywhere in the design space, we restrict ourselves to 9 points, roughly equally spaced inside the chaotic areas (green in figure~\ref{fig:rijke:bifurcation}), which are shown in figure~\ref{fig:rijke:hyp:probes} (labels A to I). The probability density functions (PDFs) of the angles between the three pairs of elements from $E^u, E^n, E^s$ are calculated at each of these design points (figure~\ref{fig:rijke:hyp:pdf}). On the one hand, design points D and E are not hyperbolic, since the PDF of $\theta_{n,s}$ is non-zero at $\theta=0$, demonstrating that the system exhibits tangencies between these two subspaces. On the other hand, the PDFs of the remaining 7 points indicate that these are hyperbolic, which, physically means that time-averaged cost functionals respond smoothly to small changes in the design parameters, i.e. their sensitivities exist.
%
(For completeness, we also report that there is evidence that shadowing-based methods work well in some non-hyperbolic systems~\citep{ni_2019}.)
In conclusion, we found that thermoacoustic systems can physically exhibit both hyperbolic and non-hyperbolic chaos, depending on the design point.
Notwithstanding, starting from design point B, we will employ shadowing techniques to compute sensitivities, which are employed in an optimisation routine to minimise the acoustic energy (\S\ref{sec:optim}).
\section{Sensitivity and optimisation of chaotic acoustic oscillations}\label{sec:sens}
\subsection{Time-averaged cost functionals}
\label{sec:cost_functionals}
\begin{figure}[tb]
	\centering
    \includegraphics[width=0.49\textwidth]{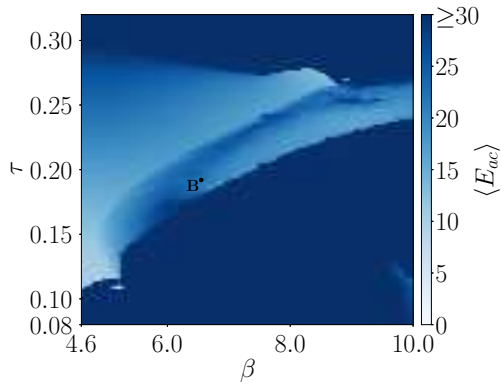}
    \caption{Colour map of time averaged acoustic energy, $\langle E_{ac} \rangle$. To increase colour resolution near the starting point, \textbf{B}, points with $\langle E_{ac} \rangle \geq 30$ have the same colour.}
    \label{fig:optim:eacMap}
\end{figure}

We analyse the chaotic acoustic oscillations of point B shown in figure~\ref{fig:rijke:hyp:probes}. 
Because thermoacoustics is a multi-physical phenomenon, there are different norms~\citep{Chu1965,George2012}, semi-norms~\citep{MagriPhD,Blumenthal2016} and functionals to define a physical measure. 
For thermoacoustic systems with negligible mean flow, which cannot advect flow inhomogeneities like entropy spots, the acoustic energy is a suitable quantity of interest. 
The instantaneous acoustic energy of the whole system is defined as
\begin{align}
    \label{eq:Eac}
    &E_{ac}(t) \triangleq \frac{1}{2}\int_0^1 \left(u^2(t) + p^2(t)\right) dx, 
\end{align}
which is the sum of the acoustic kinetic and potential energies, i.e. it is the Hamiltonian (constant of motion) of the natural acoustic system. 
Because of Parseval's theorem, the acoustic energy is related to the Galerkin modes as $E_{ac}(t)= \frac{1}{4} \sum_{j=1}^{N_g} \left(\eta_j^2(t) + \mu_j^2(t) \right)$. 
%
The acoustic energy, $E_{ac}$, is (half) the Euclidean norm of the thermoacoustic system under investigation. 
In chaotic acoustic oscillations, we are interested in calculating the sensitivity of the time-averaged acoustic energy, $\langle E_{ac} \rangle$. 
Figure~\ref{fig:optim:eacMap} shows the acoustic energy in the refined area of Fig.~\ref{fig:rijke:bifurcation}, with the optimisation starting point, B, marked. Regions similar to those depicted by different colours in Fig.~\ref{fig:rijke:hyp:probes} are visible. Notably, a sharp discontinuity exists to the right of the chaotic regions, which corresponds to a line of bifurcation points. Furthermore, the time-averaged acoustic energy is multi-modal, exhibiting multiple local extrema. Interestingly, continuous regions of the same type of attractor (Fig.~\ref{fig:rijke:bifurcation_le}) do not have extrema in their interior. Instead, the extrema are found at the edges of the regions, which suggests that gradient-based optimisation algorithms will be capable of finding the boundaries that separate attractors. 

%
%

\subsection{Minimisation of acoustic energy in chaotic acoustic oscillations by optimal design}
\label{sec:optim}
\begin{figure}[tb]
	\centering
    \begin{subfigure}[b]{0.49\textwidth}
         \subcaptionOverlay{\includegraphics[width=\textwidth]{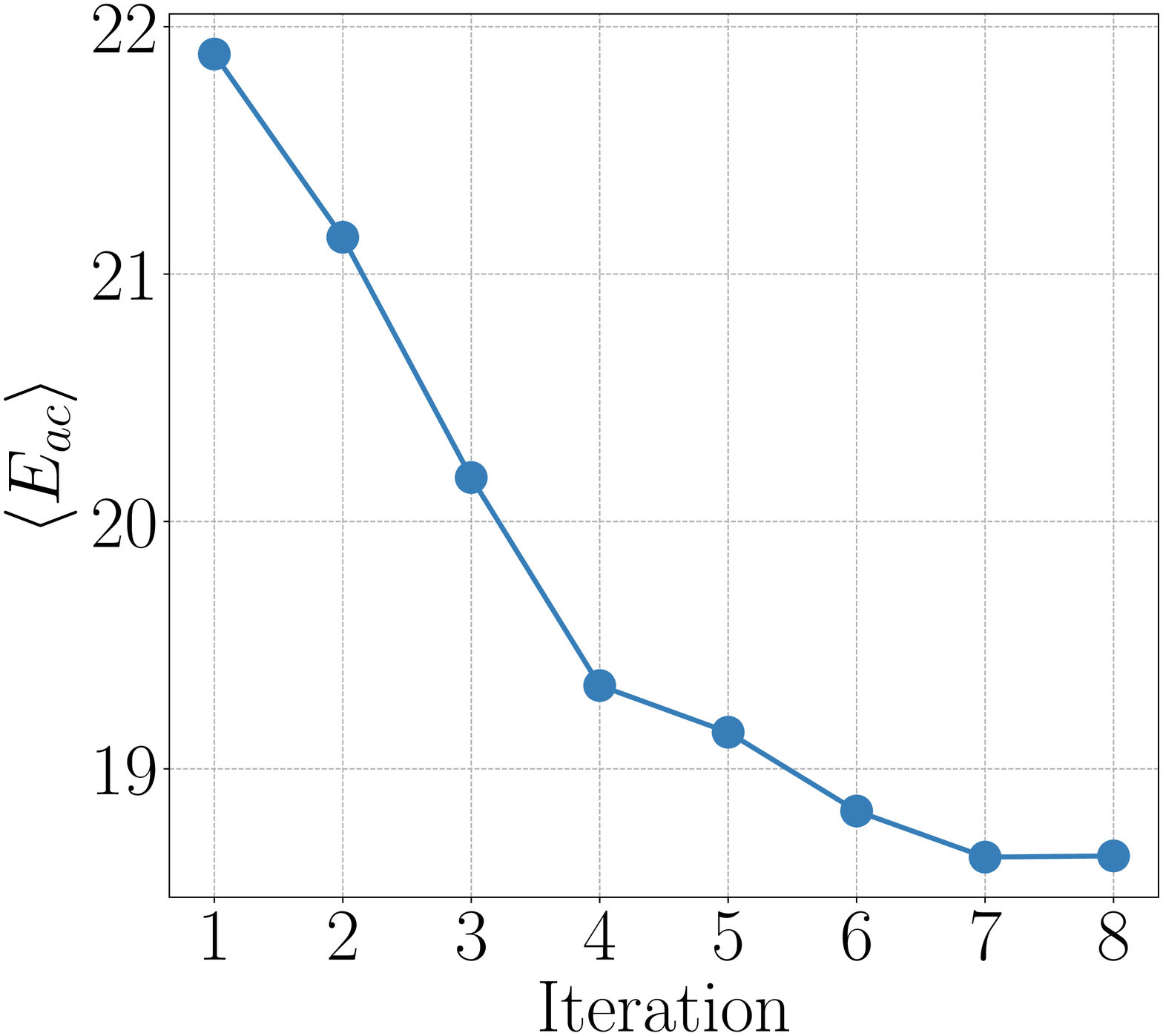}}
        \label{fig:optim:optimIter}
    \end{subfigure}
    \begin{subfigure}[b]{0.49\textwidth}
        \subcaptionOverlay{\includegraphics[width=\textwidth]{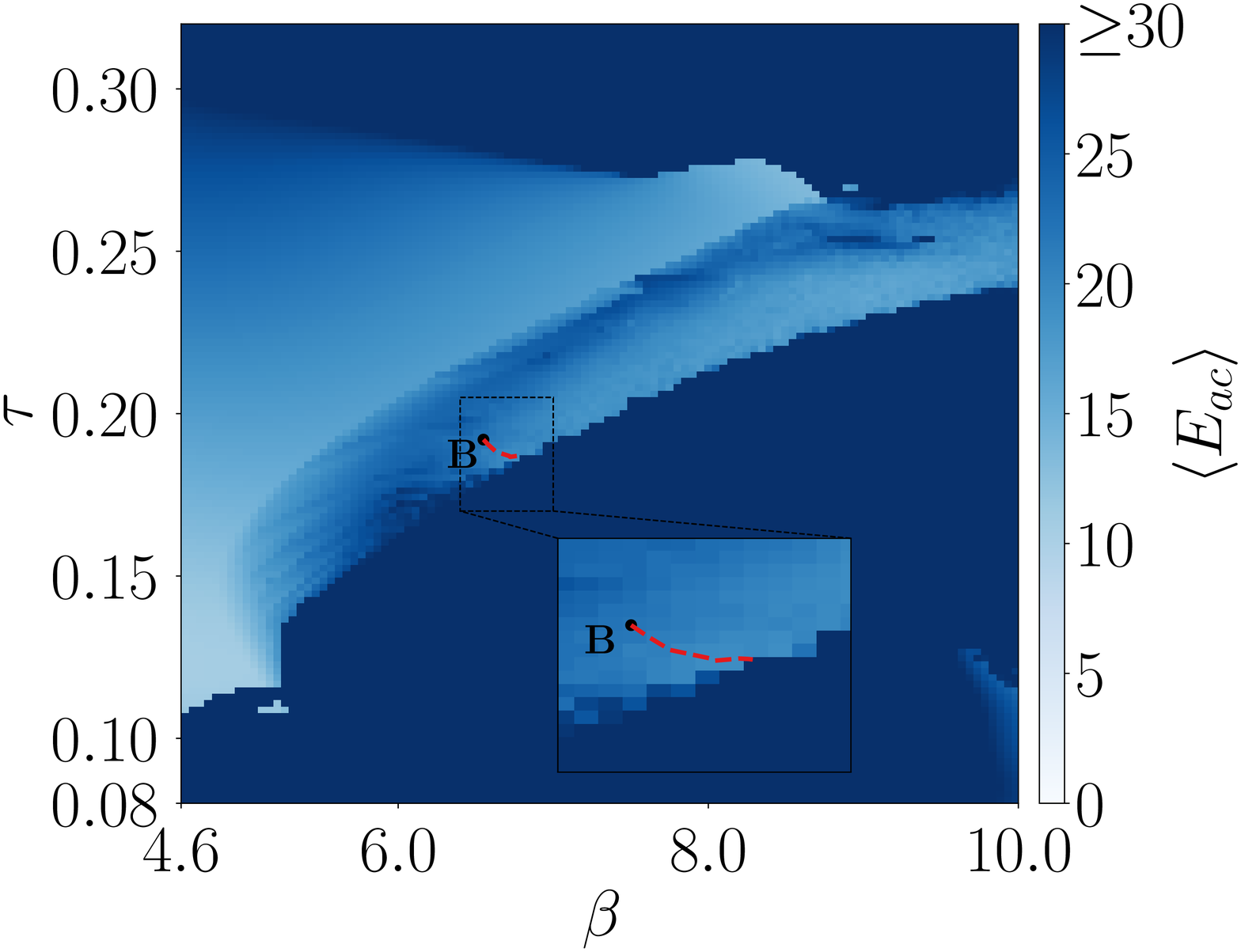}}
        \label{fig:optim:eacMapPath}
    \end{subfigure}
    \caption{\protect\subref{fig:optim:optimIter} Time-averaged acoustic energy versus optimisation iteration. The optimisation algorithm consecutively reduces the acoustic energy until iteration 7, where no progress is made. \protect\subref{fig:optim:eacMapPath} Time averaged acoustic energy, $\langle E_{ac} \rangle$ -- colour map with optimisation path superimposed (\tikzline{c2}).}
\end{figure}

To minimise the acoustic energy, either the bifurcation diagram in the multidimensional parameter space is calculated (Fig.~\ref{fig:optim:eacMap}), which is computationally cumbersome, or a nonlinear optimisation problem of a time-averaged cost functional is solved, which is computationally affordable. Following the latter route, the optimisation problem is formulated as 
\begin{equation}
    \label{eq:optim_problem}
    \begin{aligned}
        & \underset{\beta, \tau}{\text{minimise}} & & \left\langle E_{ac} (\beta, \tau)\right\rangle \\
        & \text{subject to} & & \eqref{eq:rijke:mom}, \eqref{eq:rijke:ene}, \eqref{eq:modified_kings}, \eqref{eq:advection}, \eqref{eq:advection_bc}, \eqref{eq:galerkin_decomp_u}, \eqref{eq:galerkin_decomp_p}
    \end{aligned} .
\end{equation}
%
%
%
The set of parameters is updated via the sequential least squares programming method of the \texttt{SciPy} library.
The optimisation stops when the condition 
\begin{equation}
    \label{eq:optim_condition}
    \frac{\langle E_{ac} \rangle_{j} - \langle E_{ac} \rangle_{j+1}}{\langle E_{ac} \rangle_{j}} < 0.01
\end{equation}
is met, that is, when the improvement between optimisation iterations is less than 1\% of the previous value. The usual gradient vanishing condition of extrema is not applied because of the existence of discontinuities. Since the gradient does not exist at these points, its numerical value cannot be trusted close to such points, which is why condition \eqref{eq:optim_condition} is used instead.
Figure~\ref{fig:optim:optimIter} shows the the cost functional as a function of the iteration in the optimisation algorithm. The acoustic energy decreases rapidly until iteration $7$, where no progress is made and Eq.~\eqref{eq:optim_condition} is verified, indicating that the algorithm has converged to a local minimum ($\beta = 6.79889$, $\tau = 0.18685$).
Overall, the optimisation achieves a reduction of 14.8\% in acoustic energy in 7 iterations. Figure~\ref{fig:optim:eacMapPath} shows the path that the optimisation procedure takes in the design space, showing that the final design is indeed a local minimum and that it is located at the edge of the region of chaotic solutions, as hypothesised in \S\ref{sec:cost_functionals}. In conclusion, we found the set of parameters that produce a local minimum of the time-averaged acoustic energy of a chaotic thermoacoustic system. A similar algorithm can be used to find local maxima for maximal energy extraction in the design of thermoacoustic engines. 
\subsection{Future directions}
In \S\ref{sec:cost_functionals}, we found that the time-averaged acoustic energy, $\langle E_{ac} \rangle$, displays an intricate behaviour: it is both discontinuous and multimodal. These two facts have strong implications when tackling an optimisation problem with a gradient-based approach. First, gradients are not defined at discontinuous points; second, gradient-based algorithms might converge to local minima, instead of the global optimum, as in \S\ref{sec:optim}. While pure gradient-free algorithms might be computationally expensive, hybrid techniques, which couple gradient-based with gradient-free algorithms, can be a suitable compromise between overcoming the aforementioned issues of gradient-based algorithms and exploiting local gradient information to find the shortest path to an optimal design. For example, a Monte Carlo gradient-based optimisation, where multiple gradient-based optimisations are launched from various randomised initial design points and the best of the resulting (local) minima is chosen. Since thermoacoustic systems can exhibit rich dynamics and admit several types of solutions, a general technique for calculating sensitivities, including being capable of handling chaotic attractors, is required. As discussed in this work, covariant Lyapunov vector (CLV) analysis is the most general framework of analysis of these dynamical systems. Currently, shadowing methods, which exploit shadowing orbits via covariant Lyapunov vectors, are the leading candidate for sensitivity calculation in chaotic attractors.

For a practitioner, whose objective is to suppress oscillations, CLV techniques are required because chaos is always present in realistic turbulent systems. 
For example, in turbulent flows the hydrodynamic field modulates the heat released by the flame in a chaotic way~\citep{Lieuwen2012}. 
In the bistable region of a subcritical bifurcation, where the system is eigenvalue stable but a finite-amplitude periodic solution exists~\citep{Subramanian2010a}, the turbulent hydrodynamic field chaotically forces the solution to oscillate around the stable fixed point or the upper branch limit cycle, depending on the initial condition. To eliminate the possibility of such chaotic limit cycles, the operating design point must be outside the hysteresis region. The boundary that separates these two regions (known as the fold point with one parameter) is marked by a discontinuity similar to the ones in Fig.~\ref{fig:optim:eacMap}. Because of the presence of chaos, it would not be possible to identify the fold point with traditional eigenvalue and Floquet analyses. However, it would be possible to identify it by using the CLV technique and optimisation we proposed (as shown in \S\ref{sec:optim}).

%
%
%
\section{Conclusions}
\label{sec:conclusion}
Traditional tools in flow instability, such as eigenvalue and Floquet analyses, fail when the solution is chaotic.
We propose to use covariant Lyapunov vector analysis as a general tool to calculate the stability and sensitivity of unsteady solutions with chaotic behaviour. 
First, the connections between covariant Lyapunov vectors, eigenfunctions and Floquet modes are mathematically shown. 
We analytically recover the limits of eigenvalue analysis when the attractor is a fixed point, and Floquet analysis when the attractor is a limit cycle. 
Second, we explain the importance of testing the hyperbolicity of the chaotic solution for the calculation of sensitivities. 
Third, we apply the theoretical analysis to a chaotic acoustic system with a heat source.
We show that the system admits both hyperbolic and non-hyperbolic chaotic attractors, which means that sensitivities might not exist for some sets of parameters.
Departing from a hyperbolic point, and by exploiting the shadow trajectory via the non-intrusive least-squares shadowing method to calculate sensitivities, we minimise the acoustic energy of the chaotic oscillations by changing the heat-source parameters in an optimisation routine.
This work opens up new possibilities for the control of unsteady acoustic oscillations by  optimal design. Because the theoretical framework is general, the techniques presented can be used in other unsteady fluid dynamics problems with virtually no modification.
\section*{Acknowledgments}
F. H. is supported by Fundação para a Ciência e Tecnologia under the Research Studentship no. SFRH/BD/134617/2017. L. Magri gratefully acknowledges financial support from the Royal Academy of Engineering Research Fellowships. Fruitful discussions with Professor Q. Wang, Dr P. Blonigan, N. Chandramoorthy and A. Ni are gratefully acknowledged. 
%
 
\bibliographystyle{jfm}

\begin{thebibliography}{0}
\expandafter\ifx\csname natexlab\endcsname\relax\def\natexlab#1{#1}\fi
\def\au#1{#1} \def\ed#1{#1} \def\yr#1{#1}\def\at#1{#1}\def\jt#1{\textit{#1}}
  \def\bt#1{#1}\def\bvol#1{\textbf{#1}} \def\vol#1{#1} \def\pg#1{#1}
  \def\publ#1{#1}\def\arxiv#1{#1}\def\org#1{#1}\def\st#1{\textit{#1}}

\end{thebibliography}


\begin{thebibliography}{81}
\expandafter\ifx\csname natexlab\endcsname\relax\def\natexlab#1{#1}\fi
\def\au#1{#1} \def\ed#1{#1} \def\yr#1{#1}\def\at#1{#1}\def\jt#1{\textit{#1}}
  \def\bt#1{#1}\def\bvol#1{\textbf{#1}} \def\vol#1{#1} \def\pg#1{#1}
  \def\publ#1{#1}\def\arxiv#1{#1}\def\org#1{#1}\def\st#1{\textit{#1}}

\bibitem[Aguilar \& Juniper(2018)]{Aguilar2018_asme}
{\sc \au{Aguilar, J} \& \au{Juniper, M~P}} \yr{2018} {Adjoint methods for
  elimination of thermoacoustic oscillations in a model annular combustor via
  small geometry modifications}.  \bt{In {\em ASME Turbo Expo, Oslo,
  Norway\/}},  \pg{pp. GT2018--75692}.

\bibitem[Balasubramanian \& Sujith(2008)]{Balasubramanian2008a}
{\sc \au{Balasubramanian, K.} \& \au{Sujith, R.~I.}} \yr{2008}
  \at{{Thermoacoustic instability in a Rijke tube: Non-normality and
  nonlinearity}}.  \jt{Physics of Fluids}  \bvol{20}~(4),  \pg{044103}.

\bibitem[Birkhoff(1931)]{Birkhoff1931}
{\sc \au{Birkhoff, George~D.}} \yr{1931}  \at{Proof of the ergodic theorem}.
  \jt{Proceedings of the National Academy of Sciences}  \bvol{17}~(12),
  \pg{656--660},  \arxiv{arXiv:
  https://www.pnas.org/content/17/12/656.full.pdf}.

\bibitem[Blonigan(2017)]{Blonigan2017}
{\sc \au{Blonigan, Patrick~J.}} \yr{2017}  \at{{Adjoint sensitivity analysis of
  chaotic dynamical systems with non-intrusive least squares shadowing}}.
  \jt{Journal of Computational Physics}  \bvol{348},  \pg{803--826}.

\bibitem[Blonigan {\em et~al.\/}(2016)Blonigan, Fernandez, Murman, Wang, Rigas
  \& Magri]{Blonigan2016ctr}
{\sc \au{Blonigan, P.~J.}, \au{Fernandez, P.}, \au{Murman, S.~M.}, \au{Wang,
  Q.}, \au{Rigas, G.} \& \au{Magri, L.}} \yr{2016} {Towards a chaotic adjoint
  for LES}.  \bt{In {\em Center for Turbulence Research, Summer Program\/}}.

\bibitem[Blonigan \& Wang(2014)]{Blonigan2014a}
{\sc \au{Blonigan, Patrick~J.} \& \au{Wang, Qiqi}} \yr{2014}  \at{{Least
  squares shadowing sensitivity analysis of a modified Kuramoto-Sivashinsky
  equation}}.  \jt{Chaos, Solitons and Fractals}  \bvol{64}~(1),  \pg{16--25},
  \arxiv{arXiv: arXiv:1307.8197v2}.

\bibitem[Blumenthal {\em et~al.\/}(2016)Blumenthal, Tangirala, Sujith \&
  Polifke]{Blumenthal2016}
{\sc \au{Blumenthal, R.~S.}, \au{Tangirala, A.~K.}, \au{Sujith, R.} \&
  \au{Polifke, W.}} \yr{2016}  \at{{A systems perspective on non-normality in
  low-order thermoacoustic models: Full norms, semi-norms and transient
  growth}}.  \jt{International Journal of Spray and Combustion Dynamics}
  \bvol{9}~(1),  \pg{19--43}.

\bibitem[Bowen \& Ruelle(1975)]{Bowen1975}
{\sc \au{Bowen, Rufus} \& \au{Ruelle, David}} \yr{1975}  \at{{The ergodic
  theory of Axiom A flows}}.  \jt{Inventiones Mathematicae}  \bvol{29},
  \pg{181--202}.

\bibitem[Chu(1965)]{Chu1965}
{\sc \au{Chu, B.~T.}} \yr{1965}  \at{{On the Energy Transfer to Small
  Disturbances in Fluid Flow (Part I)}}.  \jt{Acta Mechanica}  \bvol{1}~(3),
  \pg{215--234}.

\bibitem[Culick(2006)]{Culick2006}
{\sc \au{Culick, F. E.~C.}} \yr{2006} {\em {Unsteady motions in combustion
  chambers for propulsion systems}\/}.  \publ{RTO AGARDograph AG-AVT-039, North
  Atlantic Treaty Organization}.

\bibitem[Curtain \& Zwart(1995)]{Curtain1995}
{\sc \au{Curtain, Ruth~F} \& \au{Zwart, Hans}} \yr{1995} {\em {An introduction
  to infinite-dimensional linear systems theory}\/}, ,  \vol{vol.~21}.
  \publ{New York: Springer-Verlag New York}.

\bibitem[Dowling(1997)]{Dowling1997}
{\sc \au{Dowling, A.~P.}} \yr{1997}  \at{{Nonlinear self-excited oscillations
  of a ducted flame}}.  \jt{Journal of Fluid Mechanics}  \bvol{346},
  \pg{271--290}.

\bibitem[Dowling(1999)]{Dowling1999}
{\sc \au{Dowling, A.~P.}} \yr{1999}  \at{{A kinematic model of a ducted
  flame}}.  \jt{Journal of Fluid Mechanics}  \bvol{394},  \pg{51--72}.

\bibitem[Dowling \& Mahmoudi(2015)]{Dowling2015}
{\sc \au{Dowling, Ann~P.} \& \au{Mahmoudi, Yasser}} \yr{2015}  \at{{Combustion
  noise}}.  \jt{Proceedings of the Combustion Institute}  \bvol{35}~(1),
  \pg{65--100}.

\bibitem[Dowling \& Morgans(2005)]{Dowling2005}
{\sc \au{Dowling, A.~P.} \& \au{Morgans, A.~S.}} \yr{2005}  \at{{Feedback
  Control of Combustion Oscillations}}.  \jt{Annual Review of Fluid Mechanics}
  \bvol{37},  \pg{151--182}.

\bibitem[Eckmann(1981)]{Eckmann1981}
{\sc \au{Eckmann, J.-P.}} \yr{1981}  \at{{Roads to turbulence in dissipative
  dynamical systems}}.  \jt{Reviews of Modern Physics}  \bvol{53}~(4),
  \pg{643--654}.

\bibitem[Eckmann \& Ruelle(1985)]{Eckmann1985}
{\sc \au{Eckmann, J.~P.} \& \au{Ruelle, D.}} \yr{1985}  \at{{Ergodic theory of
  chaos and strange attractors}}.  \jt{Reviews of Modern Physics}  \bvol{57},
  \pg{617--656}.

\bibitem[Eyink {\em et~al.\/}(2004)Eyink, Haine \& Lea]{Eyink2004}
{\sc \au{Eyink, G.~L.}, \au{Haine, T.~W.N.} \& \au{Lea, D.~J.}} \yr{2004}
  \at{{Ruelle's linear response formula, ensemble adjoint schemes and
  L{\'{e}}vy flights}}.  \jt{Nonlinearity}  \bvol{17}~(5),  \pg{1867--1889}.

\bibitem[Fernandez \& Wang(2017)]{Fernandez2017}
{\sc \au{Fernandez, P.} \& \au{Wang, Q.}} \yr{2017}  \at{{Lyapunov spectrum of
  the separated flow around the NACA 0012 airfoil and its dependence on
  numerical discretization}}.  \jt{Journal of Computational Physics}
  \bvol{350},  \pg{453--469},  \arxiv{arXiv: 1612.07409}.

\bibitem[Gallavotti(2006)]{Gallavotti2006}
{\sc \au{Gallavotti, Giovanni}} \yr{2006}  \at{{Entropy, thermostats, and
  chaotic hypothesis}}.  \jt{Chaos: An Interdisciplinary Journal of Nonlinear
  Science}  \bvol{16}~(4),  \pg{043114}.

\bibitem[Gallavotti \& Cohen(1995)]{Gallavotti1995}
{\sc \au{Gallavotti, G.} \& \au{Cohen, E. G.~D.}} \yr{1995}  \at{{Dynamical
  ensembles in stationary states}}.  \jt{Journal of Statistical Physics}
  \bvol{80}~(5-6),  \pg{931--970}.

\bibitem[George \& Sujith(2012)]{George2012}
{\sc \au{George, K.~J.} \& \au{Sujith, R.I.}} \yr{2012}  \at{{Disturbance
  energy norms: A critical analysis}}.  \jt{Journal of Sound and Vibration}
  \bvol{331}~(7),  \pg{1552--1566}.

\bibitem[Ginelli {\em et~al.\/}(2013)Ginelli, Chaté, Livi \&
  Politi]{ginelli_2013}
{\sc \au{Ginelli, Francesco}, \au{Chaté, Hugues}, \au{Livi, Roberto} \&
  \au{Politi, Antonio}} \yr{2013}  \at{Covariant lyapunov vectors}.
  \jt{Journal of Physics A: Mathematical and Theoretical}  \bvol{46}~(25),
  \pg{254005}.

\bibitem[Ginelli {\em et~al.\/}(2007)Ginelli, Poggi, Turchi, Chat\'e, Livi \&
  Politi]{Ginelli2007}
{\sc \au{Ginelli, F.}, \au{Poggi, P.}, \au{Turchi, A.}, \au{Chat\'e, H.},
  \au{Livi, R.} \& \au{Politi, A.}} \yr{2007}  \at{{Characterizing dynamics
  with covariant lyapunov vectors}}.  \jt{Physical Review Letters}
  \bvol{99}~(13),  \pg{130601},  \arxiv{arXiv: 0706.0510}.

\bibitem[Gotoda {\em et~al.\/}(2012)Gotoda, Ikawa, Maki \& Miyano]{Gotoda2012}
{\sc \au{Gotoda, H.}, \au{Ikawa, T.}, \au{Maki, K.} \& \au{Miyano, T.}}
  \yr{2012}  \at{{Short-term prediction of dynamical behavior of flame front
  instability induced by radiative heat loss.}}  \jt{Chaos}  \bvol{22}~(1),
  \pg{033106}.

\bibitem[Gotoda {\em et~al.\/}(2011)Gotoda, Nikimoto, Miyano \&
  Tachibana]{Gotoda2011}
{\sc \au{Gotoda, H.}, \au{Nikimoto, H.}, \au{Miyano, T.} \& \au{Tachibana, S.}}
  \yr{2011}  \at{{Dynamic properties of combustion instability in a lean
  premixed gas-turbine combustor.}}  \jt{Chaos}  \bvol{21}~(1),  \pg{013124}.

\bibitem[Guckenheimer \& Holmes(1983)]{Guck1983}
{\sc \au{Guckenheimer, J.} \& \au{Holmes, P.}} \yr{1983} {\em {Nonlinear
  oscillations, dynamical systems, and bifurcations of vector fields}\/}.
  \publ{Springer-Verlag New York}.

\bibitem[Heckl(1988)]{Heckl1988}
{\sc \au{Heckl, M.~A.}} \yr{1988}  \at{{Active control of the noise from a
  Rijke tube}}.  \jt{Journal of Sound and Vibration}  \bvol{124}~(1),
  \pg{117--133}.

\bibitem[Heckl(1990)]{Heckl1990}
{\sc \au{Heckl, M.~A.}} \yr{1990}  \at{{Non-linear acoustic effects in the
  Rijke tube}}.  \jt{Acustica}  \bvol{72},  \pg{63--71}.

\bibitem[Holmes {\em et~al.\/}(1996)Holmes, Lumley \& Berzook]{Holmes1996}
{\sc \au{Holmes, P.}, \au{Lumley, J.~L.} \& \au{Berzook, G.}} \yr{1996} {\em
  {Turbulence, Coherent Structures, Dynamical Systems and Symmetry}\/}.
  \publ{Cambridge: Cambridge University Press}.

\bibitem[Inubushi {\em et~al.\/}(2015)Inubushi, Takehiro \&
  Yamada]{Inubushi2015}
{\sc \au{Inubushi, Masanobu}, \au{Takehiro, Shin-ichi} \& \au{Yamada, Michio}}
  \yr{2015}  \at{Regeneration cycle and the covariant lyapunov vectors in a
  minimal wall turbulence}.  \jt{Phys. Rev. E}  \bvol{92},  \pg{023022}.

\bibitem[Jarlebring(2008)]{Jarlebring2008}
{\sc \au{Jarlebring, E.}} \yr{2008}  \at{{The spectrum of delay-differential equations: numerical methods, stability and perturbation}}. KTH Royal Institute of Technology.
  
\bibitem[Juniper(2011)]{juniper_2011}
{\sc \au{Juniper, Matthew~P.}} \yr{2011}  \at{Triggering in the horizontal
  rijke tube: non-normality, transient growth and bypass transition}.
  \jt{Journal of Fluid Mechanics}  \bvol{667},  \pg{272--308}.

\bibitem[Juniper \& Sujith(2018)]{juniper_sujith_2018}
{\sc \au{Juniper, Matthew~P.} \& \au{Sujith, R.I.}} \yr{2018}  \at{Sensitivity
  and nonlinearity of thermoacoustic oscillations}.  \jt{Annual Review of Fluid
  Mechanics}  \bvol{50}~(1),  \pg{661--689},  \arxiv{arXiv:
  https://doi.org/10.1146/annurev-fluid-122316-045125}.

\bibitem[Kabiraj {\em et~al.\/}(2012)Kabiraj, Saurabh, Wahi \&
  Sujith]{Kabiraj2012}
{\sc \au{Kabiraj, Lipika}, \au{Saurabh, Aditya}, \au{Wahi, Pankaj} \&
  \au{Sujith, R.~I.}} \yr{2012}  \at{{Route to chaos for combustion instability
  in ducted laminar premixed flames}}.  \jt{Chaos}  \bvol{22}~(2),  \pg{0--12}.

\bibitem[Kabiraj {\em et~al.\/}(2011)Kabiraj, Sujith \& Wahi]{Kabiraj2011}
{\sc \au{Kabiraj, Lipika}, \au{Sujith, R~I} \& \au{Wahi, Pankaj}} \yr{2011}
  \at{{Bifurcations of Self-Excited Ducted Laminar Premixed Flames}}.
  \jt{Journal of Engineering for Gas Turbines and Power}  \bvol{134}~(3),
  \pg{31502}.

\bibitem[Kashinath {\em et~al.\/}(2013{\natexlab{{\em a\/}}})Kashinath,
  Hemchandra \& Juniper]{Kashinath2013c}
{\sc \au{Kashinath, Karthik}, \au{Hemchandra, Santosh} \& \au{Juniper,
  Matthew~P.}} \yr{2013{\natexlab{{\em a\/}}}}  \at{{Nonlinear Phenomena in
  Thermoacoustic Systems With Premixed Flames}}.  \jt{Journal of Engineering
  for Gas Turbines and Power}  \bvol{135}~(6),  \pg{061502}.

\bibitem[Kashinath {\em et~al.\/}(2013{\natexlab{{\em b\/}}})Kashinath,
  Hemchandra \& Juniper]{Kashinath2013b}
{\sc \au{Kashinath, K.}, \au{Hemchandra, S.} \& \au{Juniper, M.~P.}}
  \yr{2013{\natexlab{{\em b\/}}}}  \at{{Nonlinear thermoacoustics of ducted
  premixed flames: The influence of perturbation convection speed}}.
  \jt{Combustion and Flame}  \bvol{160}~(12),  \pg{2856--2865}.

\bibitem[Kashinath {\em et~al.\/}(2014)Kashinath, Waugh \&
  Juniper]{Kashinath2013}
{\sc \au{Kashinath, Karthik}, \au{Waugh, Iain~C.} \& \au{Juniper, Matthew~P.}}
  \yr{2014}  \at{{Nonlinear self-excited thermoacoustic oscillations of a
  ducted premixed flame: bifurcations and routes to chaos}}.  \jt{Journal of
  Fluid Mechanics}  \bvol{761},  \pg{399--430}.

\bibitem[Katok \& Hasselblatt(1995)]{katok_hasselblat}
{\sc \au{Katok, Anatole} \& \au{Hasselblatt, Boris}} \yr{1995} {\em
  Introduction to the Modern Theory of Dynamical Systems\/}.  \publ{Cambridge
  University Press}.

\bibitem[King(1914)]{King1914}
{\sc \au{King, L.~V.}} \yr{1914}  \at{{On the convection of heat from small
  cyclinders in a stream of fluid: determination of the convection constants of
  small platinum wires with applications to hot-wire anemometry}}.
  \jt{Proceedings of the Royal Society of {\ldots}}  \bvol{214}~(8),
  \pg{373--434}.

\bibitem[Landau \& Lifshitz(1987)]{Landau1987}
{\sc \au{Landau, L.~D.} \& \au{Lifshitz, E.~M.}} \yr{1987} {\em {Fluid
  Mechanics}\/}, 2nd edn.  \publ{Pergamon Press}.

\bibitem[Lea {\em et~al.\/}(2000)Lea, Allen \& Haine]{lea_2000}
{\sc \au{Lea, Daniel~J.}, \au{Allen, Myles~R.} \& \au{Haine, Thomas~W.N.}}
  \yr{2000}  \at{Sensitivity analysis of the climate of a chaotic system}.
  \jt{Tellus A: Dynamic Meteorology and Oceanography}  \bvol{52}~(5),
  \pg{523--532},  \arxiv{arXiv: https://doi.org/10.3402/tellusa.v52i5.12283}.

\bibitem[Lieuwen(2012)]{Lieuwen2012}
{\sc \au{Lieuwen, T.}} \yr{2012} {\em {Unsteady Combustor Physics}\/}.
  \publ{Cambridge University Press}.

\bibitem[Lieuwen \& Yang(2005)]{Lieuwen2005}
{\sc \au{Lieuwen, T.~C.} \& \au{Yang, V.}} \yr{2005} {\em {Combustion
  Instabilities in Gas Turbine Engines: Operational Experience, Fundamental
  Mechanisms, and Modeling}\/}.  \publ{American Institute of Aeronautics and
  Astronautics, Inc.}

\bibitem[Magri(2015)]{MagriPhD}
{\sc \au{Magri, L.}} \yr{2015}  \at{{Adjoint methods in thermo-acoustic and
  combustion instability}}. PhD thesis, University of Cambridge.

\bibitem[Magri(2018)]{Magri2019_amr}
{\sc \au{Magri, L.}} \yr{2018}  \at{{Adjoint methods as design tools in
  thermoacoustics}}.  \jt{Applied Mechanics Reviews} \bvol{71}~(2),  \pg{020801}.

\bibitem[Magri {\em et~al.\/}(2016)Magri, Bauerheim, Nicoud \&
  Juniper]{Magri2016c}
{\sc \au{Magri, L.}, \au{Bauerheim, M.}, \au{Nicoud, F.} \& \au{Juniper,
  M.~P.}} \yr{2016}  \at{{Stability analysis of thermo-acoustic nonlinear
  eigenproblems in annular combustors. Part II. Uncertainty quantification}}.
  \jt{Journal of Computational Physics}  \bvol{325},  \pg{411--421},
  \arxiv{arXiv: 1602.08440}.

\bibitem[Magri \& Juniper(2013)]{magri_juniper_2013}
{\sc \au{Magri, Luca} \& \au{Juniper, Matthew~P.}} \yr{2013}  \at{Sensitivity
  analysis of a time-delayed thermo-acoustic system via an adjoint-based
  approach}.  \jt{Journal of Fluid Mechanics}  \bvol{719},  \pg{183–202}.

\bibitem[Magri \& Juniper(2014)]{Magri2013c}
{\sc \au{Magri, L.} \& \au{Juniper, M.~P.}} \yr{2014}  \at{{Global modes,
  receptivity, and sensitivity analysis of diffusion flames coupled with duct
  acoustics}}.  \jt{Journal of Fluid Mechanics}  \bvol{752},  \pg{237--265}.

\bibitem[Mensah {\em et~al.\/}(2018)Mensah, Magri \& Moeck]{Mensah2018}
{\sc \au{Mensah, G.~A.}, \au{Magri, L.} \& \au{Moeck, J.~P.}} \yr{2018}
  \at{{Methods for the calculation of thermoacoustic stability margins and
  Monte-Carlo free uncertainty quantification}}.  \jt{Journal of Engineering
  for Gas Turbines and Power}  \bvol{140}~(6),  \pg{061501}.

\bibitem[Mensah \& Moeck(2017)]{Mensah2017a}
{\sc \au{Mensah, Georg~A.} \& \au{Moeck, Jonas~P.}} \yr{2017}  \at{{Acoustic
  Damper Placement and Tuning for Annular Combustors: An Adjoint-Based
  Optimization Study}}.  \jt{Journal of Engineering for Gas Turbines and Power}
   \bvol{139}~(6),  \pg{061501}.

\bibitem[Miles(1984)]{Miles1984}
{\sc \au{Miles, J}} \yr{1984}  \at{{Strange attractors in fluid dynamics}}.
  \jt{Advances in Applied Mechanics}  \bvol{24},  \pg{189--214}.

\bibitem[Nair \& Sujith(2015)]{Nair2015}
{\sc \au{Nair, V.} \& \au{Sujith, R.~I.}} \yr{2015}  \at{{A reduced-order model
  for the onset of combustion instability: Physical mechanisms for
  intermittency and precursors}}.  \jt{Proceedings of the Combustion Institute,
  in press}  \bvol{35}~(3),  \pg{3193--3200}.

\bibitem[Nair {\em et~al.\/}(2014)Nair, Thampi \& Sujith]{Nair2014}
{\sc \au{Nair, Vineeth}, \au{Thampi, Gireeshkumaran} \& \au{Sujith, R~I}}
  \yr{2014}  \at{{Intermittency route to thermoacoustic instability in
  turbulent combustors}}.  \jt{Journal of Fluid Mechanics}  \bvol{756},
  \pg{470--487}.

\bibitem[Ni(2019)]{ni_2019}
{\sc \au{Ni, Angxiu}} \yr{2019}  \at{Hyperbolicity, shadowing directions and
  sensitivity analysis of a turbulent three-dimensional flow}.  \jt{Journal of
  Fluid Mechanics}  \bvol{863},  \pg{644–669}.

\bibitem[Ni \& Wang(2017)]{ni_2017}
{\sc \au{Ni, Angxiu} \& \au{Wang, Qiqi}} \yr{2017}  \at{Sensitivity analysis on
  chaotic dynamical systems by non-intrusive least squares shadowing (nilss)}.
  \jt{Journal of Computational Physics}  \bvol{347},  \pg{56 -- 77}.

\bibitem[Orchini {\em et~al.\/}(2015)Orchini, Illingworth \&
  Juniper]{Orchini2015a}
{\sc \au{Orchini, a.}, \au{Illingworth, S.~J.} \& \au{Juniper, M.~P.}}
  \yr{2015}  \at{{Frequency domain and time domain analysis of thermoacoustic
  oscillations with wave-based acoustics}}.  \jt{Journal of Fluid Mechanics}
  \bvol{775},  \pg{387--414}.

\bibitem[Orchini {\em et~al.\/}(2016)Orchini, Rigas \& Juniper]{orchini_2016}
{\sc \au{Orchini, Alessandro}, \au{Rigas, Georgios} \& \au{Juniper,
  Matthew~P.}} \yr{2016}  \at{Weakly nonlinear analysis of thermoacoustic
  bifurcations in the rijke tube}.  \jt{Journal of Fluid Mechanics}
  \bvol{805},  \pg{523--550}.

\bibitem[Oseledets(1968)]{oseledets_1968}
{\sc \au{Oseledets, Valery~Ivan}} \yr{1968}  \at{A multiplicative ergodic
  theorem: Lyapunov characteristic numbers for dynamical systems.}  \jt{Trans.
  Moscow Math. Soc.}  \bvol{19},  \pg{197--231}.

\bibitem[Petzold(1983)]{petzold_1983}
{\sc \au{Petzold, L.}} \yr{1983}  \at{Automatic selection of methods for
  solving stiff and nonstiff systems of ordinary differential equations}.
  \jt{SIAM Journal on Scientific and Statistical Computing}  \bvol{4}~(1),
  \pg{136--148},  \arxiv{arXiv: https://doi.org/10.1137/0904010}.

\bibitem[Pilyugin(2006)]{pilyugin2006shadowing}
{\sc \au{Pilyugin, Sergei~Yu}} \yr{2006} {\em {Shadowing in dynamical
  systems}\/}.  \publ{Springer}.

\bibitem[Poinsot(2017)]{Poinsot2017}
{\sc \au{Poinsot, T.}} \yr{2017}  \at{{Prediction and control of combustion
  instabilities in real engines}}.  \jt{Proceedings of the Combustion
  Institute}  \bvol{36}~(1),  \pg{1--28}.

\bibitem[Polifke {\em et~al.\/}(2001)Polifke, Poncet, Paschereit \&
  D{\"{o}}bbeling]{Polifke2001}
{\sc \au{Polifke, W.}, \au{Poncet, A.}, \au{Paschereit, C.~O.} \&
  \au{D{\"{o}}bbeling, K.}} \yr{2001}  \at{{Reconstruction of Acoustic Transfer
  Matrices by Instationary Computational Fluid Dynamics}}.  \jt{Journal of
  Sound and Vibration}  \bvol{245}~(3),  \pg{483--510}.

\bibitem[Rayleigh(1878)]{rayleigh_1878}
{\sc \au{Rayleigh, Lord}} \yr{1878}  \at{The explanation of certain acoustical
  phenomena}.  \jt{Nature}  \bvol{18},  \pg{319 --}.

\bibitem[Ruelle(1980)]{Ruelle1980}
{\sc \au{Ruelle, David}} \yr{1980}  \at{{Measures Describing a Turbulent
  Flow}}.  \jt{Annals of the New York Academy of Sciences}  \bvol{357}~(1),
  \pg{1--9}.

\bibitem[Ruelle(2009)]{ruelle_2009}
{\sc \au{Ruelle, David}} \yr{2009}  \at{A review of linear response theory for
  general differentiable dynamical systems}.  \jt{Nonlinearity}  \bvol{22}~(4),
   \pg{855}.

\bibitem[Schubert \& Lucarini(2015)]{Schubert2015}
{\sc \au{Schubert, Sebastian} \& \au{Lucarini, Valerio}} \yr{2015}
  \at{Covariant lyapunov vectors of a quasi-geostrophic baroclinic model:
  analysis of instabilities and feedbacks}.  \jt{Quarterly Journal of the Royal
  Meteorological Society}  \bvol{141}~(693),  \pg{3040--3055},  \arxiv{arXiv:
  https://rmets.onlinelibrary.wiley.com/doi/pdf/10.1002/qj.2588}.

\bibitem[Silva {\em et~al.\/}(2016)Silva, Magri, Runte \& Polifke]{Silva2016}
{\sc \au{Silva, C.~F.}, \au{Magri, L.}, \au{Runte, T.} \& \au{Polifke, W.}}
  \yr{2016}  \at{{Uncertainty Quantification of Growth Rates of Thermoacoustic
  Instability by an Adjoint Helmholtz Solver}}.  \jt{Journal of Engineering for
  Gas Turbines and Power}  \bvol{139}~(1),  \pg{011901}.

\bibitem[Subramanian {\em et~al.\/}(2011)Subramanian, Mariappan, Sujith \&
  Wahi]{Subramanian2010a}
{\sc \au{Subramanian, Priya}, \au{Mariappan, Sathesh}, \au{Sujith, R.~I.} \&
  \au{Wahi, Pankaj}} \yr{2011}  \at{{Bifurcation analysis of thermoacoustic
  instability in a horizontal Rijke tube}}.  \jt{International Journal of Spray
  and Combustion Dynamics}  \bvol{2}~(4),  \pg{325--355}.

\bibitem[Subramanian \& Sujith(2011)]{Subramanian2011}
{\sc \au{Subramanian, P.} \& \au{Sujith, R.~I.}} \yr{2011}  \at{{Non-normality
  and internal flame dynamics in premixed flame-acoustic interaction}}.
  \jt{Journal of Fluid Mechanics}  \bvol{679}~(2011),  \pg{315--342}.

\bibitem[Takeuchi {\em et~al.\/}(2011)Takeuchi, Yang, Ginelli, Radons \&
  Chat\'e]{takeuchi_2011}
{\sc \au{Takeuchi, Kazumasa~A.}, \au{Yang, Hong-liu}, \au{Ginelli, Francesco},
  \au{Radons, G\"unter} \& \au{Chat\'e, Hugues}} \yr{2011}  \at{Hyperbolic
  decoupling of tangent space and effective dimension of dissipative systems}.
  \jt{Phys. Rev. E}  \bvol{84},  \pg{046214}.

\bibitem[Trefethen(2000)]{trefethen2000spectral}
{\sc \au{Trefethen, Lloyd~N}} \yr{2000} {\em {Spectral methods in MATLAB}\/}, ,
   \vol{vol.~10}.  \publ{Siam}.

\bibitem[Trevisan \& Pancotti(1998)]{Trevisan1998}
{\sc \au{Trevisan, Anna} \& \au{Pancotti, Francesco}} \yr{1998}  \at{Periodic
  orbits, lyapunov vectors, and singular vectors in the lorenz system}.
  \jt{Journal of the Atmospheric Sciences}  \bvol{55}~(3),  \pg{390--398},
  \arxiv{arXiv:
  https://doi.org/10.1175/1520-0469(1998)055<0390:POLVAS>2.0.CO;2}.

\bibitem[Tyagi {\em et~al.\/}(2007)Tyagi, Jamadar \& Chakravarthy]{Tyagi2007b}
{\sc \au{Tyagi, M.}, \au{Jamadar, N.} \& \au{Chakravarthy, S.}} \yr{2007}
  \at{{Oscillatory response of an idealized two-dimensional diffusion flame:
  Analytical and numerical study}}.  \jt{Combustion and Flame}  \bvol{149}~(3),
   \pg{271--285}.

\bibitem[Wang(2013)]{wang_2013}
{\sc \au{Wang, Qiqi}} \yr{2013}  \at{Forward and adjoint sensitivity
  computation of chaotic dynamical systems}.  \jt{Journal of Computational
  Physics}  \bvol{235},  \pg{1 -- 13}.

\bibitem[Wang {\em et~al.\/}(2014)Wang, Hu \& Blonigan]{wang_2014}
{\sc \au{Wang, Qiqi}, \au{Hu, Rui} \& \au{Blonigan, Patrick}} \yr{2014}
  \at{Least squares shadowing sensitivity analysis of chaotic limit cycle
  oscillations}.  \jt{Journal of Computational Physics}  \bvol{267},  \pg{210
  -- 224}.

\bibitem[Waugh {\em et~al.\/}(2014)Waugh, Kashinath \& Juniper]{Waugh2014}
{\sc \au{Waugh, I.~C.}, \au{Kashinath, K.} \& \au{Juniper, M.~P.}} \yr{2014}
  \at{{Matrix-free continuation of limit cycles and their bifurcations for a
  ducted premixed flame}}.  \jt{Journal of Fluid Mechanics}  \bvol{759},
  \pg{1--27}.

\bibitem[Xu \& Paul(2016)]{Xu2016}
{\sc \au{Xu, M.} \& \au{Paul, M.~R.}} \yr{2016}  \at{{Covariant Lyapunov
  vectors of chaotic Rayleigh-Benard convection}}.  \jt{Physical Review E -
  Statistical, Nonlinear, and Soft Matter Physics}  \bvol{93}~(6),  \pg{1--12}.
  
  \bibitem[Yu, Juniper \& Magri(2019)]{Yu2019_jcp}
{\sc \au{Yu, H.}, \au{Juniper, M.~P.} \& \au{Magri, L.}} \yr{2019}  \at{{Combined State and Parameter Estimation in Level-Set Methods}}.  \jt{Journal of Computational Physics} \pg{doi:10.1016/j.jcp.2019.108950}.
  

\bibitem[Zinn \& Lores(1971)]{Zinn1971}
{\sc \au{Zinn, Ben~T.} \& \au{Lores, Manuel~E.}} \yr{1971}  \at{{Application of
  the Galerkin Method in the Solution of Non-linear Axial Combustion
  Instability Problems in Liquid Rockets}}.  \jt{Combustion Science and
  Technology}  \bvol{4}~(1),  \pg{269--278}.

\end{thebibliography}

\end{document}